\documentclass[onecolumn,noshowpacs,superscriptaddress,nobibnotes,nofootinbib,12pt,showkeys,preprintnumbers]{revtex4-1}
\UseRawInputEncoding
\usepackage{float}
\usepackage{amsmath,amssymb}
\usepackage{bm}
\usepackage{slashed}
\usepackage{braket}
\usepackage{epsfig}
\usepackage{graphicx}
\usepackage{hyperref}
\usepackage{xcolor}
\hypersetup{
    colorlinks=true,
    linkcolor=blue,
    filecolor=magenta,      
    urlcolor=blue,
    citecolor=blue
}
\newcommand{\comment}[1]{}

\begin{document}
\title{Quark production and thermalization of the quark-gluon plasma}
\author{Sergio Barrera Cabodevila}
\email{sergio.barrera.cabodevila@usc.es}
\affiliation{Instituto Galego de F\'isica de Altas Enerx\'ias IGFAE, Universidade de Santiago de Compostela,
E-15782 Galicia-Spain}

\author{Carlos A. Salgado}
\email{carlos.salgado@usc.es}
\affiliation{Instituto Galego de F\'isica de Altas Enerx\'ias IGFAE, Universidade de Santiago de Compostela,
E-15782 Galicia-Spain}
\affiliation{Axencia Galega de Innovaci\'on (GAIN), Xunta de Galicia, Galicia-Spain}

\author{Bin Wu}
\email{bin.wu@usc.es}
\affiliation{Instituto Galego de F\'isica de Altas Enerx\'ias IGFAE, Universidade de Santiago de Compostela,
E-15782 Galicia-Spain}

\begin{abstract}
We first assemble a full set of the Boltzmann Equation in Diffusion Approximation (BEDA) for studying thermalization/hydrodynamization as well as the production of massless quarks and antiquarks in out of equilibrium systems. In the BEDA, the time evolution of a generic system is characterized by the following space-time dependent quantities: the jet quenching parameter, the effective temperature, and two more for each quark flavor that describe the conversion between gluons and quarks/antiquarks via the $2\leftrightarrow2$ processes. Out of the latter two quantities, an effective net quark chemical potential is defined, which equals the net quark chemical potential after thermal equilibration. We then study thermalization and the production of three flavors of massless quarks and antiquarks in spatially homogeneous systems initially filled only with gluons. A parametric understanding of thermalization and quark production is obtained for either initially very dense or dilute systems, which are complemented by detailed numerical simulations for intermediate values of initial gluon occupancy $f_0$. For a wide range of $f_0$, the final equilibration time is determined to be about one order of magnitude longer than that in the corresponding pure gluon systems. Moreover, during the final stage of the thermalization process for $f_0\geq 10^{-4}$, gluons are found to thermalize earlier than quarks and antiquarks, undergoing the top-down thermalization.

\end{abstract}

\maketitle
\section{Introduction}

Intricate quantum many-body systems in QCD are observed to be produced in heavy-ion collisions at RHIC and the LHC. So far, a complete description of real-time dynamics of such systems is still beyond the reach of first-principles calculations in QCD. On the other hand, a promising, consistent understanding of this issue in the weak-coupling/high-energy limit is emerging (see refs.~\cite{Schlichting:2019abc, Berges:2020fwq, Gelis:2021zmx} for recent reviews). The validity of such a perturbative description is built upon the theoretical idea of parton saturation, which predicts that the partons initially released in high-energy nuclear collisions are mostly saturated gluons that carry a semi-hard, perturbative scale of the order of $Q_s$, the saturation momentum~\cite{Jalilian-Marian:1996mkd, Kovchegov:1998bi, Mueller:1999fp}.

Assuming that $Q_s$ lies in the perturbative regime, a unified method in QCD to describe the libration of partons from nuclear wave functions during a heavy-ion collision and the ensuing thermalization/hydrodynamization process is yet to be established. Alternatively, classical field simulations, which sum over tree-level diagrams up to all orders in the strong coupling, have been employed to study the earl-time dynamics in heavy-ion collisions~\cite{Berges:2013eia, Epelbaum:2013ekf}. However, the classical field approximation is known to break down before the system thermalizes as the classical theory is, in essence, non-renormalizable~\cite{Epelbaum:2014yja} and the results under this approximation eventually become dependent of the ultra-violet (UV) cutoff~\cite{Berges:2013lsa,Epelbaum:2014mfa}.  Based on the quasi-particle approximation, it has been argued that the classical field theory and the Boltzmann equation are equivalent when the gluon occupation number is higher than unity but lower than $1/\alpha_s$ with $\alpha_s$ the strong coupling~\cite{Mueller:2002gd}. Although such a transition cannot be confirmed by detailed calculations at the lowest order beyond classical diagrams~\cite{Wu:2017rry, Kovchegov:2017way}, switching from classical field simulations to the Boltzmann equation has become a common practice in the studies of pre-equilibrium dynamics~\cite{Schlichting:2019abc, Berges:2020fwq, Gelis:2021zmx}.

The Boltzmann equation remains the primary weak-coupling method in QCD for investigating thermalization/hydrodynamization. The leading-order (LO) Boltzmann equation  incorporates collision kernels for both the $2\leftrightarrow2$ processes and the $1\leftrightarrow2$ processes. Two variations of the LO Boltzmann equation have been presented in the literature~\cite{Baier:2000sb, Arnold:2002zm}. In ref.~\cite{Baier:2000sb}, the diffusion approximation~\cite{Mueller:1999pi} has been utilized to the $2\leftrightarrow2$ kernel. This variation is equivalent to that in ref.~\cite{Arnold:2002zm}, referred to as Effective Kinetic Theory (EKT), under the leading logarithmic approximation.

The $2\leftrightarrow2$ collision kernel in
diffusion approximation~\cite{Mueller:1999pi}, after being completed with quantum statistics, has been used to investigate the onset and formation of the Bose-Einstein condensate (BEC) of gluons in initially over-populated systems~\cite{Blaizot:2013lga, Blaizot:2014jna}. Including only the $2\leftrightarrow2$ kernel, the Boltzmann equation in diffusion approximation (BEDA) admits unique solutions with the gluon distribution composed of a regular distribution for momentum $p>0$ and a BEC: a macroscopic accumulation of gluons as a result of non-vanishing gluon flux at $p=0$~\cite{Blaizot:2014jna}. Such solutions describe the evolution of initially over-populated systems: the excessive number of gluons compared to that can be accommodated by the final thermal distributions eventually form a BEC. Similar solutions were known to exist generically for the Boltzmann equation of bosons when only the $2\leftrightarrow2$ processes are taken into account~\cite{Semikoz:1994zp, Semikoz:1995rd, Scardina:2014gxa, Xu:2014ega, Epelbaum:2015vxa}. When the number-changing processes are considered, different theories could behave differently. In the spin-0 scalar theory studied in ref.~\cite{Lenkiewicz:2019glw}, 
the BEC could form transiently before it vanishes during the thermalization process. Unexpected from the parametric estimates in ref.~\cite{Blaizot:2011xf}, this is, however, not the case in QCD. In pure gluon systems, the gluon flux at $p=0$ is observed to always vanish in diffusion approximation and no gluons are deposited in a BEC when the $1\leftrightarrow 2$ processes are included~\cite{Blaizot:2016iir, BarreraCabodevila:2022jhi}. Moreover, more elaborated analyses of infrared modes beyond the scope of the kinetic theory have showed no evidence for a BEC either~\cite{Kurkela:2012hp}.

 Inside a dense QCD medium, the most efficient mechanism for a high-energy parton to lose all its energy is known to be radiative energy loss driven by the Landau-Pomeranchuk-Migdal (LPM) effect~\cite{Baier:1996kr, Zakharov:1996fv, Baier:1998kq}. When the $1 \to 2$ processes due to the LPM effect dominate the evolution of a system, one intriguing phenomenon could arise: comparing to the typical time scale for the $1\to 2$ splitting, the thermalization time for soft partons produced from hard ones in the system is so brief that these soft partons establish thermal equilibration among themselves. Then, the thermal bath of soft gluons is gradually heated up by quenching hard partons until the system achieves full thermalization. This phenomenon is referred to as {\it the bottom-up thermalization}, which was originally discovered in longitudinally boost-invariant systems of gluons in ref.~\cite{Baier:2000sb} (and first confirmed by detailed numerical simulations in ref.~\cite{Kurkela:2015qoa}). In a spatially homogeneous, non-expanding system, the bottom-up thermalization also emerges as the last stage of thermalization in initially very dilute cases~\cite{Kurkela:2014tea, BarreraCabodevila:2022jhi}.

Allowing the production of quarks and antiquarks delays the thermalization process~\cite{Blaizot:2014jna, Kurkela:2018oqw, Kurkela:2018xxd, Du:2020zqg, Du:2020dvp}. In spatially homogeneous systems, the equilibration time for the case with three flavors of massless quarks and antiquarks was found to be typically about 5 to 6 times longer than that in the pure gluon case when only the $2\leftrightarrow2$ processes are considered in the BEDA~\cite{Blaizot:2014jna}.  The thermalization process is still delayed when the $1\leftrightarrow2$ processes are additionally taken into account as studied using EKT~\cite{Kurkela:2018xxd, Du:2020dvp}. Moreover, gluons are found to achieve kinetic equilibrium among themselves before quarks and antiquarks approach thermal equilibrium~\cite{Kurkela:2018xxd}. In longitudinally boost-invariant cases, the delay due to quark production also occurs, and the system is observed to hydrodynamize before it achieves chemical equilibration~\cite{Kurkela:2018oqw, Kurkela:2018xxd, Du:2020zqg, Du:2020dvp}.

In this paper, we present, for the first time, a full set of BEDA with both the $2\leftrightarrow2$ and $1\leftrightarrow2$ kernels for gluons and massless quarks and antiquarks, and carry out a comprehensive study of thermalization and quark production in spatially homogeneous systems by solving the BEDA. In our studies, we carry out parametrical estimates and obtain numerical solutions of the full theory as well as some analytic results under certain approximations. With this, we obtain a more detailed understanding of both thermalization and quark production in spatially homogeneous systems initially occupied by gluons, as well as the fact that top-down thermalization of gluons always appear due to the slow production rate of quarks and antiquarks if the plasma is not extremely under-occupied initially. Below, we outline the structure of this paper.

In sec.~\ref{sec:BEDA}, we incorporate the $1\leftrightarrow2$ kernel using the LPM splitting rates~\cite{Baier:1996kr, Zakharov:1996fv, Baier:1998kq, Arnold:2008zu} into the BEDA while only the $2\leftrightarrow2$ kernel for gluons and $N_f$ flavors of massless quarks and antiquarks were included in ref.~\cite{Blaizot:2014jna}. In this way, we assemble a full set of BEDA as an extension of that for pure gluon systems in ref.~\cite{Baier:2000sb}. Since this paper focuses on investigating spatially homogeneous systems, we streamline the BEDA for this scenario in sec.~\ref{sec:BEDAhomo}. We also study the low momentum limit of the BEDA in order to examine whether one can consistently impose a boundary condition with a vanishing gluon flux at $p=0$ onto the $2\leftrightarrow2$ kernel. This extends previous discussions on the onset of a BEC in pure gluon systems~\cite{Blaizot:2011xf, Kurkela:2012hp, Blaizot:2013lga, Blaizot:2014jna, Blaizot:2016iir, BarreraCabodevila:2022jhi} to generic systems including additional $N_f$ flavors of massless quarks and antiquarks. 

By solving the BEDA, the rest part of this paper focuses on the exploration of thermalization and quark production in spatially homogeneous systems. Especially, we investigate whether and to what extent quark production would modify each stage of thermalization in pure gluon systems as previously discussed by the authors in ref.~\cite{ BarreraCabodevila:2022jhi}. For this purpose, the initial gluon distribution is chosen to be the same as those in~\cite{Blaizot:2013lga, Blaizot:2014jna, BarreraCabodevila:2022jhi} while the distributions for $N_f=3$ flavors of quarks and antiquarks vanish initially, as detailed in sec.~\ref{sec:initFinal}. In this case, the time evolution of the system is characterized solely by the jet quenching parameter $\hat{q}$ and the screening mass squared $m_D^2$ or, equivalently, the effective temperature $T_*$. This subsection also covers the corresponding thermal equilibrium states, and our definition of the equilibration time. 
 
In sec.~\ref{sec:parametricDilute}, we carry out parametric estimates for the systems initially composed only of gluons typically carrying a hard momentum $Q$ with an occupation number $f_0\ll 1$. A similar analysis is carried out in sec.~\ref{sec:overPara} for initially over-populated systems of gluons with $f_0\gg 1$. In neither case do we observe any qualitative difference in the thermalization process when compared with the pure gluon scenario~\cite{Baier:2000sb, Kurkela:2014tea, Kurkela:2015qoa, Kurkela:2018oqw, Kurkela:2018xxd, Du:2020zqg, Du:2020dvp, BarreraCabodevila:2022jhi, Schlichting:2019abc}. Complementary quantitative studies are provided in sec.~\ref{sec:underQuan} for initially under-populated systems with $f_0=10^{-2}, 10^{-4}$ and $10^{-6}$ and  in sec.~\ref{sec:overQuan} for initially over-populated systems with $f_0=1, 10$ and $100$. 

Besides, the $1\to2$ splitting rates used in our studies are documented in Appendix~\ref{sec:1to2}, and the numeric methods for our GPU and CPU simulations are outlined in Appendix~\ref{sec:numerics}. The code is available in~\cite{Barrera_Cabodevila_hBEDA_2023}.

\section{The QCD Boltzmann equation in diffusion approximation}
In this section, we assemble a full set of the Boltzmann equation in diffusion approximation (BEDA) for gluons and $N_f$ flavors of massless quarks and antiquarks as an extension of that for pure gluon systems~\cite{Baier:2000sb, BarreraCabodevila:2022jhi}.

\subsection{The BEDA for gluons, quarks and antiquarks}
\label{sec:BEDA}
The leading-order QCD Boltzmann equation includes not only the $2\leftrightarrow2$ processes but also the medium-induced $1\leftrightarrow2$ splitting~\cite{Baier:2000sb, Arnold:2002zm}
\begin{align}
\left(\partial_t + {\bm v}\cdot \nabla_{\bm x}\right) f^a =&C^a_{2\leftrightarrow2} + C^a_{1\leftrightarrow2},
\label{eq:Boltzmann}
\end{align}
where $a=g,q^i$ and $\bar{q}^i$ respectively standing for gluons, quarks and antiquarks of the $i^{\text{th}}$ flavor. The distributions are so normalized that the entropy density, number density, and energy density for gluons, quarks and antiquarks are respectively given by
\begin{subequations}
\begin{align}
    &s = s_g + s_{q} + s_{\bar{q}},\\
    &n=2 N_c \int\frac{d^3p}{(2\pi)^3}\bigg[ 2C_F f^g + \sum\limits_{i=1}^{N_f}(f^{q_i}+f^{\bar{q}_i}) \bigg] \equiv n_g + n_q + n_{\bar{q}},\label{eq:n}\\
&\epsilon=2 N_c \int\frac{d^3p}{(2\pi)^3}p\left[ 2C_F f^g + \sum\limits_{i=1}^{N_f}(f^{q_i}+f^{\bar{q}_i}) \right] \equiv \epsilon_g + \epsilon_q + \epsilon_{\bar{q}},\label{eq:e}
\end{align}
\end{subequations}
where the number of colors $N_c=3$, $C_F=(N_c^2-1)/(2N_c)$, and the entropy densities of gluons, quarks and antiquarks are respectively defined as
\begin{align}
    &s_{g}\equiv - 4 N_c C_F\int \frac{d^3p}{(2\pi)^3}\left[ f^g \log f^g - (1+f^g)\log(1+f^g) \right],\\
    &s_{q}\equiv - 2 N_c \sum\limits_{i=1}^{N_f}\int \frac{d^3p}{(2\pi)^3}\left[  f^{q^i} \log f^{q^i}  + (1- f^{q^i})\log(1- f^{q^i})\right],\qquad s_{\bar{q}}=s_{q}\bigg|_{f^{q^i}\to f^{\bar{q}^i}}\,.
\end{align}

In the diffusion approximation the $2\leftrightarrow2$ kernel takes the form~\cite{Mueller:1999pi, Hong:2010at, Blaizot:2013lga, Blaizot:2014jna}
\begin{align}\label{eq:C2to2}
C^a_{2\leftrightarrow2}=\frac{1}{4}\hat{q}_a(t)\nabla_{{\bm p}} \cdot\left[  \nabla_{{\bm p}}f^a  + \frac{{\bm v}}{T^*(t)}f^a(1+\epsilon_a f^a)\right]+\mathcal{S}_a,
\end{align}
where $\epsilon_a=1$ for bosons and  $\epsilon_a=-1$ for fermions. Here, the jet quenching parameter~\cite{Baier:1996sk} for species $a$ is defined as $\hat{q}_a = C_a \hat{\bar{q}}$, and
\begin{align}
    \hat{\bar{q}} \equiv 4\pi \alpha_s^2\mathcal{L}\int\frac{d^3\bm p}{(2\pi)^3}\bigg\{ 2 N_c
    f^g\left(1+f^g\right)+\sum\limits_{i=1}^{N_f}[f^{q^i}(1-f^{q^i})+ f^{\bar{q}^i}(1-f^{\bar{q}^i})]\bigg\},
\end{align}
where $C_a=\lbrace C_A=N_c, C_F \rbrace$ respectively for $g$ (adjoint) and $q^i/\bar{q}^i$ (fundamental), $\mathcal{L}\equiv  \ln\frac{\langle p_t^2\rangle}{m_D^2}$, and the typical transverse momentum broadening $\langle p_t^2\rangle = \frac{2}{3}\langle p^2\rangle$ for isotropic systems. The effective temperature is given by~\cite{Blaizot:2013lga}
\begin{align}\label{eq:Ts}
    T_*(t)\equiv\frac{\hat{q}_A}{\alpha_s N_c \mathcal{L} m_D^2},
\end{align}
where the screening mass squared is defined as
\begin{align}
    m_D^2 = m_{D, g}^2+  m_{D, q}^2 + m_{D, \bar{q}}^2\equiv 8\pi\alpha_s\int\frac{d^3 \bm{p}}{(2\pi)^3}\frac{1}{|\bm{p}|}\bigg[2N_c f^g + \sum\limits_{i=1}^{N_f}(f^{q^i} + f^{\bar{q}^i})\bigg]
\end{align}
with $m_{D, a}^2$ standing for the contribution from parton $a$. Here, $m_D^2$ reduces exactly to the Debye mass squared at $O(\alpha_s)$ for a QGP in thermal equilibrium~\cite{Bellac:2011kqa}. It is given by two times the effective mass squared $m^2_{\text{eff,g}}$ in ref.~\cite{Arnold:2002zm}.

By extending the source terms in ref.~\cite{Blaizot:2014jna}, one has, for each quark flavor $i$,
\begin{align}
   &\mathcal{S}_{q^i} =\frac{2\pi\alpha_s^2  C_F^2 \mathcal{L}}{ p}\bigg[\mathcal{I}^i_c f^g (1 - f^{q^i} ) - \bar{\mathcal{I}}^i_c f^{q^i} ( 1 + f^g ) \bigg],\notag\\
   &\mathcal{S}_{\bar{q}^i} =\frac{2\pi\alpha_s^2  C_F^2 \mathcal{L}}{ p}\bigg[\bar{\mathcal{I}}^i_c f^g (1 - f^{\bar{q}^i} ) - \mathcal{I}^i_c f^{\bar{q}^i} ( 1 + f^g ) \bigg],\notag\\
   &\mathcal{S}_g=-\frac{1}{2C_F}\sum\limits_{i=1}^{N_f}({\mathcal{S}_{q^i}+\mathcal{S}_{\bar{q}^i}}),
   \label{eq:drag_kern}
\end{align}
where one needs to define two more parameters for each quark flavor:
\begin{align}\label{eq:IcIcb}
    &\mathcal{I}^i_c=\int \frac{d^3p}{(2\pi)^3} \frac{1}{p} [ f^g + f^{q^i} + f^g( f^{q^i} -f^{\bar{q}^i})],\notag\\  &\bar{\mathcal{I}}^i_c=\int \frac{d^3p}{(2\pi)^3} \frac{1}{p} [ f^g + f^{\bar{q}^i} + f^g( f^{\bar{q}^i} -f^{q^i})].
\end{align}
In terms of these two quantities, we define an effective net quark chemical potential for the $i^{\text{th}}$ flavor
\begin{align}\label{eq:mus}
    {\mu_*^i}={T_*}\ln\frac{\mathcal{I}_c^i}{\mathcal{\bar{I}}^i_c},
\end{align}
which reduces to the corresponding net quark potential in thermal equilibrium, denoted by $\mu^i$, as a result of the conservation of the net quark number. Besides, the $2\leftrightarrow2$ kernel also conserves the total energy and the total parton number.\footnote{
Without the $1\leftrightarrow2$ processes, one may need to take into account the number of the BEC for initially over-populated systems~\cite{Blaizot:2014jna} in order to preserve this conservation.
} 

The $1\leftrightarrow2$ processes are described by
\begin{align}
    C^a_{1\leftrightarrow2} = &\int_0^1 dx\sum\limits_{b,c}\bigg[\frac{1}{x^3}\frac{\nu_c}{\nu_a}C^c_{ab}({\bm p}/{x};{\bm p},{\bm p}(1-x)/x)-\frac{1}{2}C^a_{bc}(\bm p;x{\bm p},(1-x){\bm p})\bigg],
    \label{eq:inel_kern}
\end{align}
where $\nu_a$ is the number of spin times color degrees of freedom for parton $a$: $\nu_a=2 N_c$ for $q$ or $\bar{q}$ and $\nu_a=2 (N_c^2-1)$ for $g$, and
\begin{align}
    C^{a}_{bc}(\bm p;x{\bm p},(1-x){\bm p})\equiv\frac{dI_{a\to bc}(p)}{dxdt}\mathcal{F}^a_{bc}(\bm p;x{\bm p},(1-x){\bm p})
    \label{eq:inel_kern2}
\end{align}
with
\begin{align}
    \mathcal{F}^a_{bc}(\bm p;{\bm l},{\bm k})= f^a_{\bm p}(1+\epsilon_b f^b_{\bm l})(1+\epsilon_c f^c_{\bm k})-f^b_{\bm l}f^c_{\bm k}(1+\epsilon_a f^a_{\bm p}).
    \label{eq:inel_kern3}
\end{align}
Here, $x$ is the momentum fraction carried by particle $b$ for the process $a\to\,bc$, ${\bm l}\equiv x {\bf p}$ and ${\bm k}\equiv (1-x){\bf p}$. Following ref.~\cite{Baier:2000sb}, the splitting rates in the deep LPM regime~\cite{Baier:1998kq, Arnold:2008zu} (see also Appendix~\ref{sec:1to2}) are used in this paper:
\begin{align}
   &\frac{d I_{g\to g g}(p)}{dxdt} = \frac{\alpha_s N_c}{\pi}\frac{(1-x+x^2)^{5/2}}{(x-x^2)^{3/2}}\sqrt{\frac{\hat{q}_A}{p}},\notag\\
   &\frac{d I_{g\to q\bar{q}}(p)}{dxdt}= \frac{\alpha_s}{4\pi}[x^2+(1-x)^2]\bigg[\frac{\frac{C_F}{C_A}-x(1-x)}{x(1-x)}\bigg]^{\frac{1}{2}}
   \sqrt{\frac{\hat{q}_A}{p}},\notag\\
   &\frac{d I_{q\to g q}(p)}{dxdt} =\frac{d I_{\bar{q}\to g \bar{q}}(p)}{dxdt}= \frac{\alpha_s C_F}{2\pi}\frac{1 + (1-x)^2}{x}\bigg[\frac{1-x+\frac{C_F}{C_A}x^2}{x(1-x)}\bigg]^{\frac{1}{2}}
   \sqrt{\frac{\hat{q}_A}{p}},\notag\\
   &\frac{d I_{q\to qg}(p)}{dxdt} =\frac{d I_{\bar{q}\to \bar{q}g}(p)}{dxdt}= \frac{\alpha_s C_F}{2\pi}\frac{1 + x^2}{1-x}\bigg[\frac{x+\frac{C_F}{C_A}(1-x)^2}{x(1-x)}\bigg]^{\frac{1}{2}}
   \sqrt{\frac{\hat{q}_A}{p}},
   \label{eq:splitting_rates}
\end{align}
where $q$ and $\bar{q}$ are understood to be of the same flavor, and the quark flavor index has been dropped as the splitting rates are the same for each flavor of massless quarks and antiquarks. 

Explicitly, one has
\begin{align}
    C^g_{1\leftrightarrow2}=&\int_0^1\frac{dx}{x^3}\bigg[C^g_{gg}+\frac{1}{2C_F}\sum\limits_{i=1}^{N_f}(C^{q^i}_{gq^i}+C^{\bar{q}^i}_{g\bar{q}^i})\bigg]({\bm p}/{x};{\bm p},{\bm p}(1-x)/x)\notag\\
    &-\int_0^1dx\bigg[\frac{1}{2}C^g_{gg}+\sum\limits_{i=1}^{N_f} C^g_{q^i\bar{q}^i}\bigg](\bm p;x{\bm p},(1-x){\bm p}),\\
    C^{q^i}_{1\leftrightarrow2}=&\int_0^1 dx\bigg[\frac{1}{x^3}\bigg(C^{q^i}_{q^ig}+ 2C_F C^g_{q^i \bar{q}^i}\bigg)({\bm p}/{x};{\bm p},{\bm p}(1-x)/x)-C^{q^i}_{q^ig}(\bm p;x{\bm p},(1-x){\bm p})\bigg],\\
    C^{\bar{q}^i}_{1\leftrightarrow2}=&\int_0^1 dx\bigg[\frac{1}{x^3}\bigg(C^{\bar{q}^i}_{\bar{q}^i g}+ 2C_F C^g_{\bar{q}^i q^i}\bigg)({\bm p}/{x};{\bm p},{\bm p}(1-x)/x)-C^{\bar{q}^i}_{\bar{q}^ig}(\bm p;x{\bm p},(1-x){\bm p})\bigg].
\end{align}
It is straightforward to check that the $1\leftrightarrow2$ kernel above conserves the total energy and the net quark number for each flavor. As a result, the BEDA with both the $2\leftrightarrow2$ and $1\leftrightarrow2$ kernels admits the following thermal fixed-point solutions:
\begin{align}\label{eq:fsThermal}
    f^g_{eq}=\frac{1}{e^{\frac{p}{T}}-1},\qquad f^{q^i}_{eq}=\frac{1}{e^{\frac{p-\mu^i}{T}}+1},\qquad f^{\bar{q}^i}_{eq}=\frac{1}{e^{\frac{p+\mu^i}{T}}+1},
\end{align}
where $T$ and $\mu^i$ are the thermal equilibrium temperature and the thermal net quark chemical potential, respectively. In the case where the energy density and the net quark number density for flavor $i$ are conserved, $T$ and $\mu^i$ can be obtained by the conservation of these two quantities.

\subsection{The BEDA for spatially homogeneous systems}
\label{sec:BEDAhomo}
For spatially homogeneous systems, the BEDA reduces to
\begin{align}\label{boltzHomo}
\dot{f}^a =&\frac{1}{p^2}(p^2J^a)' +\mathcal{S}_a + C^a_{1\leftrightarrow2},
\end{align} 
where the overdot and the prime denote the derivatives with respect to $t$ and $p$, respectively, and we have introduced the current $J^a$ as
\begin{align}
    \frac{1}{p^2}(p^2J^a)' \equiv \frac{1}{4}\frac{\hat{q}_a}{p^2}\bigg[p^2\bigg({f^a}' + \frac{1}{T_*}f^a(1+\epsilon_a f^a)\bigg)\bigg]'.
\end{align}

As the $2\leftrightarrow2$ kernel contains second-order derivatives with respect to $p$, we need to impose two boundary conditions to solve the BEDA in eq.~(\ref{boltzHomo}), as described in detail in Appendix~\ref{sec:numerics}. One natural choice is to impose
\begin{align}
    \lim\limits_{p\to0}p^2J^a=0=\lim\limits_{p\to\infty}p^2J^a.
\end{align}
For gluons, the boundary condition $\lim\limits_{p\to0}p^2J^g=0$ may not be imposed consistently: If $f^g\propto 1/p$ at low $p$, it requires $\lim\limits_{p\to0}pf^g\to T_*$~\cite{Blaizot:2014jna}. In order to check the consistency of imposing this boundary condition, let us first investigate the low-$p$ behavior of $f^a$ below.

Following the pure gluon case~\cite{Blaizot:2016iir, BarreraCabodevila:2022jhi}, we expand the right-hand side of eq.~(\ref{boltzHomo}) around $p=0$. Here, one needs only to keep the first term on the right-hand side of eq.~(\ref{eq:inel_kern}) for the $1\leftrightarrow2$ kernel, evaluated with the splitting rates at leading order in $x$ and $\mathcal{F}$ at leading order in $p$. To be more specific, by assuming $p\ll p/x$ one has
\begin{align}
    \mathcal{F}^a_{bc}\big(\frac{\bm p}{x};{\bm p},\frac{1-x}{x}{\bm p}\big)\approx f^a_{\frac{\bm p}{x}}\big(1+
    \epsilon_c f^c_{\frac{\bm p}{x}}\big)+ f^b_{\bm p}\big(\epsilon_b f^a_{\frac{\bm p}{x}}-f^c_{\frac{\bm p}{x}}\big) + pf^b_{\bm p}{f^c_{\frac{\bm p}{x}}}' - \epsilon_c p f^a_{\frac{\bm p}{x}} {f^c_{\frac{\bm p}{x}}}',
\end{align}
where the last two terms on the right-hand side can be neglected for quarks and antiquarks while only the last term is subleading for gluons as $f^g\propto 1/p$. In this way, one has
\begin{align}\label{eq:Clowp}
   C^g_{1\leftrightarrow2}&\approx\frac{\alpha_s }{2\pi}\sqrt{\frac{\hat{q}_A}{p}}\int\frac{dx}{x^4}\bigg[2N_c\mathcal{F}^g_{gg}+\sum\limits_{i=1}^{N_f}\big(\mathcal{F}^{q^i}_{gq^i}+\mathcal{F}^{\bar{q}^i}_{g\bar{q}^i}\big)\bigg]\bigg(\frac{\bm p}{x};{\bm p},\frac{1-x}{x}{\bm p}\bigg)\notag\\
   &\approx\frac{\alpha_s N_c}{\pi}I_a\sqrt{\frac{\hat{q}_A}{p}}\frac{1}{p^3}\bigg(1-\frac{pf^{g}}{T_*}\bigg),\notag\\
  C^{q^i}_{1\leftrightarrow2}&\approx\frac{\alpha_s C_F}{2\pi}\sqrt{\frac{\hat{q}_F}{p}}\int\frac{dx}{x^3}\bigg[\mathcal{F}^{q^i}_{q^ig}+\mathcal{F}^{g}_{q^i\bar{q}^i}\bigg]\bigg(\frac{\bm p}{x};{\bm p},\frac{1-x}{x}{\bm p}\bigg)\notag\\
  &\approx{\alpha_s C_F}{\pi}(\mathcal{I}_c^i+\bar{\mathcal{I}}_c^i)\sqrt{\frac{\hat{q}_F}{p}}\frac{1}{p^2}\left(\frac{1}{e^{-\frac{\mu^i_*}{T_*}}+1}-f^{q^i}\right),\notag\\
  C^{\bar{q}^i}_{1\leftrightarrow2}&\approx{\alpha_s C_F}{\pi}(\mathcal{I}_c^i+\bar{\mathcal{I}}_c^i)\sqrt{\frac{\hat{q}_F}{p}}\frac{1}{p^2}\left(\frac{1}{e^{\frac{\mu^i_*}{T_*}}+1}-f^{\bar{q}^i}\right)
\end{align}
with $\mu_*^i$ defined in terms of $\mathcal{I}_c^i$ and $\bar{\mathcal{I}}_c^i$ in eq.~(\ref{eq:mus}), and $I_a \equiv\hat{q}_A/({\frac{4\alpha_s^2 N_c^2}{\pi} \mathcal{L}})$. Accordingly, in the limit $p\to0$, the distribution functions take the form
\begin{align}
    pf^g\to T_*,\qquad f^{q^i}\to \frac{1}{e^{-\frac{\mu^i_*}{T_*}}+1},\qquad  f^{\bar{q}^i}\to \frac{1}{e^{\frac{\mu^i_*}{T_*}}+1}.
\end{align}
That is, they are the same as thermal distributions with the temperature and the net quark chemical potential respectively given by  $T_*$ and $\mu_*^i$. As a result, one can consistently set $\lim\limits_{p\to0}p^2J^g=0$, and there is no transient formation of a BEC according to \cite{Blaizot:2014jna}.

Neglecting the time dependence of the macroscopic quantities in eq.~(\ref{eq:Clowp}), the BEDA admits the following approximate solutions
\begin{align}\label{eq:fslowp}
    p f^g &\approx T_*\bigg[1-\bigg(1-\frac{p \left.f^g\right|_{t=0}}{T_*}\bigg)e^{-\left(\frac{p_{A*}}{p}\right)^{\frac{5}{2}}}\bigg],\\
f^{q^i} &\approx \frac{1-e^{-\big(\frac{p_{F*}}{p}\big)^{\frac{5}{2}}}}{e^{-\frac{\mu^i_*}{T_*}}+1}+\left.f^{q^i}\right|_{t=0}e^{-\big(\frac{p_{F*}}{p}\big)^{\frac{5}{2}}}\notag,\\
f^{\bar{q}^i}  &\approx \frac{1-e^{-\big(\frac{p_{F*}}{p}\big)^{\frac{5}{2}}}}{e^{\frac{\mu^i_*}{T_*}}+1}+\left.f^{\bar{q}^i}\right|_{t=0}e^{-\big(\frac{p_{F*}}{p}\big)^{\frac{5}{2}}}\notag
\end{align}
with
\begin{align}
    p_{A*}\equiv (\hat{q}_A m_D^4 t^2/16)^\frac{1}{5},\qquad p_{F*}\equiv [\alpha_s C_F\pi(\mathcal{I}_c+\bar{\mathcal{I}}_c)t]^\frac{2}{5}\hat{q}_F^\frac{1}{5}.
\end{align}

\subsection{The initial-state and final-state phase-space distributions}
\label{sec:initFinal}
In the rest part of this paper, we study thermalization and the production of three flavors of massless quarks ($N_f=3$) in a spatially homogeneous system initially filled only with gluons. In this case, all the quarks and antiquarks share identical distributions. For brevity, the distribution functions for different species $f^a$ are respectively denoted by
\begin{align}
f=f^g,\qquad F=f^q,\qquad \bar{F}=f^{\bar{q}}
\end{align}
with the quark flavor index neglected.

\subsubsection{The distributions at $t=0$}

As we are interested to study how the whole evolution history in the pure gluon system, as studied in ref.~\cite{BarreraCabodevila:2022jhi}, is affected by allowing quark production, the initial gluon distribution is chosen to be the same as that inspired by saturation physics in refs. \cite{Blaizot:2013lga, Blaizot:2014jna, BarreraCabodevila:2022jhi}:
\begin{align}\label{eq:fs0}
f(t=0) = {f_0}\theta(Q-|\vec{p}|),\qquad F(t=0)=F_0=0,\qquad \bar{F}(t=0)=\bar{F}_0=0.
\end{align}
Correspondingly, the initial number density and energy density are respectively given by
\begin{align}
n(0) = \frac{N_c^2-1}{3\pi^2}{f_0}Q^3,\quad \epsilon(0) = \frac{N_c^2-1}{4\pi^2}{f_0}Q^4.
\end{align}
Since the BEDA is invariant under $F\leftrightarrow\bar{F}$, the system preserves $F=\bar{F}$ in the subsequent evolution given $F=\bar{F}$ at the initial time. Accordingly, one has $\mathcal{I}_c = \bar{\mathcal{I}}_c$ and $\mu_*^i=0$. Furthermore, for $N_f = N_c = 3$, one has
\begin{align}
  m_D^2 =  48\pi\alpha_s \mathcal{I}_c= 48\pi\alpha_s \bar{\mathcal{I}}_c.
\end{align}
That is, in this case the time evolution of the system is governed by the same two quantities: the jet quenching parameter and $m_D^2$ as in pure gluon systems.

For the initial distributions in eq.~(\ref{eq:fs0}), the approximate solutions in eq.~(\ref{eq:fslowp}) can be expressed as
\begin{align}\label{eq:fslowpB0}
    pf_{p}(t)&\approx
\left\{\begin{array}{ll}
T_*     &  \text{for $p\lesssim p_{A*}$}\\
{pf_0\theta(Q-p) + [T_*-pf_0\theta(Q-p)]\left(\frac{p_{A*}}{p}\right)^{\frac{5}{2}}} & \text{for $p\gtrsim p_{A*}$} 
\end{array}
\right.,\notag\\
F_{p}(t)&=\bar{F}_{p}(t)=\frac{1}{2}[1-e^{-\left(\frac{p_{F*}}{p}\right)^\frac{5}{2}}]\approx
\left\{\begin{array}{ll}
\frac{1}{2}     &  \text{for $p\lesssim p_{F*}$}\\
\frac{1}{2}\left(\frac{p_{F*}}{p}\right)^{\frac{5}{2}} & \text{for $p\gtrsim p_{F*}$} 
\end{array}
\right.
\end{align}
with
\begin{align}
    p_{F*}\equiv (\hat{q}_A m_
D^4 t^2/729)^\frac{1}{5}=({16}/{729})^\frac{1}{5} p_{A*}\approx0.466 p_{A*}.
\end{align}
As $p_{F*}\sim p_{A*}$, we use
\begin{align}
    p_*\equiv (\hat{q}_A m_D^4 t^2/4)^\frac{1}{5}\sim p_{F*}\sim p_{A*}
\end{align}
for the parametric estimates in the following sections, 

\subsubsection{The thermal equilibrium states and the equilibration time}

As $\mu^i=0$, one can straightforwardly use energy conservation:
\begin{align}\epsilon(0)=\epsilon_{eq} \equiv \epsilon_{eq,g} + \epsilon_{eq,q} + \epsilon_{eq,\bar{q}} = \bigg[2(N_c^2-1) +\frac{7}{4}N_c N_f+\frac{7}{4}N_c N_f\bigg]\frac{\pi^2}{30}T ^4% = \bigg[16 +\frac{21}{2} N_f\bigg]\frac{\pi^2}{30}T ^4
\end{align}
to obtain the thermal equilibrium temperature
\begin{align}
    T  = \left[ \frac{15(N_c^2-1) f_0}{4(N_c^2-1) + 7 N_c N_f} \right]^\frac{1}{4} \frac{Q}{\pi}\, = \left( \frac{60 f_0}{16 + \frac{21}{2}N_f} \right)^\frac{1}{4} \frac{Q}{\pi}\, .
    \label{eq:temperature_eq}
\end{align}
Then, by using the thermal distributions in eq.~(\ref{eq:fsThermal}), one has
\begin{align}\label{eq:nseq}
    &n_{eq} = n_{eq, g} + n_{eq, q} + n_{eq,\bar{q}} \equiv  \big[ 2(N_c^2-1) + \frac{3}{2}N_c N_f + \frac{3}{2} N_c N_f \big] \frac{\zeta (3)}{\pi^2} T^3\,,\notag\\
    &s_{eq} = s_{eq, g} + s_{eq, q} + s_{eq, \bar{q}} \equiv \big[ 2(N_c^2 - 1) + \frac{7}{4} N_c N_f + \frac{7}{4} N_c N_f \big] \frac{2 \pi^2}{45}  T^3,
\end{align}
and
\begin{align}\label{eq:qhatmD2eq}
    \hat{q}_A=\frac{4\pi}{3} \alpha_s^2 N_c\mathcal{L}\bigg(N_c + \frac{N_f}{2}\bigg)T^3,\qquad m_D^2 =  \frac{4\pi\alpha_s}{3}\bigg(N_c + \frac{N_f}{2}\bigg)T^2.
\end{align}
And one can check that they satisfy $\epsilon_{eq}=-P + Ts_{eq} + \sum\limits_{i=1}^{N_f} \mu^i (n_{eq, q^i}-n_{eq, \bar{q}^i})$ with $P=\epsilon_{eq}/3$. Note that for $N_f=3$ quarks and antiquarks together account for over half of the amount of $\epsilon, n$ and $s$ while they only contribute one third of the overall values of $\hat{q}_A$ and $m_D^2$ in thermal equilibrium.

In order to compare quantitatively the thermalization times for $N_f=0$ and $N_f=3$, we define the equilibration time $t_{eq}$ in terms of a macroscopic quantity $\mathcal{M}$ as
\begin{equation}
    \big | 1 - \frac{\mathcal{M}(t_{eq})}{\mathcal{M}_{eq}} \big | = W \, ,
    \label{eq:thermalization_condition}
\end{equation}
with $W$ a constant. Following~\cite{Blaizot:2014jna}, we choose $W = 0.05$.

\section{Thermalization in initially under-populated systems}
\label{sec:under}
In initially under-populated systems, the total parton number at the initial time, by definition, is less than that in the corresponding thermal equilibrium state. For $N_f=N_c = 3$, an initially under-populated system has $f_0 < f_{0c}$ = 0.308 for the initial distributions in eq.~(\ref{eq:fs0}).

\subsection{Parametric estimates for $f_0\ll 1$}\label{sec:parametricDilute}
In the pure gluon case, the system establishes thermal equilibrium through three distinct stages for $f_0\ll 1$~\cite{Schlichting:2019abc, BarreraCabodevila:2022jhi}. Below, we investigate whether and how the three stages are modified when quark production is not ignored.

\subsubsection*{Stage 1. Overheating in the soft sector: $0\ll\,Qt\ll \alpha_s^{-2} f_0$} 
During this stage, the properties of the system are dominated by hard gluons with $p\sim Q$. And one has
\begin{align}\label{eq:quantities0_under}
n_g\sim n_{h, g}\sim f_0 Q^3,~~m_{D}^2\sim \alpha_s f_0 Q^2,~~\hat{q}\sim \alpha_s^2 f_0 Q^3,~~T_*\sim Q,
\end{align}
where the number density of hard gluons is denoted by $n_{h,g}$ and $\hat{q}_A$ is simply denoted by $\hat{q}$ here and below. Quarks, antiquarks and soft gluons are predominantly generated by hard gluons through the $1\leftrightarrow2$ processes while, as justified below, the $2\leftrightarrow2$ processes play a parametrically negligible role in their production.

Like the pure gluon case~\cite{BarreraCabodevila:2022jhi}, soft gluons are mostly produced via the $g\leftrightarrow\;gg$ processes. From the $g\to gg$ splitting, the number density of gluons typically carrying momentum $p\ll Q$ can be estimated as
\begin{align}
    n_{s,g}(p)\sim \alpha_s \sqrt{\frac{\hat{q}t^2}{p}} n_{h, g}\sim \alpha_s^2 f_0^{\frac{3}{2}} Q^4 t \bigg(\frac{Q}{p}\bigg)^{\frac{1}{2}},
    %\alpha_s^\frac{8}{5}f_0^\frac{1}{5} (Qt)^\frac{4}{5} \sim T_* p_*^2,
\end{align}
and, equivalently,
\begin{align}\label{eq:fscaling}
    f(p)\sim \frac{n_{s,g}(p)}{p^3} \sim \alpha_s^2 f_0^{\frac{3}{2}} Q t \bigg(\frac{Q}{p}\bigg)^{\frac{7}{2}}.
\end{align}
The above estimate for $f$ starts to break down when the reverse process:  $gg\to g$ becomes equally important. According to eq.~(\ref{eq:Clowp}), this occurs when $p f(p)/T_*\sim 1$, corresponding to $p \sim p_* =  (\hat{q}_A m_D^4 t^2)^\frac{1}{5}\sim \alpha_s^{\frac{4}{5}} f_0^{\frac{3}{5}} (Qt)^{\frac{2}{5}}Q$. At $p\lesssim p_*$, the $g\to\;gg$ splitting and its reverse process balance out, leading to thermalization of soft gluons with an overheated temperature given by $T_*$. This is consistent with the approximate solution in eq.~(\ref{eq:fslowpB0}). Accordingly, the system witnesses a pronounced accumulation of soft gluons typically carrying momentum $p_s\sim p_*$ with their number density given by
\begin{align}
    n_{s,g}\equiv n_{s,g}(p_s)\sim \alpha_s \sqrt{\frac{\hat{q}t^2}{p_*}}n_{h, g}\sim\alpha_s^\frac{8}{5}f_0^\frac{6}{5} (Qt)^\frac{4}{5} Q^3\sim T_* p_*^2.
\end{align}
The $2\leftrightarrow2$ processes can be ignored in the above parametric analysis: From multiple elastic scattering, the gluons typically pick up some momentum broadening $\Delta p\sim(\hat{q}t)^{\frac{1}{2}}\sim \alpha_s  f_0^{\frac{1}{2}} (Qt)^{\frac{1}{2}} Q$, which is parametrically negligible as $\Delta p\ll p_*$ at $Qt\ll\alpha_s^{-2} f_0$.

Quarks and antiquarks can be produced via either the $g\to\;q/\bar{q}$ conversion in the $2\leftrightarrow2$ scattering or the $g\to q\bar{q}$ process. According to the source terms in eq.~(\ref{eq:drag_kern}), the $g\to\;q/\bar{q}$ conversion yields
\begin{align}\label{eq:nqelI}
    n_q^{g\to q}\sim m_D^4 t\sim \alpha_s^2 f_0^2 Q^4 t.
\end{align}
The quarks and antiquarks produced in this process typically carry hard momentum $Q$ as hard gluons dominate $m_D^2$. From the $g\to\;q\bar{q}$ splitting, the number density of produced $q$ or $\bar{q}$ carrying a typical momentum $p$ can be estimated as
\begin{align}
    n^{g\to q\bar{q}}_q(p) \sim& x\frac{d^2I_{g\to q\bar{q}}(Q)}{dxdt} t n_{h,g}\sim \alpha_s \sqrt{x\frac{\hat{q}t^2}{Q}} n_{h, g} \sim \alpha_s^2 f_0^\frac{3}{2} x^{\frac{1}{2}}Q^4 t \qquad\text{with $x\sim\frac{p}{Q}$},
\end{align}
and, equivalently,
\begin{align}
    F(p)\sim \frac{n^{g\to q\bar{q}}_q(p)}{p^3}\sim \frac{1}{2}\left(\frac{p_{*}}{p}\right)^{\frac{5}{2}}.
\end{align}
The above estimate using single branching starts to break down at $p\sim p_*$ when $F(p_*)\sim 1/(e^{-\frac{\mu^i_*}{T_*}}+1)=1/2$. For $p\lesssim p_*$, the reverse process $q\bar{q}\to g$ becomes non-negligible and its rivalry with the $g\to\;q\bar{q}$ process leads to thermalization of this sector: $F\to 1/2$, enforcing the Pauli principle. By taking into account all the typical values of $p$, one can conclude that the quarks and antiquarks produced by the $g \to q\bar{q}$ splitting are mostly hard with a number density  given by
\begin{align}\label{eq:nqPara}
    n_q^{g\to q\bar{q}}\sim \alpha_s \sqrt{\frac{\hat{q}t^2}{Q}} n_{h, g} \ll n_{s,g}\sim \alpha_s \sqrt{\frac{\hat{q}t^2}{p_*}}n_{h, g}.
\end{align}
That is, in this process a gluon tends to radiate a quark/antiquark as hard as possible while, in contrast, the softest gluon with $p\sim p_*$ is the most likely to be emitted via $g\to gg$. This difference is due to the soft divergence of the latter process, which is absent in the former. As $f_0\ll 1$, the $g\to q\bar{q}$ splitting gives the dominant contribution, leading to a linear growth of the quark/antiquark number over time:
\begin{align}
    n_q\sim  n_q^{g\to q\bar{q}}\sim \alpha_s^2 f_0^\frac{3}{2} Q^4 t.
\end{align}

We now verify the dominance of hard gluons during this stage. As the quark number and the soft gluon number are both parametrically smaller than that of hard gluons, $\hat{q}$ receives dominant contributions from hard gluons only. For $m_D^2$, soft gluons and soft quarks with typical momentum $p_s\sim p_*$ make the following contributions:
\begin{align}
    m_{Ds,g}^2\sim \alpha_s T_* p_*\sim \alpha_s^{\frac{9}{5}}f_0^{\frac{3}{5}} (Q t)^{\frac{2}{5}} Q^2,\qquad m_{Ds,q}^2\sim \alpha_s p_*^2 \sim \alpha_s^\frac{13}{5} f_0^\frac{6}{5} (Qt)^\frac{4}{5} Q^2.
\end{align}
They are both parametrically smaller than the contributions from hard gluons with $m_D^2\sim \alpha_s f_0 Q^2$ while the contribution from hard quarks $\sim\alpha_s n_q/Q\sim \alpha_s^3 f_0^\frac{3}{2} Q^3 t$ is parametrically even smaller.

\subsubsection*{Stage 2. Cooling and overcooling in the soft sector: $\alpha_s^{-2} f_0\ll\,Qt\ll \alpha_s^{-2} f_0^{-\frac{1}{3}}$}

\begin{figure}
    \centering
    \includegraphics[scale=0.7]{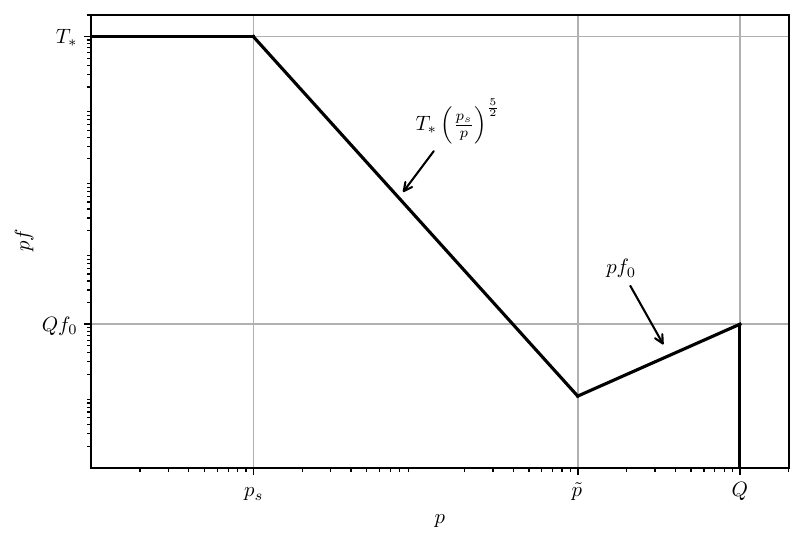}
    \caption{The qualitative feature of the gluon distribution at early times. For parametric estimates, the qualitative features of the system in Stages 1 and 2 can be derived from these three distinct portions of $f$.}
    \label{fig:pfI}
\end{figure}

During this stage, most of the partons in the system are still hard gluons, leaving both  $n_g\sim n_{h, g} \sim f_0 Q^3$ and $\hat{q}\sim\alpha_s^2 n_g\sim \alpha_s^2 f_0 Q^3$ unmodified. On the other hand, $m_D^2$ receives dominant contributions from soft gluons whose momenta are pushed up to $p_s\sim(\hat{q}t)^{\frac{1}{2}}$ by multiple elastic scattering. The number density of soft gluons can be estimated from the $g\to gg$ splitting:
\begin{align}
       n_{s, g}\sim \alpha_s\sqrt{\frac{\hat{q}t^2}{p_s}}n_{h, g}
       \sim f_0^{\frac{1}{2}} (Qp_s)^\frac{3}{2}
       \sim
      \alpha_s^{\frac{3}{2}}f_0^{\frac{5}{4}}(Qt)^{\frac{3}{4}} Q^3.
\end{align}
And their contribution to $m_D^2$ takes the form
\begin{align}
    m_D^2 \sim \alpha_s\frac{n_{s, g}}{p_s}\sim \hat{q}^{\frac{1}{2}} p_s^{\frac{1}{2}}\sim \alpha_s^{\frac{3}{2}}f_0^{\frac{3}{4}} (Qt)^{\frac{1}{4}} Q^2,
\end{align}
which is parametrically larger than that from hard gluons starting from $Qt\sim \alpha_s^{-2} f_0$. Accordingly, one has 
\begin{align}
T_*\sim \hat{q}/(\alpha_s m_D^2) \sim \alpha_s^{-\frac{1}{2}}f_0^{\frac{1}{4}}(Qt)^{-\frac{1}{4}}Q.     
\end{align}

Similar to Stage 1, $f$ can be roughly divided into three distinct portions illustrated in Fig.~\ref{fig:pfI}, which are parametrically the same as pure gluon case~\cite{BarreraCabodevila:2022jhi}. In the range of $p$ dominated by the $g\to gg$ splitting, one has
\begin{align}\label{eq:fIaII}
    f(p)\sim  \alpha_s \sqrt{\frac{\hat{q}t^2}{p}} \frac{n_{h,g}}{p^3}\sim \frac{T_*}{p_s}\bigg(\frac{p_s}{p}\bigg)^{\frac{7}{2}}.
\end{align}
It holds down to $p\sim p_s\sim (\hat{q} t)^{\frac{1}{2}}$, where ${p_s f(p_s)}/{T_*} \sim 1$ and the reverse process $gg\to g$ becomes equally important. Consequently, detailed balance is established in the soft sector with $p\lesssim p_s$. At higher $p$, the estimate in eq.~(\ref{eq:fIaII}) breaks down at $p\sim \tilde{p}$ with 
\begin{align}
    \Tilde{p} = p_s \left( \frac{p_s}{T_*} f_0 \right)^{-\frac{2}{7}}\sim \alpha_s^{\frac{4}{7}} f_0^{\frac{1}{7}} (Qt)^{\frac{2}{7}}Q,
\end{align}
where the initial distribution $f_0$ starts to dominate.

The qualitative features of the quark and antiquark distributions are predominantly determined by the gluon distribution. For $Q\gtrsim p \gtrsim \tilde{p}_F\equiv\alpha_s^{\frac{4}{3}} f_0^{\frac{1}{3}}(Qt)^{\frac{2}{3}} Q$, quarks and antiquarks are mostly produced by hard gluons via the $g\to q\bar{q}$ splitting with
\begin{align}\label{eq:FI}
    F(p)\sim  \alpha_s\sqrt{\frac{\hat{q} t^2}{Q}}\bigg(\frac{p}{Q}\bigg)^\frac{1}{2} \frac{n_{h, g}}{p^3}\sim f_0^{\frac{1}{2}} \bigg(\frac{Q}{p}\bigg)^{\frac{1}{2}} \bigg(\frac{p_s}{p}\bigg)^{2} \sim\alpha_s^2 f_0^{\frac{3}{2}} \bigg(\frac{Q}{p}\bigg)^{\frac{5}{2}} Qt.
\end{align}
For $p\lesssim \tilde{p}$, the intermediate portion of $f\propto\;p^{-\frac{7}{2}}$ contributes to $F$ as well. However, its contribution is parametrically negligible for $p\gtrsim \tilde{p}_F$. This can be justified as follows. Since it is the most likely for a gluon to radiate a quark/antiquark carrying about half of its momentum, this portion of $f$ gives a contribution
\begin{align}\label{eq:FII}
    F(p)\sim \alpha_s \sqrt{\frac{\hat{q} t^2}{p}}\frac{n_g(p)}{p^3}\sim \alpha_s^2 \frac{\hat{q} t^2}{p} \frac{n_{h,g}}{p^3}\sim \frac{1}{2}\bigg(\frac{p_s}{p}\bigg)^{4}.
\end{align}
By comparing this estimate with eq.~(\ref{eq:FI}), one can see that it becomes dominant only for $p\lesssim \tilde{p}_F$ (note that one always has $p_s\lesssim\tilde{p}_F\ll \tilde{p}$ throughout this stage). And the above estimate based on single branching breaks down at $p\sim p_s$ as $F(p_s)\sim 1/2$. At $p\lesssim p_s$, the system maintains a balance between the $g\to q\bar{q}$ process and its reverse process, corresponding to $F\to 1/2$. Accordingly, unlike Stage 1, the system witnesses a pronounced accumulation of soft quarks and antiquarks with $p\sim p_s$ besides hard ones with $p\sim Q$:
\begin{align}
    n_{s,q}^{1 \leftrightarrow 2} \sim \alpha_s \sqrt{\frac{\hat{q}t^2}{p_s}} n_{s,g} \sim p_s^3,\qquad
    n_{h,q}^{1 \leftrightarrow 2} \sim\alpha_s \sqrt{\frac{\hat{q}t^2}{Q}} n_{h, g}\sim f_0^{\frac{1}{2}} Q p_s^2.
  \end{align}
Moreover, starting at $Qt \sim \alpha_s^{-2}$ with $p_s\sim f_0^{\frac{1}{2}} Q$, soft quarks and antiquarks with $p\sim p_s$ become parametrically more abundant than hard ones.

Let us then look into quark production via the $2\leftrightarrow 2$ process. From the conversion term in \eqref{eq:drag_kern}, one can obtain the quark distribution
\begin{align}
    F(p)\sim \alpha_s m_D^2 t \frac{f_p}{p}
    \sim \alpha_s m_D^2 t \frac{T_*}{p_s^2} \bigg(\frac{p_s}{p}\bigg)^{\frac{9}{2}}
    \sim \frac{1}{2}\bigg(\frac{p_s}{p}\bigg)^{\frac{9}{2}},
\end{align}
which is negligible compared to the contribution from the $g\to q\bar{q}$ process for $p\gtrsim p_s$ in eq.~(\ref{eq:FII}). On the other hand, the quark number produced via this process is given by
\begin{align}
    n_q^{2 \leftrightarrow 2} \sim m_D^4 t \sim p_s^3 \sim \alpha_s^3 f_0^\frac{3}{2} (Qt)^\frac{3}{2} Q^3,
\end{align}
which is parametrically the same as the contribution from the $g\to q\bar{q}$ process. Since the quark number density is parametrically smaller than $n_{s,g}$ for $p_s \lesssim f_0^{\frac{1}{3}} Q$ at $Qt \lesssim \alpha_s^{-2} f_0^{-\frac{1}{3}}$, the quark and antiquark production does not modify the parametric form of $\hat{q}$ or $m_D^2$.

 In summary, even though soft gluons contribute dominantly to $m_D^2$ right after $Qt\sim \alpha_s^{-2} f_0$, they do not contribute significantly to quark production until $Qt\sim \alpha_s^{-2}$. Afterwards, the soft sector of the system becomes overcooled with $T_*\lesssim T$, and soft quarks and antiquarks are dominantly produced by soft gluons via both the $g\rightarrow q \bar{q}$ and $gg \rightarrow q \bar{q}$ processes although there are less soft gluons than hard ones in the system. As a result, one has
\begin{align}
    n_{q}\sim \left\{
    \begin{array}{ll}
       \alpha_s^2 f_0^\frac{3}{2} Q^4 t & \text{at $Qt\lesssim \alpha_s^{-2}$} \\
        \alpha_s^3 f_0^\frac{3}{2} (Qt)^\frac{3}{2} Q^3 & \text{at $\alpha_s^{-2}\lesssim Qt\lesssim \alpha_s^{-2} f_0^{-\frac{1}{3}}$}
    \end{array}
    \right..
\end{align}

\subsubsection*{Stage 3. Reheating and mini-jet quenching: $\alpha_s^{-2} f_0^{-\frac{1}{3}}\ll\,Qt\ll \alpha_s^{-2} f_0^{-\frac{3}{8}}$}
At $Qt\sim \alpha_s^{-2} f_0^{-\frac{1}{3}}$, one has
\begin{align}
   p_s\sim (\hat{q} t)^{\frac{1}{2}}\sim f_0^{\frac{1}{3}} Q\sim T_*,\qquad n_{q} \sim  n_{s, g} \sim n_{h,g} \sim f_0 Q^3,\qquad m_{D,q}^2\sim m_{D,g}^2\sim \alpha_s f_0^{\frac{2}{3}} Q^2,
\end{align}
which is consistent with a thermalized QGP of temperature $T_*\sim f_0^{\frac{1}{3}} Q$. This is qualitatively different from the previous stages as both $\hat{q}$ and $m_D^2$ are now determined by soft partons. This can be further justified by the fact that
the relaxation time $t_{rel}\sim \alpha_s^{-2} T_*^{-1}$ becomes comparable with $t$, meaning that soft partons have enough time to establish thermal equilibrium among themselves. Later on, each hard parton loses energy by typically radiating one gluon  with $p_{\text{br}} \sim \alpha_s^2 \hat{q} t^2$~\cite{Baier:2000sb, Baier:2001yt}, which rapidly degrades into soft partons with $p_s\sim T_*$~\cite{Iancu:2015uja} via democratic branching~\cite{Blaizot:2013hx} within a time of order $t$. As the energy density of the QGP, sourced by hard partons, is given by $T_*^4\sim\epsilon_s\sim p_{\text{br}} n_{h,g}$, one has
\begin{align}
    T_*\sim \alpha_s^4 f_0 (Qt)^2 Q,\qquad
m_D^2\sim \alpha_s T_*^2\sim \alpha_s^9 f_0^2 (Qt)^4 Q^2,\qquad \hat{q}\sim \alpha_s^2 T_*^3\sim \alpha_s^{14} f_0^3(Qt)^6 Q^3
\end{align}
and
\begin{align}
    n_q\sim n_g \sim \alpha_s^{12} f_0^3 (Qt)^6 Q^3.
\end{align}
 In the above parametric estimate, one can safely neglect hard quarks and antiquarks as their number density $n_{h,q}\sim \alpha_s \sqrt{\frac{\hat{q}t^2}{Q}} n_{h, g} \sim \alpha_s^8 f_0^{\frac{5}{2}} (Qt)^4 Q^3$ is parametrically smaller than $n_{h,g}\sim f_0 Q^3$ for $Qt \lesssim \alpha_s^{-2} f_0^{-\frac{3}{8}}$. And the soft sector remains a thermalized QGP with a time-dependent temperature $T_*$ as the relaxation time $t_{rel}\sim 1/(\alpha_s^{2} T_*)\sim\alpha_s^{-6} f_0^{-1} (Qt)^{-2}Q^{-1}\lesssim t$ at $Qt\gtrsim \alpha_s^{-2} f_0^{-\frac{1}{3}}$.
 
 Eventually, the system thermalizes at $Qt\sim \alpha_s^{-2} f_0^{-\frac{3}{8}}$ when all the relevant quantities have the same parametric forms as those in thermal equilibrium. Therefore, in Stage 3, the system undergoes the bottom-up thermalization, parametrically the same as the final stage in the pure gluon case with~\cite{Baier:2000sb} or without~\cite{Kurkela:2014tea, BarreraCabodevila:2022jhi} longitudinal expansion. The only distinction is that the soft thermal bath is now a QGP, composed of gluons as well as quarks and antiquarks.

\subsection{Quantitative studies}
\label{sec:underQuan}

In this subsection, we carry out some detailed simulations for initially under-populated systems, focusing on features that are not revealed in the above parametric estimates. Following \cite{Blaizot:2014jna}, $\mathcal{L}$ is kept as a constant in our simulations in order to be consistent with the harmonic oscillator approximation used to derive the splitting rates in the deep LPM regime. And we show the results for $\mathcal{L}=1$ and $\alpha_s=0.1$ below.\footnote{This choice for $\alpha_s$ is arbitrary as long as it remains within the perturbative regime and satisfies $\alpha_s f_0\lesssim 1$. A change of the value for the coupling constant translates into a re-scaling of the time variable, since both collision kernels are proportional to $\alpha_s^2$.
}

\subsubsection{Initially dilute systems with $f_0=0.01$}
\label{sec:f0_0p01}

\begin{figure}
    \centering
    \includegraphics[height = 0.45\textwidth]{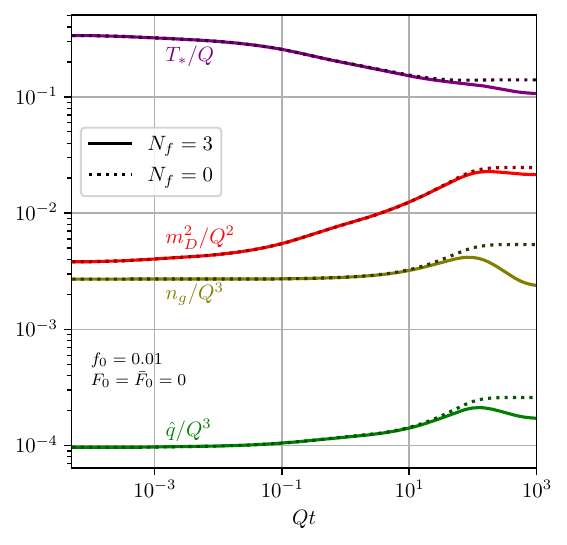}
    \includegraphics[height=0.45\textwidth]{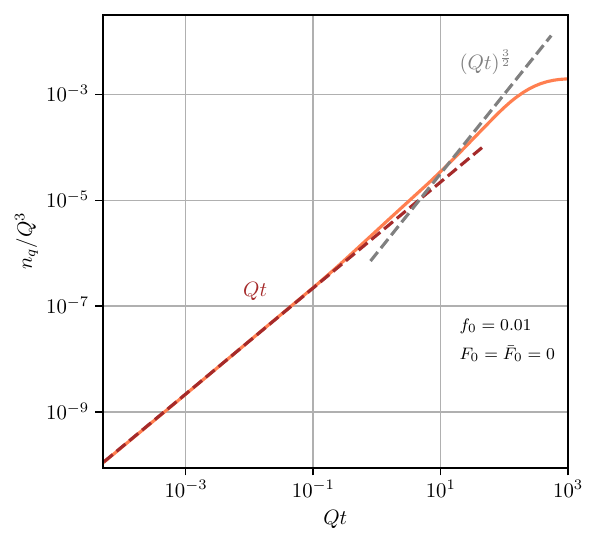}%temperature.pdf}
    \caption{The time evolution of different macroscopic quantities for $f_0=0.01$. The left panel shows the results of $T_*$, $m_D^2$, $\hat{q}$ and $n_g$ with $N_f=0$ (dotted) and $N_f=3$ (solid). The right plot shows the result of $n_q$ in comparison with the scaling laws obtained by parametric estimates.}
    \label{fig:f0_01_macs}
\end{figure}

As a direct comparison with the pure gluon case studied in ref.~\cite{BarreraCabodevila:2022jhi}, we first investigate the time evolution of the system for $f_0=0.01$. In this case, the four time scales take the following values:  $(\alpha_s^{-2} f_0, \alpha_s^{-2}, \alpha_s^{-2} f_0^{-\frac{1}{3}}, \alpha_s^{-2}f_0^{-\frac{3}{8}}) = (1, 100, 464, 562)$. Although there is no significant separation between the time scales after $Qt\sim \alpha_s^{-2}$, the three stages: overheating, cooling/overcooling, and reheating are still manifest in the time evolution of $T_*$ for $N_f = 0$ ~\cite{BarreraCabodevila:2022jhi} (see also the left panel of fig.~\ref{fig:f0_01_macs}). %In the meanwhile, the gluon number density $n_g$ always increases over time before thermalization. 
Let us examine  whether this is still the case for $N_f=3$.

Even though the system is under-populated in terms of the total parton number for $N_f=3$, there is an initial excess of gluons comparing to that in the corresponding thermal equilibrium state. Therefore, $n_g$ has to decrease to its thermal equilibrium value eventually. However, as shown in the left panel of fig.~\ref{fig:f0_01_macs}, it is observed to increase instead at early times.  Moreover, $n_g$ as well as $\hat{q}_A$, $m_D^2$ and $T_*$ are all almost indistinguishable from those in the pure gluon case before $Qt\sim 10$. That is, the early-time dynamics of the system is dominated by gluons. This is a consequence of the fact that quarks and antiquarks can only be produced slowly: $n_q$ grows linearly over $t$ up to $Qt\sim 1$ and its growth accelerates only afterwards, as shown in the right panel of fig.~\ref{fig:f0_01_macs}. Consequently, the soft sector of the system experiences almost identical overheating and cooling phases before its evolution history starts to deviate from that in the pure gluon system.

For $N_f=0$, $T_*$ undershoots the thermal temperature at the percent level and starts to increase around $Qt \sim 73$, similar to that during Stage 3 in the limit $f_0\ll 1$~\cite{BarreraCabodevila:2022jhi}, while the late-time evolution for $N_f=3$ becomes qualitatively different. At $Qt \sim 73$, one has $T_* \approx 0.130Q$, which is still above its thermal equilibrium value ($T=0.107 Q$). In the meanwhile, $\hat{q}$ reduces about 9\% compared to that for $N_f=0$ while $m_D^2$ is less sensitive to $N_f$, only dropping about 3\%. Such a trend continues later on, causing $T_*\propto \hat{q}/(\alpha_s m_D^2)$ to decrease until the system fully thermalizes. Such a qualitative difference from the dilute limit is also manifest in the fact that the growth of $n_q$ at later times, although being accelerated, is slower than the estimated scaling behavior of $(Qt)^\frac{3}{2}$, as shown in the right panel of fig.~\ref{fig:f0_01_macs}.

\begin{figure}
    \centering
    \includegraphics[width = \textwidth]{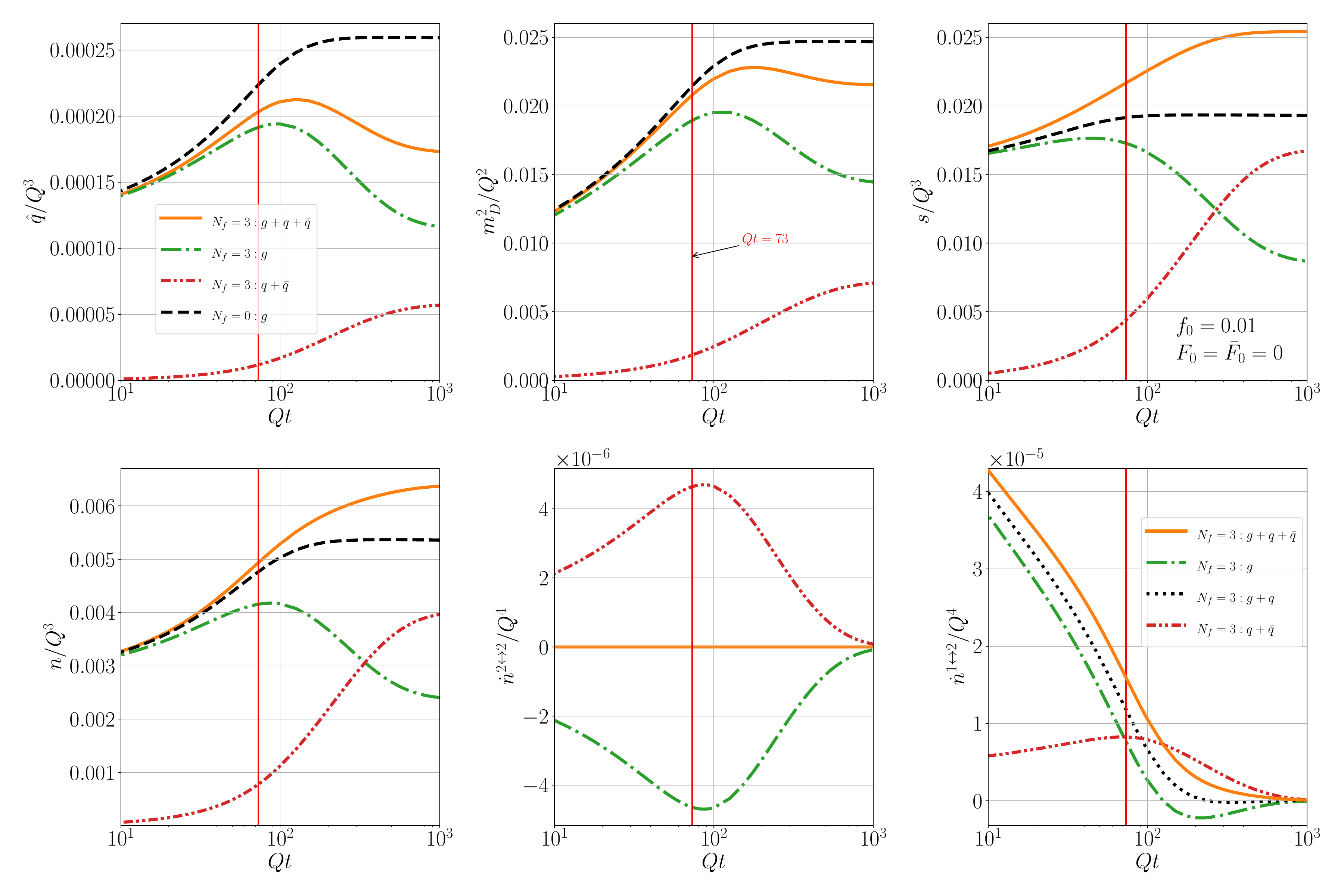}
    \caption{Contributions of gluons ($g$), quarks ($q$) and antiquarks ($\bar{q}$) to $\hat{q}$, $m_D^2$, $s$ and $n$ for $f_0=0.01$. Besides, for $N_f=3$, the time derivatives of the number densities for $g$, $q+\bar{q}$ and $g+q+\bar{q}$ via the $2\leftrightarrow2$ processes and the $1\leftrightarrow2$ processes are presented in the lower center and lower right panels, respectively. The lower right panel also shows $\dot{n}_g+\dot{n}_q=\dot{n}_g+\dot{n}_{\bar{q}}$ via the $1\leftrightarrow2$ processes.
}\label{fig:f0p01_chems}
\end{figure}

Let us delve into the details of thermalization and quark production for $N_f=3$, as further revealed by the quantities in fig.~\ref{fig:f0p01_chems}. At $Qt \sim 73$, the contributions from quarks and antiquarks to $\hat{q}$ (the upper left panel) and $m_D^2$ (the upper center panel) are both about one order of magnitude smaller than those from gluons. This means that gluons still dominate the subsequent evolution. Consequently, the system follows the trend in the pure gluon case, generating more gluons despite of the existing excess of gluons, as shown in the lower left panel. This further stokes the growth of $\hat{q}$ and $m_D^2$ and, hence, increases the likelihood of gluons splitting or converting to quarks and antiquarks in order to curb the unwanted growth of $n_g$. Accordingly, the system witnesses rapid production of quarks and antiquarks via both the $1\leftrightarrow2$ processes (the lower right panel) and the $2\leftrightarrow2$ processes (the lower center panel) around this time.

Around $Qt \sim 87$, the gluon number density starts to decrease. Shortly afterwards, the contributions from gluons to $\hat{q}$ and $m_D^2$ also start to decrease while their contribution to $s$, as depicted in the upper right panel of fig.~\ref{fig:f0p01_chems}, has already been declining. Consequently, both $\hat{q}$ and $m_D^2$ develop a peak over time as gluons always give a larger contribution than quarks and antiquarks. In contrast, both $n$ and $s$ continuously increases over time toward their thermal equilibrium values to which quarks and antiquarks contribute more significantly than gluons, as shown in eq.~(\ref{eq:nseq}). 

\begin{figure}
    \centering
    \includegraphics[height = 0.45\textwidth]{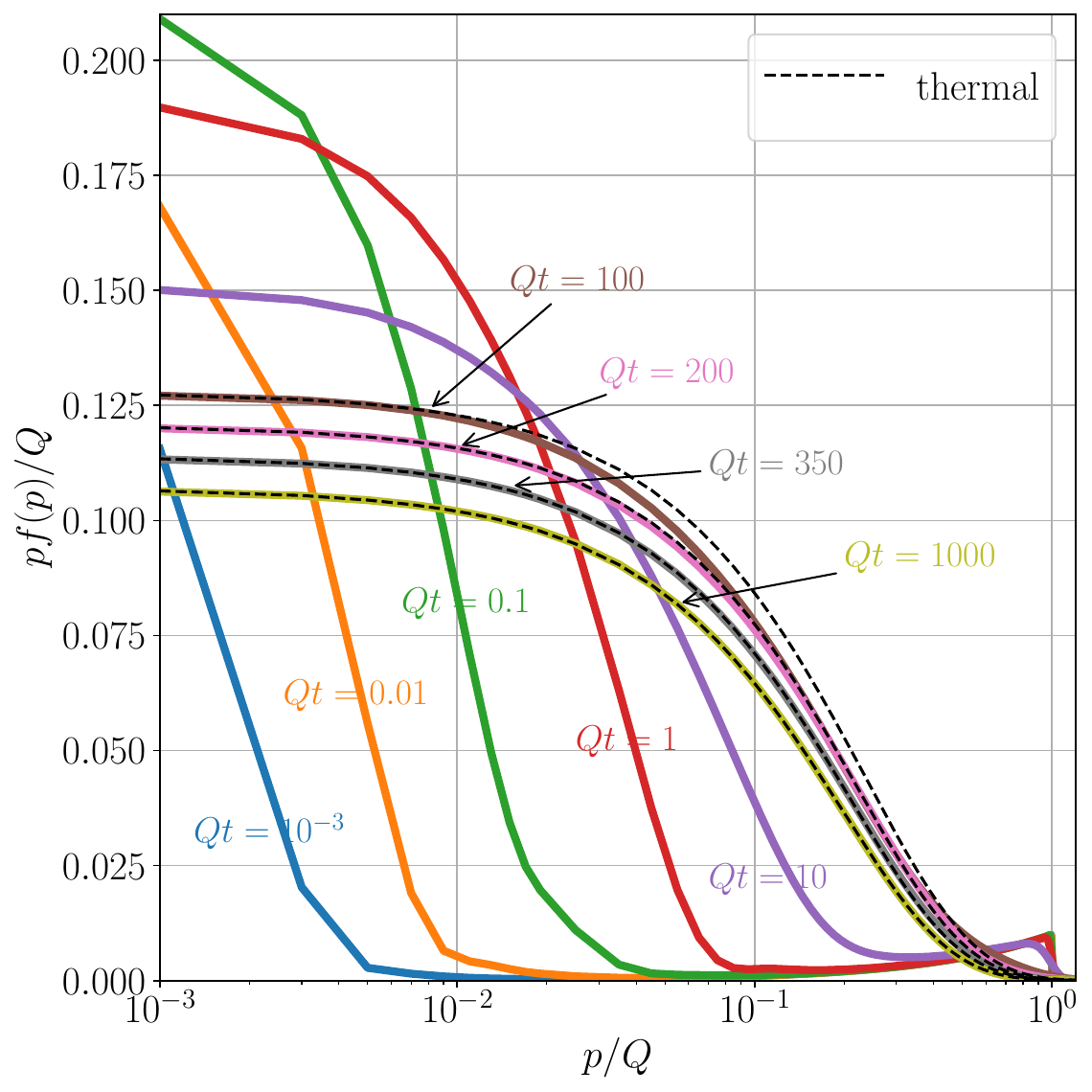}
    \includegraphics[height=0.45\textwidth]{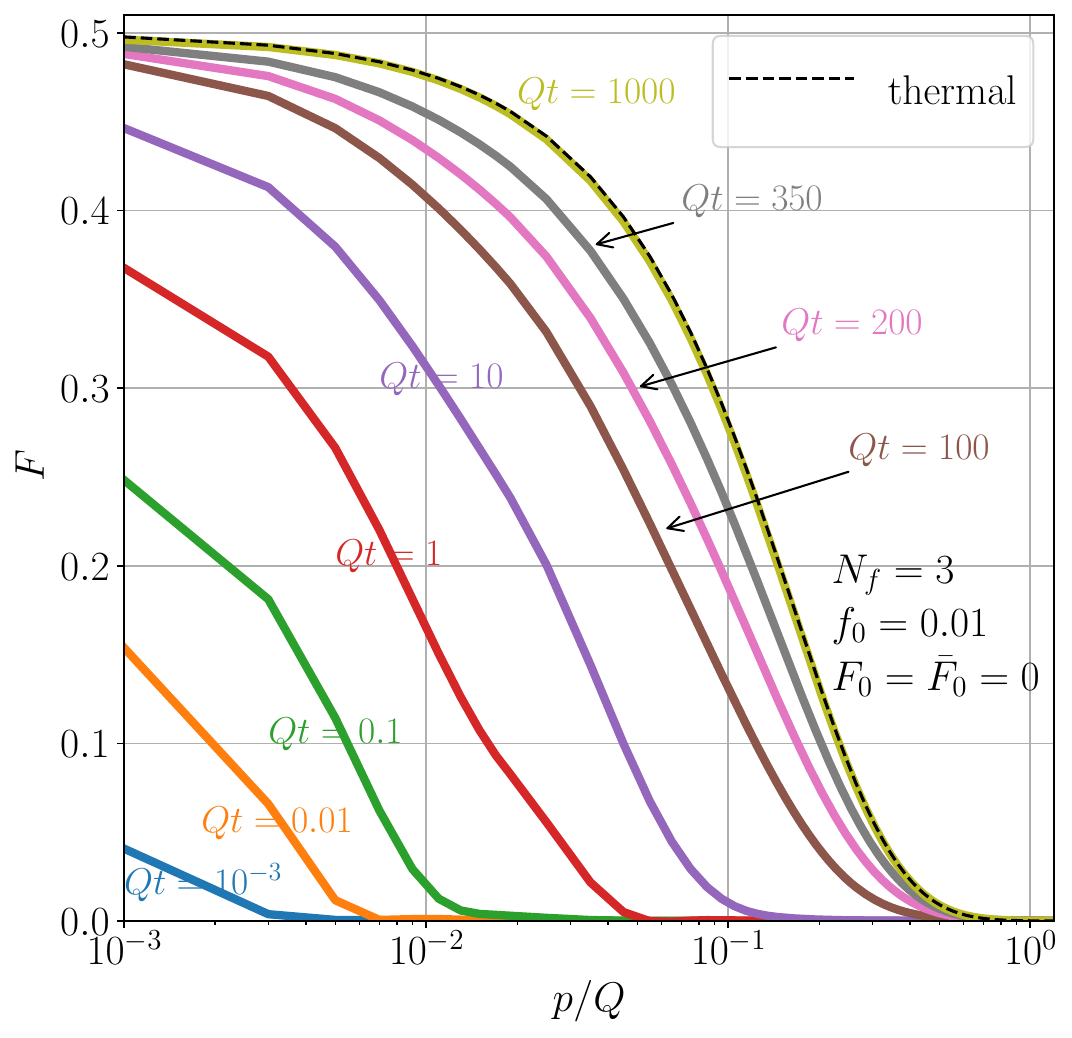}
    \caption{Gluon (left) and quark (right) distributions at different times for $f_0=0.01$. Starting around $Qt=200$, the gluon distribution can be well fitted by a Bose-Einstein distribution (thermal) with a time-dependent temperature $T_*$ up to some maximum momentum larger than $T_*$ while the quark distribution only approaches a Fermi-Dirac distribution (thermal) towards the end of the process.}
    \label{fig:f0p01_distibutions}
\end{figure}

The production of quarks and antiquarks is only partially responsible for the rapid decrease in $\dot{n}_g$ via the $1 \leftrightarrow 2$ processes shown in the lower right panel of fig.~\ref{fig:f0p01_chems}. Another contributing factor to its decrease is the increasing significance of the $gg \to g$ process, which counteracts the $g \to gg$ process primarily responsible for the earlier increase in $n_g$. Starting around $Qt=200$, one has $\dot{n}_g^{1 \leftrightarrow 2} + \dot{n}_q^{1 \leftrightarrow 2} = \dot{n}_g^{1 \leftrightarrow 2}+ \dot{n}_{\bar{q}}^{1 \leftrightarrow 2}\approx 0$. This implies that the excess of gluons is dominantly eliminated by the $g \to q/\bar{q}$ conversion in the $2 \leftrightarrow 2$ kernel and the $g \to q \bar{q}$ splitting. As for the $g \to gg$ process,  it has nearly established detailed balance with its reverse process (note that all the quantities for $N_f=0$ shown in  fig.~\ref{fig:f0p01_chems} have already reached their thermal equilibrium values). This is consistent with the observation in the left panel of fig.~\ref{fig:f0p01_distibutions}: the gluon distribution for $Qt \gtrsim 200$ can all be well fitted, up to some maximum momentum larger than $T_*$, by the Bose-Einstein distribution with a time-dependent temperature given by $T_*$. In contrast, the quark distribution, presented in the right panel of fig.~\ref{fig:f0p01_distibutions}, does not fit the Fermi-Dirac distribution (except that $F\to 1/2$ at low $p$) until the very end of the thermalization process. That is, gluons undergo the top-down thermalization, meaning that they thermalize first with a decreasing temperature $T_*$ before the system fully thermalizes. This observation is qualitatively consistent with the analysis in ref.~\cite{Kurkela:2018oqw}.

\begin{figure}
    \centering
    \includegraphics[width = 0.48\textwidth]{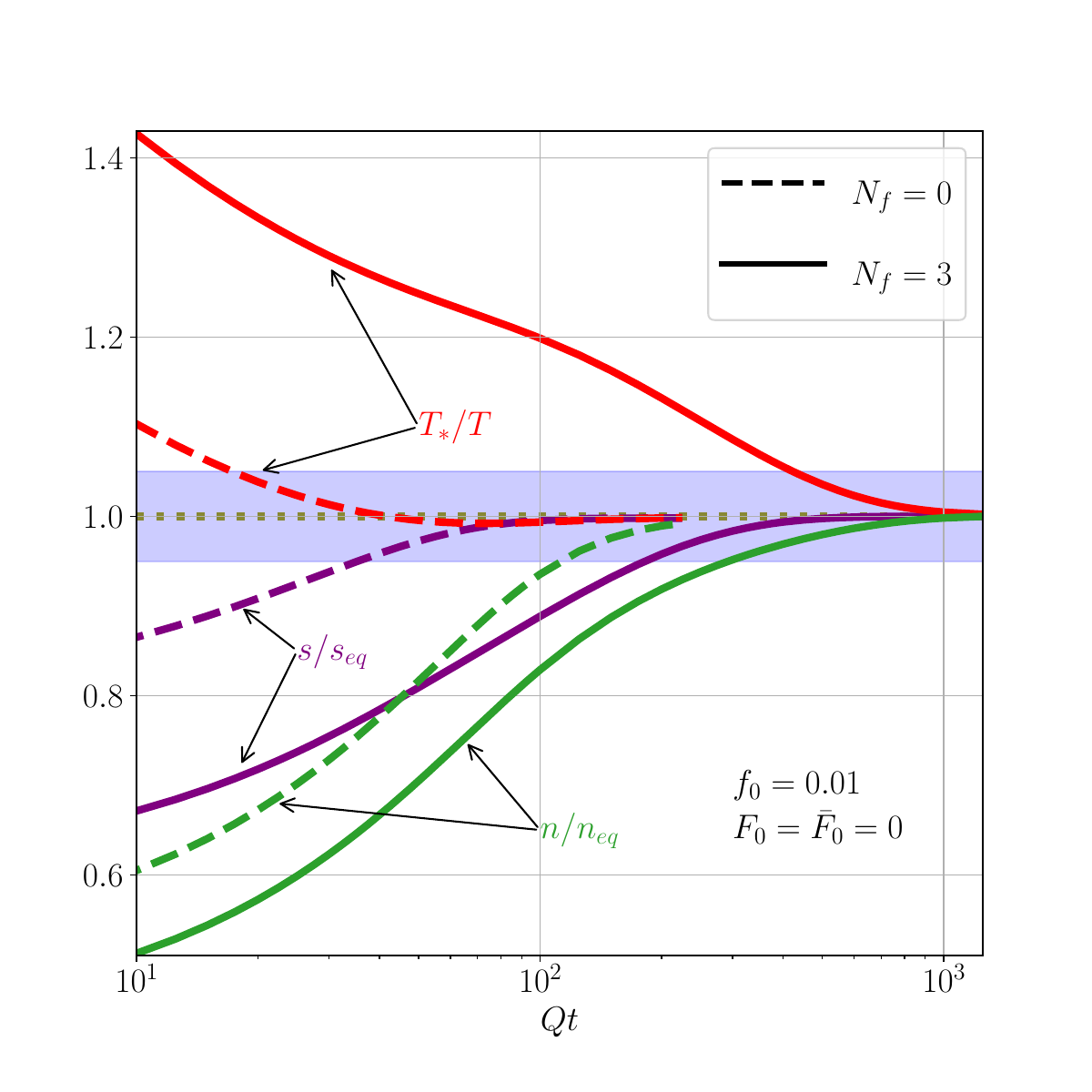}
    \includegraphics[width = 0.48\textwidth]{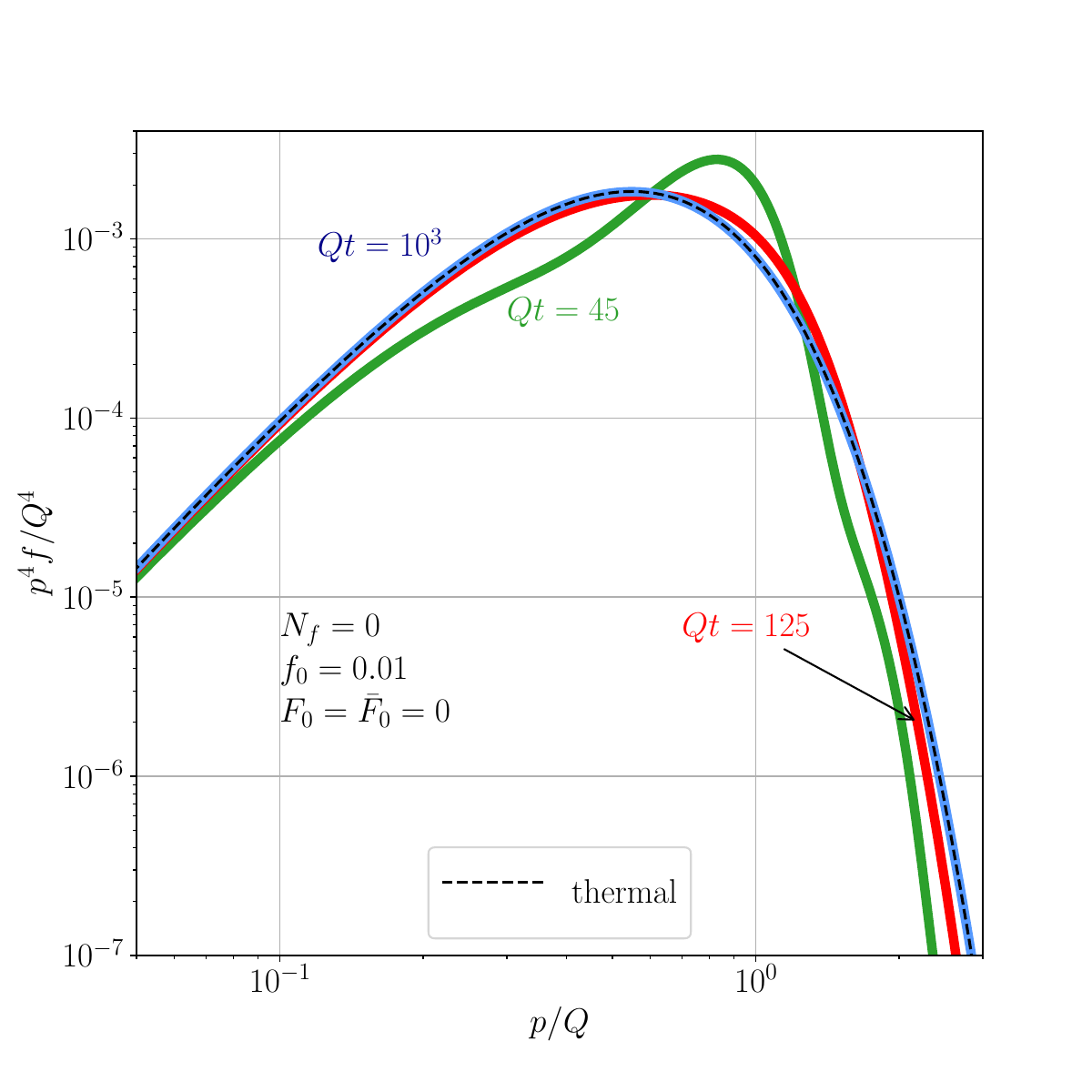}
\\              
    \includegraphics[width=0.48\textwidth]{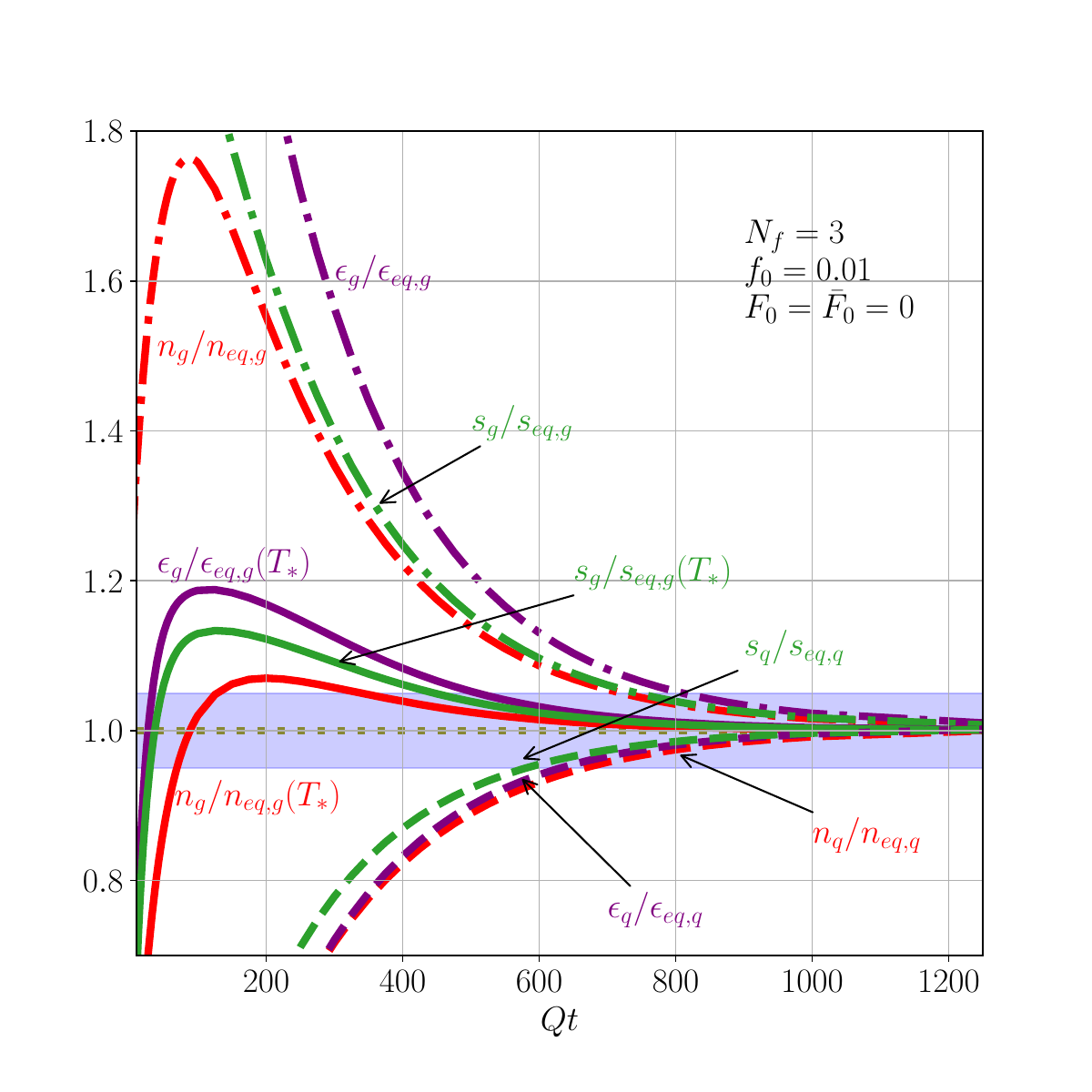}
    \includegraphics[width=0.48\textwidth]{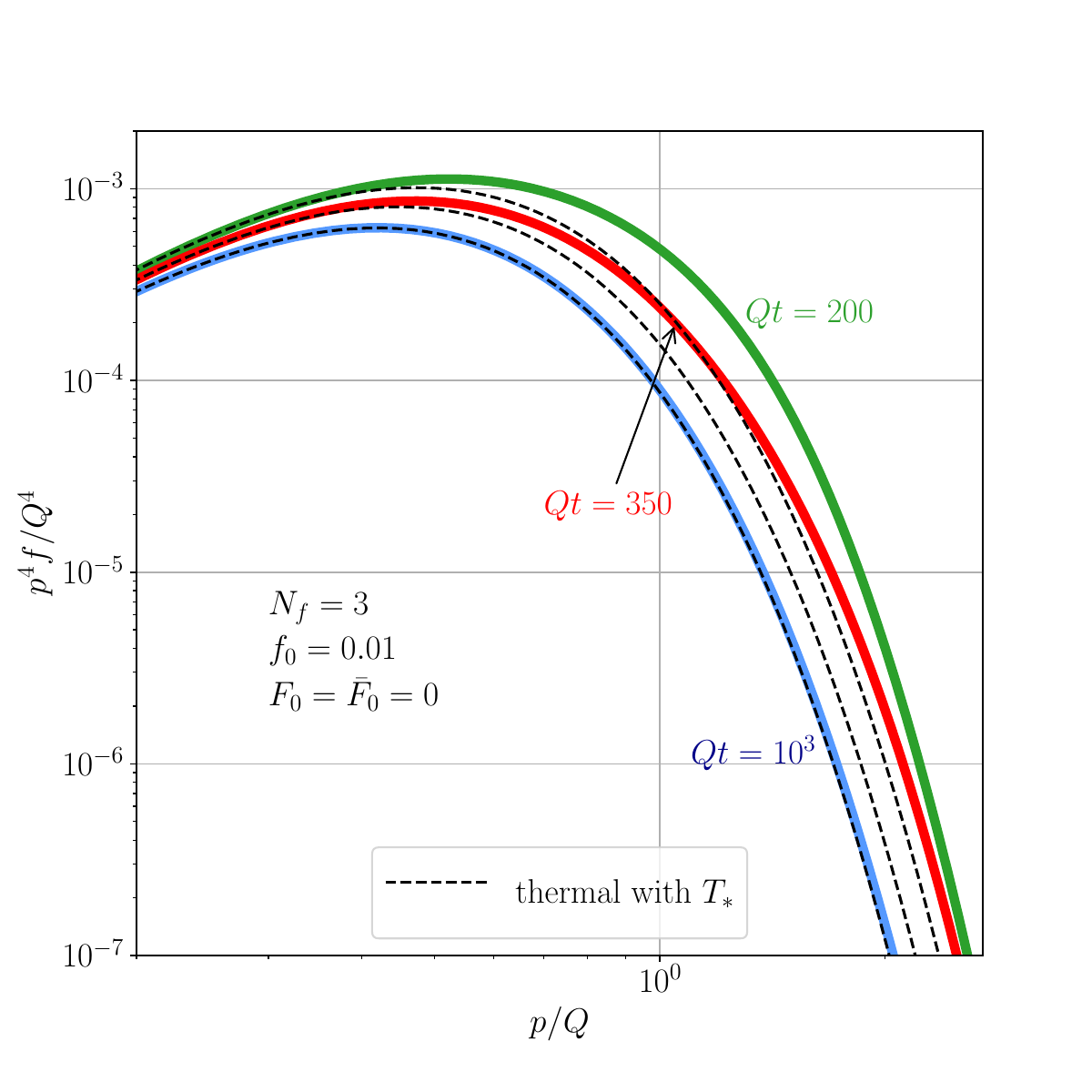}    \caption{Thermalization for $f_0=0.01$. The upper left panel shows the ratios of $T_*$, $s$ and $n$ to their final thermal equilibrium values for $N_f=0$ and $3$. The lower left panel shows the ratios of the energy density, number density and entropy density of $g$ and $q$ to their thermal equilibrium values with the temperature given by $T$ or $T_*$ (only their dependence on $T_*$ is kept explicit). The two right panels show $p^4 f$ at different times for $N_f=0$ (upper) and $N_f=3$ (lower).}
    \label{fig:f0_0p01_thermalization}
\end{figure}

Let us now discuss thermalization more quantitatively in terms of the equilibration time defined in eq.~(\ref{eq:thermalization_condition}). For $N_f=0$, it takes $t^{N_f=0}_{eq} \approx 112 Q^{-1}$ for the relative differences of $T_*, n$ and $s$ from their thermal equilibrium values to all become less than 5\% (see the upper left panel of fig.~\ref{fig:f0_0p01_thermalization}). As shown in the upper right panel of fig.~\ref{fig:f0_0p01_thermalization}, the gluon distribution matches a Bose-Einstein distribution with a relative error smaller than $5\%$ for all the values of $p<5T_*$ around this time. And it fits perfectly the thermal distribution at still later times though, as exemplified by the result at $Qt=10^{3}$.

In terms of $T_*, n$ and $s$, one has $t_{eq} \approx 424 Q^{-1}$ for $N_f=3$. However, the lower left panel of fig.~\ref{fig:f0_0p01_thermalization} shows that the energy density, entropy density and number density of gluons and quarks are still quite different from their thermal equilibrium values. The seemingly equilibration of $T_*, n$ and $s$ is a result of the contributions from gluons being much higher than their thermal equilibrium values, effectively compensating for the much lower contributions from quarks and antiquarks. On the other hand, at this time, $n_g$, $s_g$ and $\epsilon_g$ are only 3.5\%, 5.5\% and 7.5\% respectively above their thermal equilibrium values with a time-dependent temperature $T_*$.  And it takes about $t_{eq,g}=509 Q^{-1}$ for all these relative differences to drop below 5\%, marking quantitatively the equilibration of gluons according the criteria in eq.~(\ref{eq:thermalization_condition}). Eventually, all the quantities shown in the lower left panel of fig.~\ref{fig:f0_0p01_thermalization} approach their thermal equilibrium values with a relative difference of less than 5\% at a later time $t_{eq}=810 Q^{-1}$, which is 7.2 times that for $N_f=0$. Such a separation between the equilibration times justifies the top-down thermalization of gluons, which is consistent with the observation in EKT that chemical equilibration takes a longer time to achieve~\cite{Kurkela:2018oqw, Kurkela:2018xxd, Du:2020zqg, Du:2020dvp}.

The lower right panel of fig.~\ref{fig:f0_0p01_thermalization} focuses on the high-$p$ behavior of the gluon distribution at $Qt\geq 200$. At $Qt=200$, the gluon distribution fits,  with a relative deviation being equal to or less than 5\%, the Bose-Einstein distribution with temperature $T_*$ for all values of $p\leq 3.3T_*$. At higher momentum, it evidently deviates more from the thermal distribution. The maximal momentum where $f$ fits the thermal distribution with temperature $T_*$ increases over time, as illustrated by the results at $Qt=350$ and $10^{3}$ in this plot. In this aspect, the top-down thermalization in the weak-coupling limit of QCD is qualitatively different from that in the strongly-coupled CFT using the AdS/CFT correspondence~\cite{Maldacena:1997re, Witten:1998qj}, where the UV modes equilibrate first~\cite{Lin:2008rw, Balasubramanian:2010ce, Balasubramanian:2011ur, Wu:2012rib, Wu:2013qi, Steineder:2013ana, Keranen:2015mqc,
Attems:2018vuw, Bernamonti:2018vmw}.

\subsubsection{Initially very dilute systems}

The parametric estimates in sec.~\ref{sec:parametricDilute} are valid only if $f_0\ll1$ such that all the four time scales are well separated. As discussed above,  for $f_0=0.01$, gluons, after being coupled to $N_f=3$ flavors of massless quarks and antiquarks, thermalize in a top-down manner at late times. This is qualitatively different from Stage 3 in the parametric estimates. In this subsection, we investigate quantitatively whether late-time reheating could occur in initially more dilute systems.

\begin{figure}
\begin{center}
\includegraphics[width=0.48\textwidth]{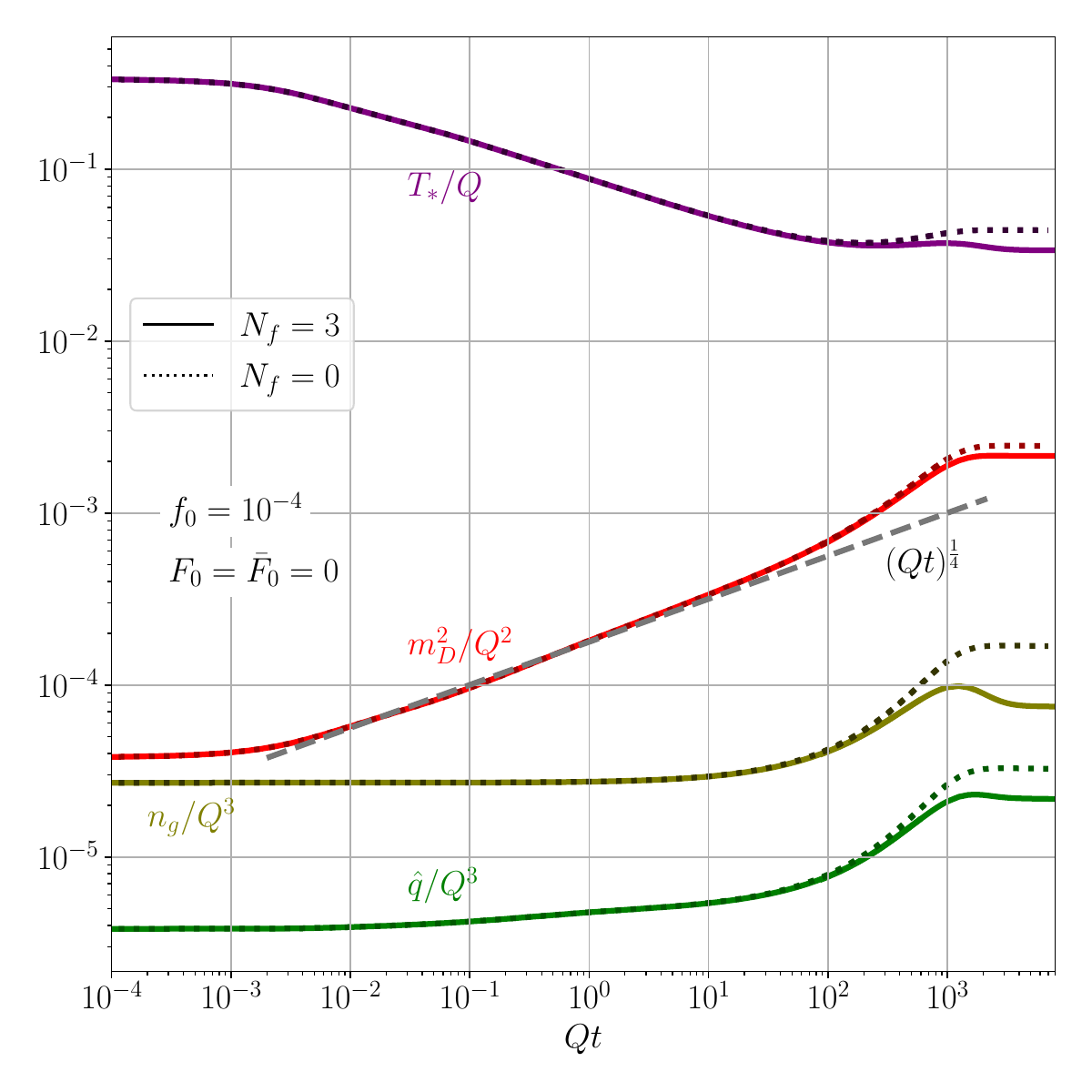}
\includegraphics[width=0.48\textwidth]{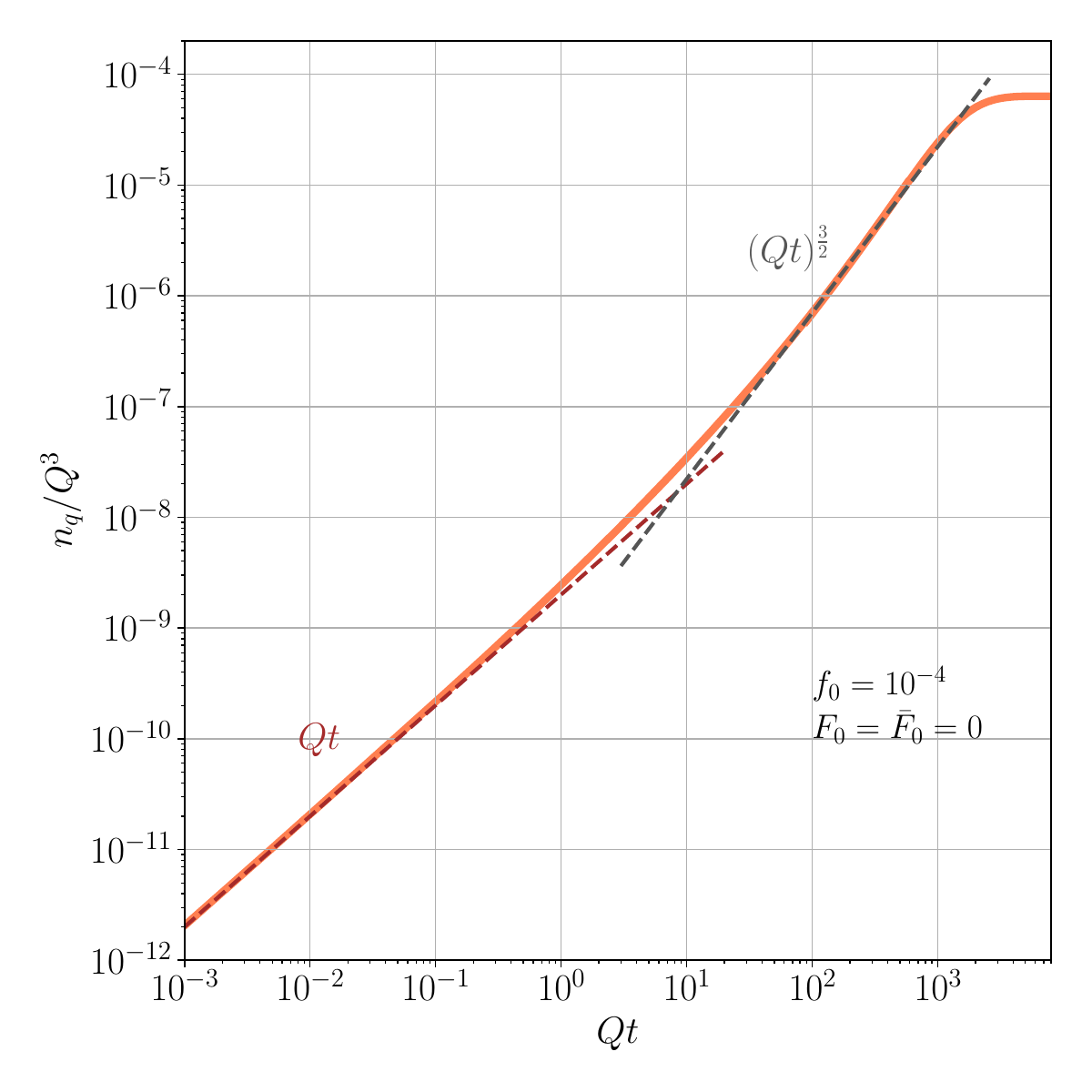}
\caption{The time evolution of different macroscopic quantities for $f_0=10^{-4}$. The left panel shows the results of $T_*$, $m_D^2$, $\hat{q}$ and $n_g$ for $N_f=0$ (dotted) and $N_f=3$ (solid). The right plot shows the result of $n_q$ in comparison with the scaling laws obtained by parametric estimates.
\label{fig:f0_1em4_macs}}
\end{center}
\end{figure}

Let us start with $f_0=10^{-4}$, which gives  ($\alpha_s^{-2} f_0$, $\alpha_s^{-2}$, $\alpha_s^{-2} f_0^{-\frac{1}{3}}$, $\alpha_s^{-2}f_0^{-\frac{3}{8}}$) = (0.01, 100, 2154, 3161). The detailed time-evolution of $T_*$, $m_D^2$, $\hat{q}$ and $n_g$ for $N_f=0$ and $N_f=3$ (the left panel) as well as $n_q$ for $N_f=3$ (the right panel) are presented in fig.~\ref{fig:f0_1em4_macs}. Similar to the case with $f_0=0.01$, these quantities are insensitive to $N_f$ at early times, showing that the soft sector of the system experiences overheating and cooling. For $N_f=0$, $T_*$ stops decreasing at $Qt \approx 196$, dropping about 16\% below the corresponding thermal equilibrium temperature. At this time, $\hat{q}$ and $m_D^2$ for $N_f=3$ are respectively 4.5\% and 1.4\% below their corresponding values for $N_f=0$. Up to this time, $m_D^2$ can be approximated by the scaling law of $(Qt)^{1/4}$ for a wide range of $t$. Meanwhile, following a linear growth, $n_q$ can be better approximated by the scaling law of $(Qt)^{3/2}$ for a noticeable range of $t$.

\begin{figure}
    \centering
    \includegraphics[width=0.48\textwidth]{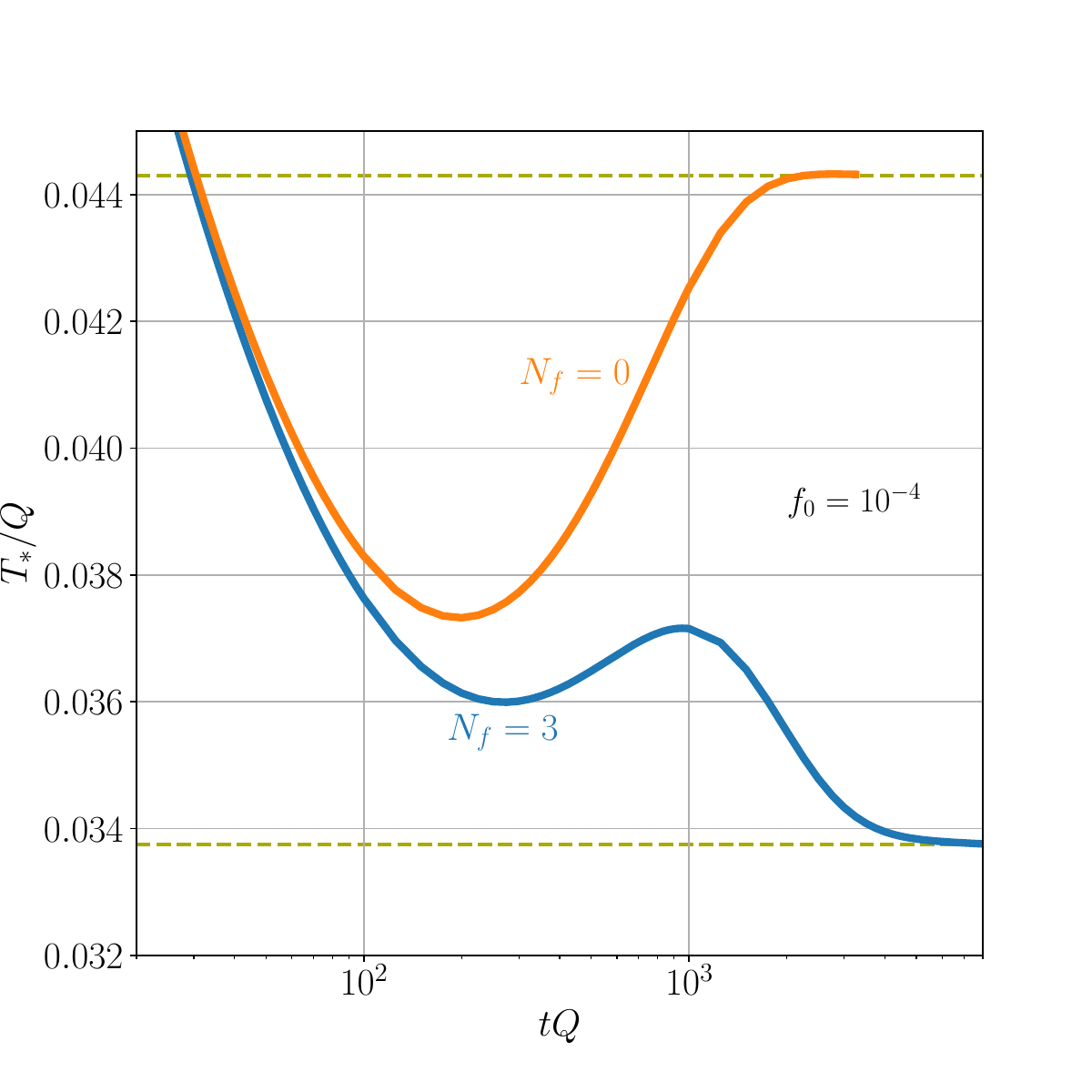}\
    \includegraphics[width=0.48\textwidth]{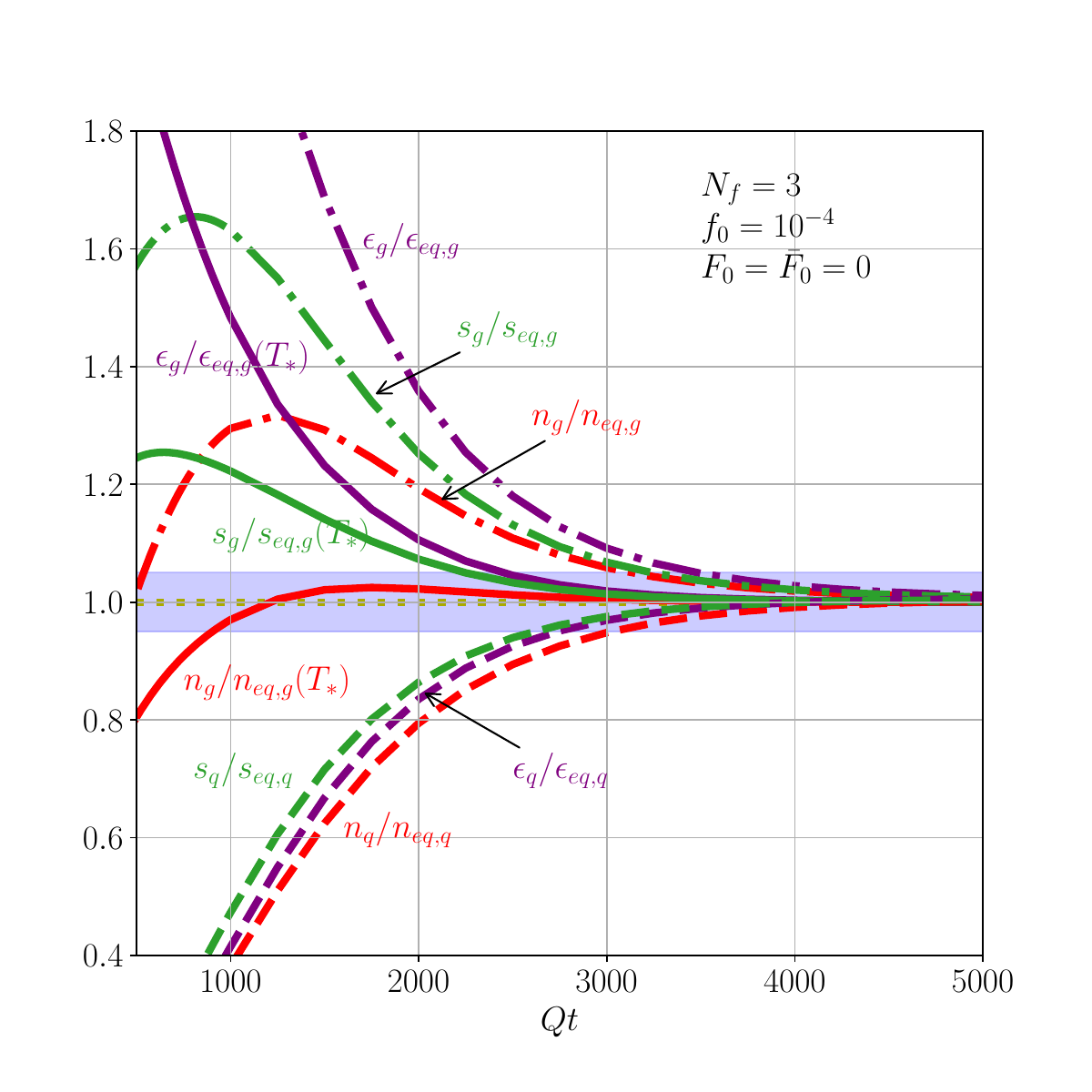}
    \caption{Evolution at late times for $f_0=10^{-4}$. The left panel shows an amplified representation of $T_*$ for both $N_f=0$ and $N_f=3$. The right panel shows thermalization of the same set of quantities as those in the lower left panels of fig.~\ref{fig:f0_0p01_thermalization}.}
    \label{fig:f0_1em4_thermalization}
\end{figure}

For $f_0=10^{-4}$, the late-time evolution process is still qualitatively different for $N_f=0$ and $N_f=3$, as shown in the left panel of fig.~\ref{fig:f0_1em4_thermalization}. After $Qt = 196$, the pure gluon system continuously reheats up to the thermal equilibrium temperature ($T=0.0443Q$), as expected in the parametric estimates. For $N_f=3$, the ensuing thermalization pattern becomes more complex. In this case, the first cooling phase stops at $Qt \approx 270$, leaving $T_*$ 6.7\% above the thermal equilibrium temperature ($T=0.0337Q$). Then, following the trend in the pure gluon case, it gradually heats up to $T_*=0.0372Q$ at $Qt = 956$. Note that $n_g$ has already overshot its thermal equilibrium value around this time, as shown by the ratio of $n_g/n_{eq,g}$ in the left panel of fig.~\ref{fig:f0_1em4_thermalization}. Afterwards, the system undergoes the second phase of cooling with $T_*$ decreasing toward its thermal equilibrium value.

At late times, the top-down thermalization persists for $f_0=10^{-4}$. According to their energy density, number density and entropy density shown in the right panel of fig.~\ref{fig:f0_1em4_thermalization}, the gluons first establish thermal equilibration among themselves with a time-dependent temperature $T_*$ around $t_{eq, g} = 2446 Q^{-1}$. Then, the energy density, number density and entropy density of gluons and quarks all converge towards their thermal equilibrium values and start to exhibit a relative deviation of 5\% or less around $t_{eq} = 3480 Q^{-1}$. This qualitatively agrees with the case for $f_0=0.01$ except that the final equilibration time is only about 42\% longer than $t_{eq,g}$. As another indicator of the shortened period of the top-down thermalization, we determine the equilibration time for $N_f=0$ in terms of $T_*$, $n$ and $s$ and obtain $t^{N_f=0}_{eq}=1489 Q^{-1}$. This tells us that the final equilibration time for $N_f=3$ is only about 2.3 times the value of $t^{N_f=0}_{eq}$. That is, thermalization is less delayed due to quark production compared to the case with $f_0=0.01$.

\begin{figure}
    \centering
    \includegraphics[width = 0.48\textwidth]{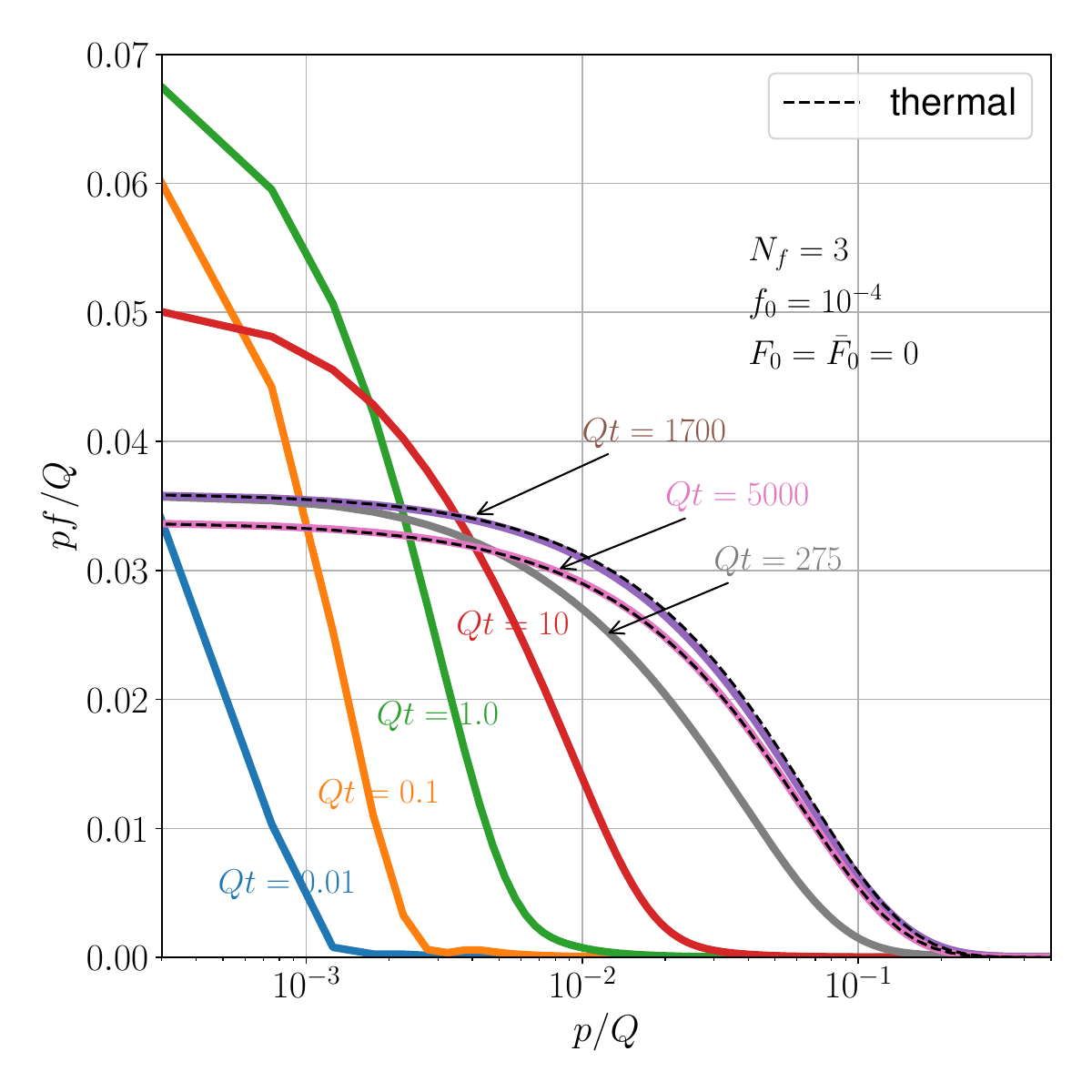}
    \includegraphics[width=0.48\textwidth]{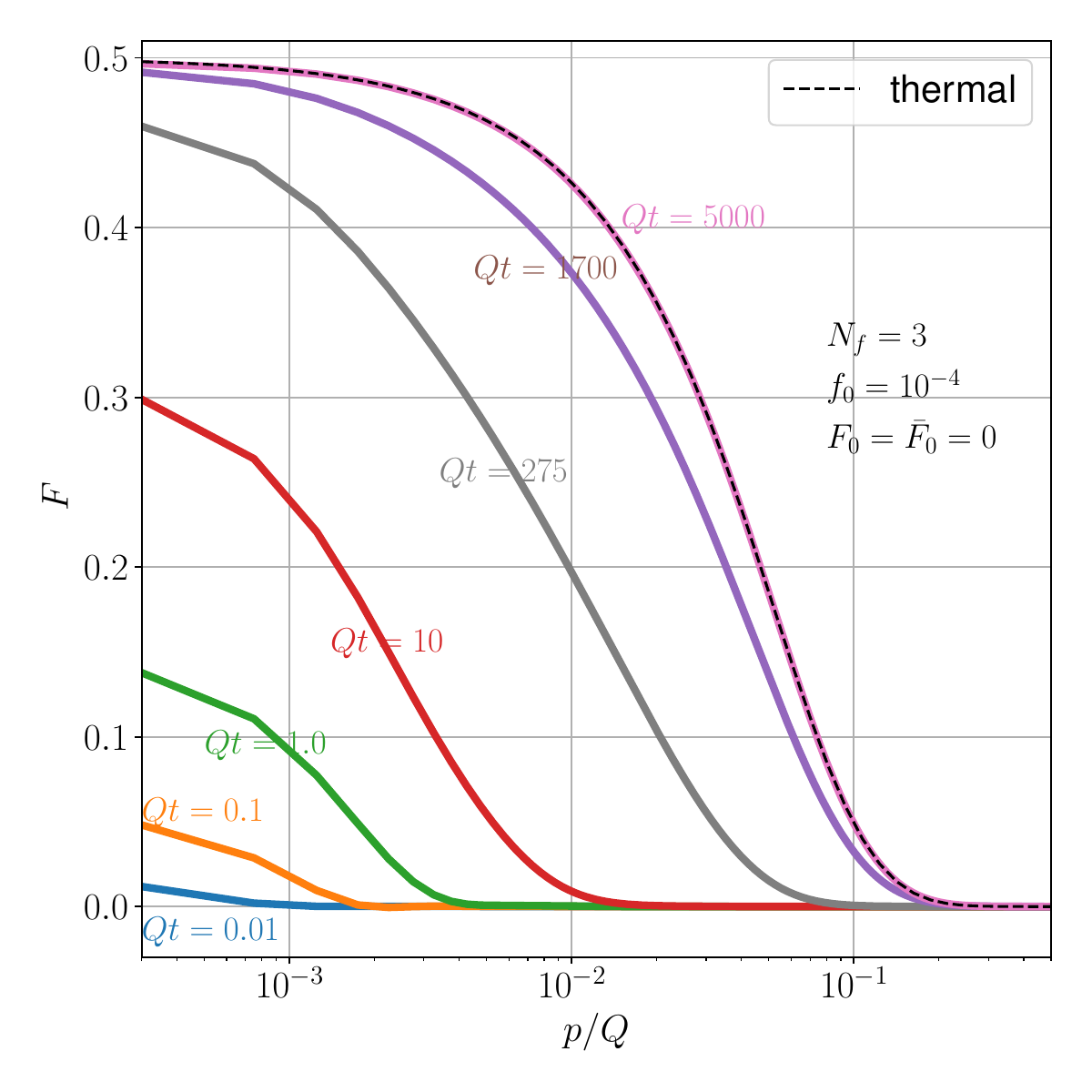}
    \caption{The gluon (left) and quark (right) distributions at different times for $f_0=10^{-4}$. At low momentum, $pf$ follows the complex evolution pattern of $T_*$: overheating, cooling, reheating and cooling except for every early times when the thermalized soft sector is out of the shown range of $p$. The quark distribution $F$, on the other hand, keeps increasing over $t$ across the shown range of $p$ towards the thermal distribution (dashed).}
    \label{fig:f0_1em4_fs}
\end{figure}

The above complex thermalization pattern is also manifest in the evolution of the distribution functions, as presented in fig.~\ref{fig:f0_1em4_fs}. The left panel shows the gluon distribution at different times. Except for the every early times when the typical momentum of soft gluons $p_s < p_{min}$ with $p_{min}$ the infrared momentum cutoff, our numerical results confirm its low-$p$ behavior with $pf\to T_*$. Moreover, we find that around $Qt=1700 < t_{eq, g}$, the gluon distribution can already be well fitted by the Bose-Einstein distribution with temperature $T_*$ for all values of $p\leq 5T_*$ (with a relative error of 5\% or less). On the other hand, the quark distribution $F$, as shown in the right panel, keeps increasing over time throughout the shown range of $p$. It approaches the thermal distribution from below only at the very late stage of the evolution.

\begin{figure}
    \centering
    \includegraphics[width=0.48\textwidth]{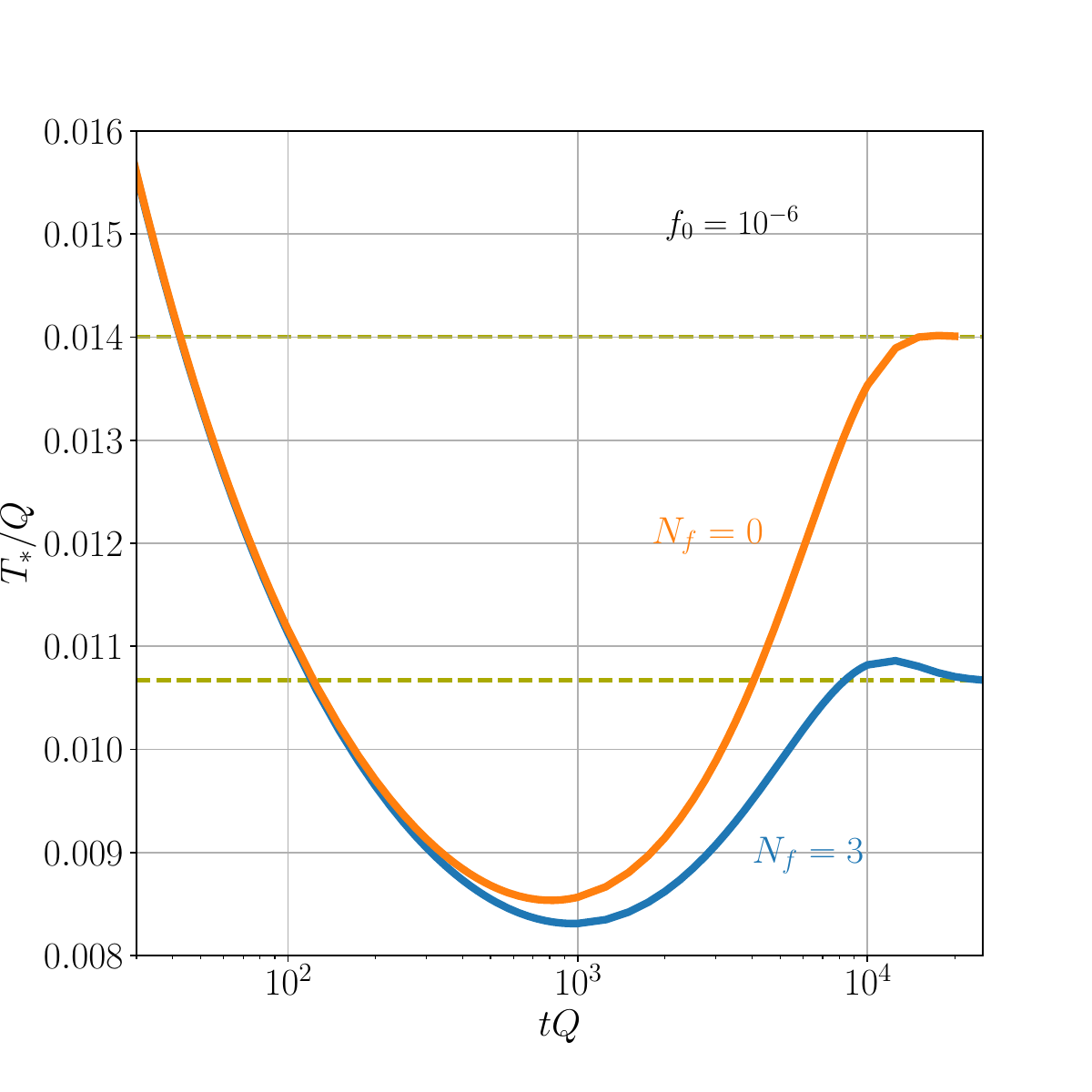}
    \includegraphics[width=0.48\textwidth]{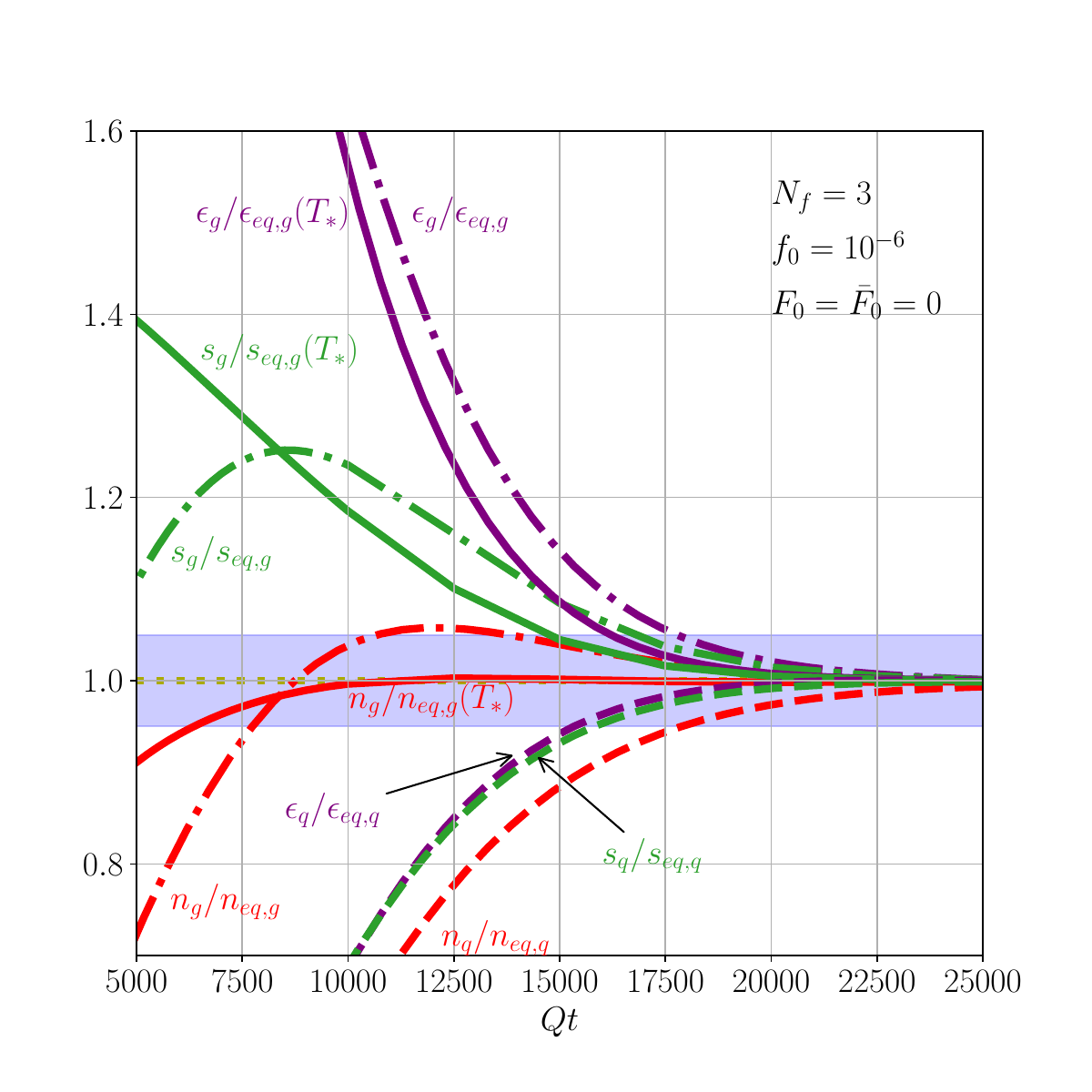}
    \caption{Thermalization for $f_0=10^{-6}$. The left panel shows $T_*$ as a function of $Qt$ for both $N_f=0$ and $N_f=3$. The right panel shows the time evolution of the same set of quantities as those in the lower left panels of fig.~\ref{fig:f0_0p01_thermalization}.}
    \label{fig:f0_1em6_thermalization}
\end{figure}

Aiming to illustrate how late-time reheating could emerge for $N_f=3$, let us study the system with $f_0=10^{-6}$, which corresponds to ($\alpha_s^{-2} f_0$, $\alpha_s^{-2}$, $\alpha_s^{-2} f_0^{-\frac{1}{3}}$, $\alpha_s^{-2}f_0^{-\frac{3}{8}}$) = ($10^{-4}$, $10^2$, $10^4$, 17783). The late-time behavior of $T_*$ is shown in the left panel of fig.~\ref{fig:f0_1em6_thermalization}.\footnote{We find that late-time results are robust to values of the infrared cutoff $p_{min}$. However, one needs to choose properly the value of $p_{min}$ in order to obtain correct early-time results, especially for $m_D^2$.
} It shows that the system overcools irrespective of the values of $N_f$ with $T_*$ reaching its minimum at $Qt\approx 809$ for $N_f=0$ and at $Qt\approx 1087$ for $N_f=3$. For $N_f=0$, the system, as expected, reheats towards full thermalization afterwrads. For $N_f=3$, reheating finally occurs. However, as a vestigial top-down thermalization, $T_*$ later on overshoots the thermal equilibrium temperature slightly at the percent level before the system fully thermalizes.

The delay of thermalization due to quark production is much milder for $f_0=10^{-6}$, as shown in the right panel of fig.~\ref{fig:f0_1em6_thermalization}. In terms of $T_*$, $n$ and $s$, one has $t^{N_f=0}_{eq}=1.19\times 10^{4} Q^{-1}$ for the pure gluon system. For $N_f=3$, it is no longer evident that gluons thermalize first and undergo the top-down thermalization. And the final equilibration time in terms of the energy density, number density and entropy density of gluons and quarks is determined to be $t_{eq}=1.79\times 10^4 Q^{-1}$, which is only about 50\% longer than $t^{N_f=0}_{eq}$. 

\section{Thermalization in initially over-populated systems}
\label{sec:over}

In this section, we study initially over-populated systems in which there are a larger number of partons at $t=0$ than that can be accommodated by the final thermal equilibrium state. Below we carry out both parametric estimates and numerical simulations for such systems.

\subsection{Parametric estimates for $f_0\gg 1$}
\label{sec:overPara}

In the limit $f_0\gg 1$, one initially has
\begin{align}\label{eq:macsOverI}
n_g\sim n_{h, g} \sim f_0 Q^3,~~m_{D}^2\sim \alpha_s f_0 Q^2,~~\hat{q}\sim \alpha_s^2 f_0^2 Q^3, ~~ T_*\sim f_0 Q.
\end{align}
Like the pure gluon case~\cite{Kurkela:2014tea, Schlichting:2019abc, BarreraCabodevila:2022jhi}, the system goes through the following two stages towards achieving thermal equilibrium.

\subsubsection*{\it Stage 1. Soft gluon radiation and overheating: $0\ll\,Qt\ll (\alpha_s f_0)^{-2}$} 

During this stage, $\hat{q}$ and $m_D^2$ are dominated by hard gluons, and they are parametrically the same as those at the initial time. Accordingly, the soft sector of the system is the same as that in an overheated system with $T_*\sim f_0Q\gg T\sim f_0^{\frac{1}{4}}Q$. This can be justified by the following parametric estimates.%, which are similar to those for Stage 1 in the initially very dilute limit.

The soft sector of the gluon distribution is rapidly populated through radiation off hard gluons. For the range of $p$ dominated by the $g\to gg$ splitting, the number density of gluons typically carrying momentum $p$ is given by
\begin{align}\label{eq:fsgIOver}
    n_g(p)\sim x\frac{dI(Q)}{dxdt} {n_{h, g}}(1+f_0)t\sim
\alpha_s\sqrt{\frac{\hat{q}}{p}} f_0 n_{h, g} t\sim \alpha_s^{-1}\sqrt{\frac{\hat{q}}{p}} \hat{q} t\qquad\text{with $x\sim p/Q$},
\end{align}
and, accordingly, the gluon distribution takes the following parametric form
\begin{align}
    f(p)\sim \frac{n(p)}{p^3}\sim \alpha_s^{-1}\sqrt{\frac{\hat{q}}{p^7}} \hat{q} t.
\end{align}
The above estimate is valid for $p_s\lesssim p\lesssim \tilde{p}\equiv (\hat{q}^3 t^2/\alpha_s^2 f_0^2)^{1/7}\sim (\alpha_s^2 f_0^2 Qt)^{2/7}Q$ with the typical soft momentum $p_s$ given by the balancing condition:
\begin{align}
    1\sim\frac{p_s f(p_s)}{T_*}\sim \sqrt{\frac{\hat{q}}{p_s^5}} m_D^2 t\Leftrightarrow
p_s\sim
p_* = (\hat{q} m_D^4 t^2)^{\frac{1}{5}}\sim (\alpha_s^2\,f_0^2\,Qt)^{\frac{2}{5}} Q.
\end{align}
For $p\lesssim p_s\sim p_*$, $f$ approaches the thermal distribution with temperature $T_*$. That is, $f$ can be expressed in the same parametric form as that illustrated in fig.~\ref{fig:pfI} (with $p_s$ and $\tilde{p}$ given above). Accordingly, besides hard gluons, there is a pronounced accumulation of soft gluons carrying a typical momentum of order $p_s$:
\begin{align}
&n_{s, g}\sim
\alpha_s\sqrt{\frac{\hat{q} t^2}{p_s}} f_0 n_{h, g}\sim T_* p_s^2\sim \alpha_s^{\frac{8}{5}} f_0^{\frac{8}{5}} (Qt)^{\frac{4}{5}} n_{h, g}.
\end{align}
Evidently, it is parametrically smaller than the hard gluon number density at $Qt\lesssim (\alpha_s f_0)^{-2}$.

Quarks and antiquarks are produced much more slowly than soft gluons. Via the $g\to q/\bar{q}$ conversion in the $2\leftrightarrow2$ processes, one has $n_q^{g\to q}$$\sim m_D^4 t$ $\sim\alpha_s^2 f_0^2 Q^4 t$. It is parametrically the same as that produced via $g\to q\bar{q}$:
\begin{align}
    n_q\sim n_q^{g\leftrightarrow q\bar{q}}\sim \alpha_s \sqrt{\frac{\hat{q}}{Q}} n_{h, g} t\sim \alpha_s^2 f_0^2 Q^4 t.
\end{align}
That is, the $1\leftrightarrow 2$ processes are not more efficient than the $2\leftrightarrow2$ processes in producing quarks and antiquarks. On the other hand, for $p\gtrsim p_s$, the quark distribution is mostly determined by the $g\to q\bar{q}$ splitting:
\begin{align}
    F(p)\sim  \alpha_s\sqrt{\frac{\hat{q} t^2}{Q}}\bigg(\frac{p}{Q}\bigg)^\frac{1}{2} \frac{n_{h,g}}{p^3}\sim \sqrt{\frac{\hat{q} t^2}{Q}}\bigg(\frac{p}{Q}\bigg)^\frac{1}{2} \frac{m_D^2 Q}{p^3}\sim \frac{1}{2}\left(\frac{p_s}{p}\right)^{\frac{5}{2}}
\end{align}
with $F(p_s)\sim 1/2$. And at lower $p$, quarks and antiquarks fill a thermal distribution with $F\to 1/2$.

During this stage, the contributions from soft gluons, quarks and antiquarks to $m_D^2$ and $\hat{q}$ are all parametrically smaller than those from hard gluons. At the end of this stage at $Qt\sim (\alpha_s f_0)^{-2}$, one has $p_s\sim \tilde{p}\sim Q$, $n_{s, g}\sim n_{h, g}\sim f_0 Q^3 \gg n_{q}\sim Q^3$ and
\begin{align}
\hat{q}^g\sim \alpha_s^2 f_0^2 Q^3\gg \hat{q}^q\sim \alpha_s^2 p_s^3\sim \alpha_s Q^3,\qquad 
m^2_{D, g}\sim \alpha_s f_0 Q^2 \gg m^2_{D, q}\sim \alpha_s p_s^2\sim \alpha_s Q^2
\end{align}
with $\hat{q}^a$ denoting the contribution from parton $a$. That is, the properties of the system are still mostly determined by gluons.

\subsubsection*{\it Stage 2. Momentum broadening and cooling: $(\alpha_s f_0)^{-2}\ll\,Qt\ll \alpha_s^{-\frac{7}{4}}(\alpha_s f_0)^{-\frac{1}{4}}$}

As justified below, the system is still dominated by gluons during this stage. In this case, the scaling laws for all the relevant quantities can be derived in the same way as the pure gluon case. That is, solving consistently the following two equations respectively given by momentum broadening due to multiple elastic scattering and energy conservation~\cite{BarreraCabodevila:2022jhi}
\begin{align}
    p^2 \sim \hat{q} t \sim\alpha_s^2 n_g^2 t/p^3,\qquad
    \epsilon\sim\,f_0 Q^4 \sim\,n_g p
\end{align}
yields
\begin{align}\label{eq:macsOverII}
&n_g\sim \frac{(\alpha_s f_0)^{\frac{5}{7}}}{\alpha_s}\frac{Q^3}{ (Qt)^\frac{1}{7}}, \qquad p\sim (\alpha_s f_0)^{\frac{2}{7}}(Qt)^{\frac{1}{7}} Q,\qquad\,f\sim \frac{(\alpha_s f_0)^{-\frac{1}{7}}}{\alpha_s}\frac{1}{ (Qt)^\frac{4}{7}},\notag\\
&\hat{q}\sim (\alpha_s f_0)^{\frac{4}{7}}\frac{Q^3}{(Qt)^\frac{5}{7}}, \qquad m_D^2 \sim (\alpha_s f_0)^{\frac{3}{7}}\frac{Q^2}{(Qt)^\frac{2}{7}},\qquad T_*\sim\frac{(\alpha_s f_0)^{\frac{1}{7}}}{\alpha_s}\frac{Q}{(Qt)^\frac{3}{7}}\sim pf.
\end{align}
Irrespective of the detailed mechanism how the number of excessive gluons are eliminated, the above derivation is valid as long as it does not provide an alternative, faster way than multiple elastic scattering to increase the typical momentum of gluons~\cite{Blaizot:2011xf, Kurkela:2012hp, BarreraCabodevila:2022jhi}.

Allowing the production of quarks and antiquarks does not modify the scaling laws derived above. As in Stage 1, the $2\leftrightarrow2$ and $1\leftrightarrow2$ processes are equally efficient to produce quarks and antiquarks, yielding
\begin{align}\label{eq:nqOverII}
    n_q \sim m_D^4 t \sim \alpha_s \sqrt{\frac{\hat{q}}{p}} n_g t \sim (\alpha_s f_0)^{\frac{6}{7}} (Qt)^{\frac{3}{7}} Q^3,
\end{align}
and
\begin{align}
    F\sim \frac{n_q}{p^3}\sim \frac{1}{2}.
\end{align}
That is, parametrically, the majority of partons in the system consistently fill thermal distributions up to $p\sim \hat{q}t \sim (\alpha_s f_0)^{\frac{2}{7}}(Qt)^{\frac{1}{7}} Q$. Accordingly, one has
\begin{align}
    \hat{q}^q\sim \alpha_s^2 n_q\sim \alpha_s^2 p^3,\qquad m_{D,q}^2\sim \alpha_s\frac{n_q}{p}\sim \alpha_s p^2.
\end{align}
At $Qt\lesssim \alpha_s^{-\frac{7}{4}}(\alpha_s f_0)^{-\frac{1}{4}}$, the typical momentum of partons $p$ is always parametrically lower than $T_*$. As a result, $\hat{q}$ and $m_D^2$ do not receive significant contributions from quarks and antiquarks, and the system is hence dominated by gluons.

At $Q t\sim \alpha_s^{-\frac{7}{4}}(\alpha_s f_0)^{-\frac{1}{4}}$, all the above quantities approach their final equilibrium values and full thermalization is established around this time.

\subsection{Quantitative studies}
\label{sec:overQuan}

In this subsection, we carry out simulations for initially over-populated systems with $f_0> 0.308$ for the initial distributions given in eq.~(\ref{eq:fs0}). We take $\alpha_s=0.1$ for $f_0=1$ and $10$, and $\alpha_s=0.01$ for $f_0=100$ with $\mathcal{L}$ fixed to 1.

\subsubsection{Initially dense system with $f_0=1$}

\begin{figure}
    \centering
    \includegraphics[height = 0.45\textwidth]{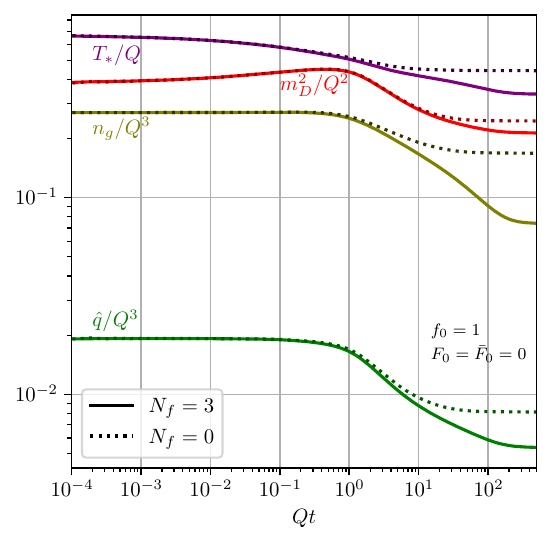}
    \includegraphics[height=0.45\textwidth]{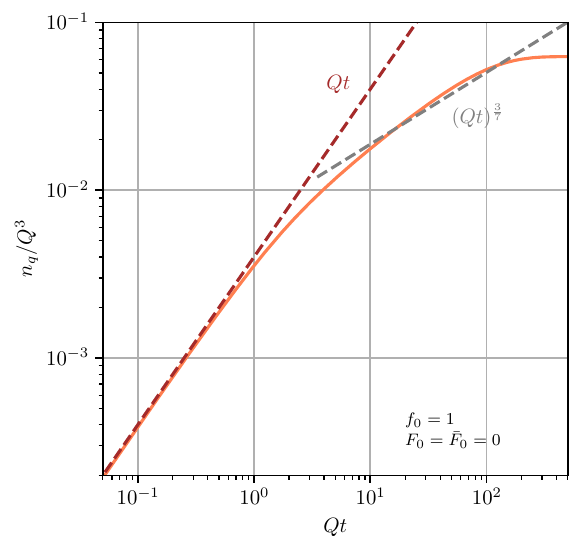}
    \caption{The time evolution of different macroscopic quantities for $f_0=1$. The left panel shows the results of $T_*$, $m_D^2$, $\hat{q}$ and $n_g$ for $N_f=0$ (dotted) and $N_f=3$ (solid). %Center panel shows a more detailed plot of the effective temperature $T_*$.
    The right panel shows $n_q$ over time in comparison with the scaling behavior in the limit $f_0\gg 1$.}
    \label{fig:f0_1_macs}
\end{figure}

\begin{figure}
    \centering
    \includegraphics[width = 0.48\textwidth]{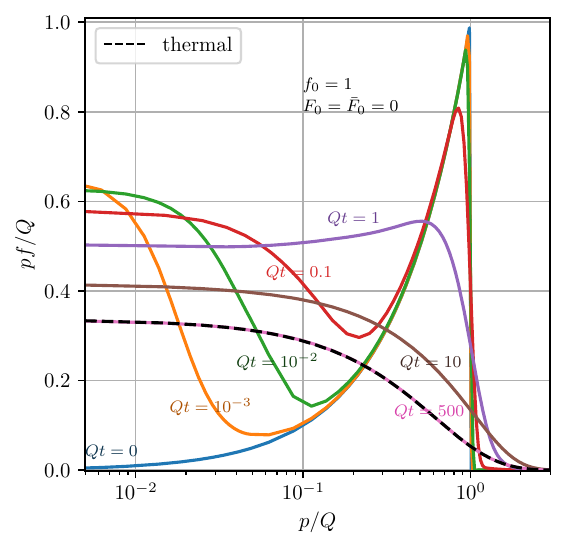}
    \includegraphics[width=0.48\textwidth]{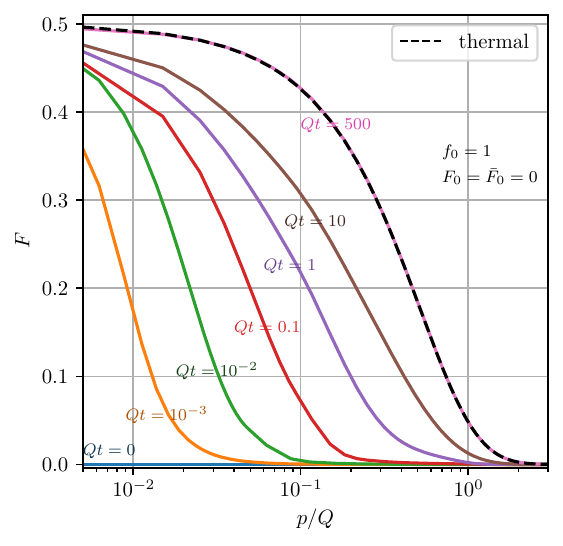}
    \caption{The gluon (left) and quark (right) distributions at different times for $f_0=1$. At low momentum, $pf$ is confirmed numerically to equal to $T_*$ (for $Qt\gtrsim 10^{-2}$) and the quark distribution $F$ shows the trend to approach 1/2. Both gluon and quark distributions fit very well the thermal distributions (thermal) at the end of the evolution. %For comparison, the approximate solutions in eq.~(\ref{eq:fslowp}) (approx) are also plotted for $Qt\leq 10^{-2}$.
    }
    \label{fig:f0_1_fs}
\end{figure}

The time evolution of $\hat{q}$, $m_D^2$, $T_*$, $n_g$ and $n_q$ for $f_0=1$ is shown in the two panels of Fig. \ref{fig:f0_1_macs}. At early times, $\hat{q}$, $m_D^2$, $T_*$ and $n_g$, as displayed in the left panel, are insensitive to $N_f$. Neither of them experience significant changes before $Qt \sim 1 $. Besides, the right panel confirms the early-time linear growth of the quark number density. These observations align with the predicted behaviors in Stage 1 according to the parametric estimates. This is expected since the early-time evolution is universally described by the approximate solutions in eq.~(\ref{eq:fslowp}) for all values of $f_0$.% A comparison between our numerical results of the distributions for $N_f=3$ and the approximate solutions at early times is presented in the two plots of fig.~\ref{fig:f0_1_fs}.

The subsequent evolution of $\hat{q}$, $m_D^2$, $T_*$ and $n_g$ appears qualitatively similar for both $N_f=0$ and $N_f=3$: they all steadily decrease  towards their thermal equilibrium values. Alternatively, the evolution history of $T_*$ can be read out from the gluon distribution at low momentum with $pf\to T_*$, as shown in the left panel of fig.~\ref{fig:f0_1_fs} for $N_f=3$. And $f$ approaches the Bose-Einstein distribution with temperature $T$ from above while the quark distribution, shown in the right panel of fig.~\ref{fig:f0_1_fs}, steadily approaches the Fermi-Dirac distribution from below. This is all seemingly consistent with a cooling process in Stage 2. However,  $\hat{q}$, $m_D^2$, $T_*$ and $n_g$ do not fit the scaling laws derived in the limit $f_0\gg1$.  Similarly, the growth of $n_q$ is shown to slow down, but it does not exhibit the scaling behavior of $(Qt)^{\frac{3}{7}}$.

\begin{figure}
    \centering
    \includegraphics[width=0.48\textwidth]{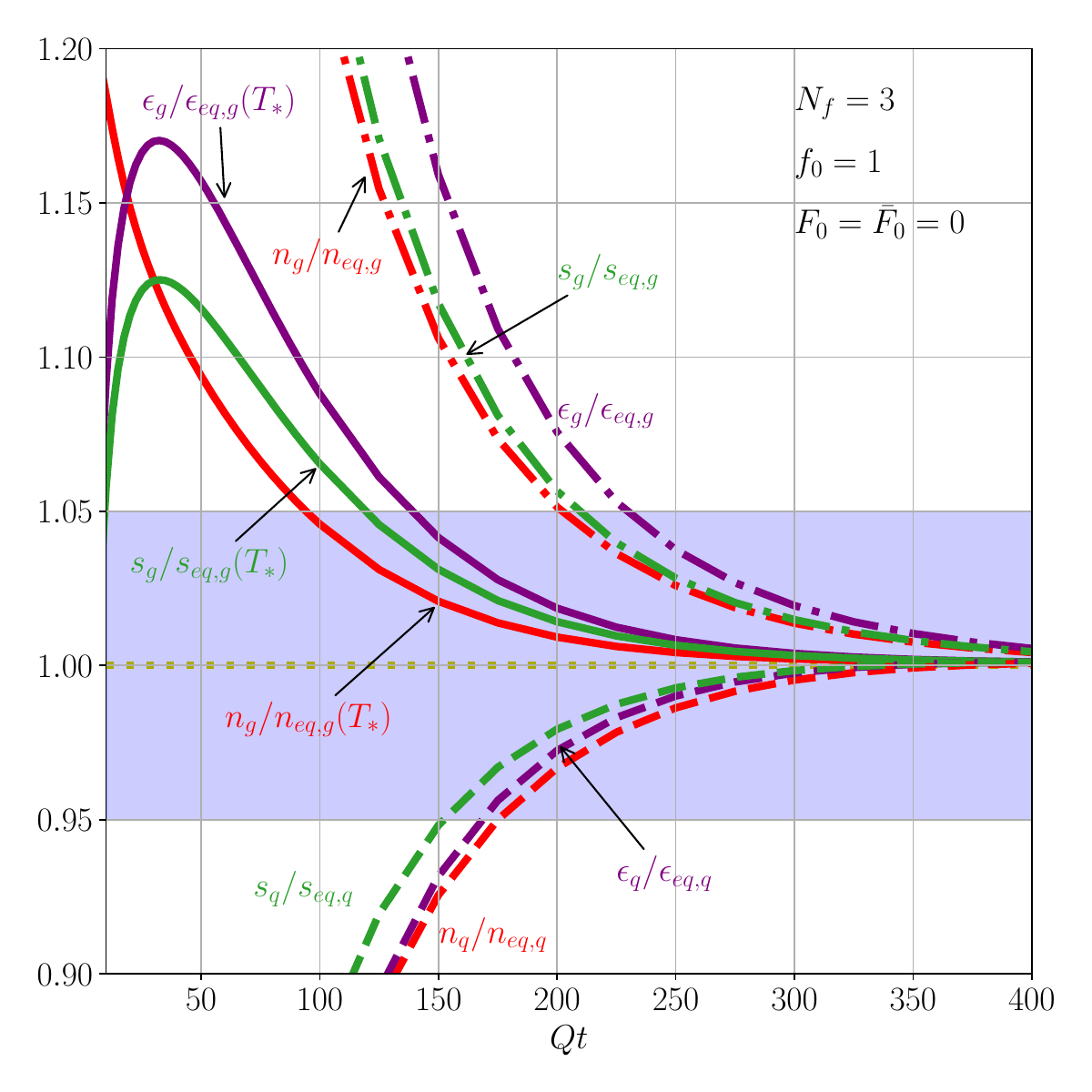}    \includegraphics[width=0.48\textwidth]{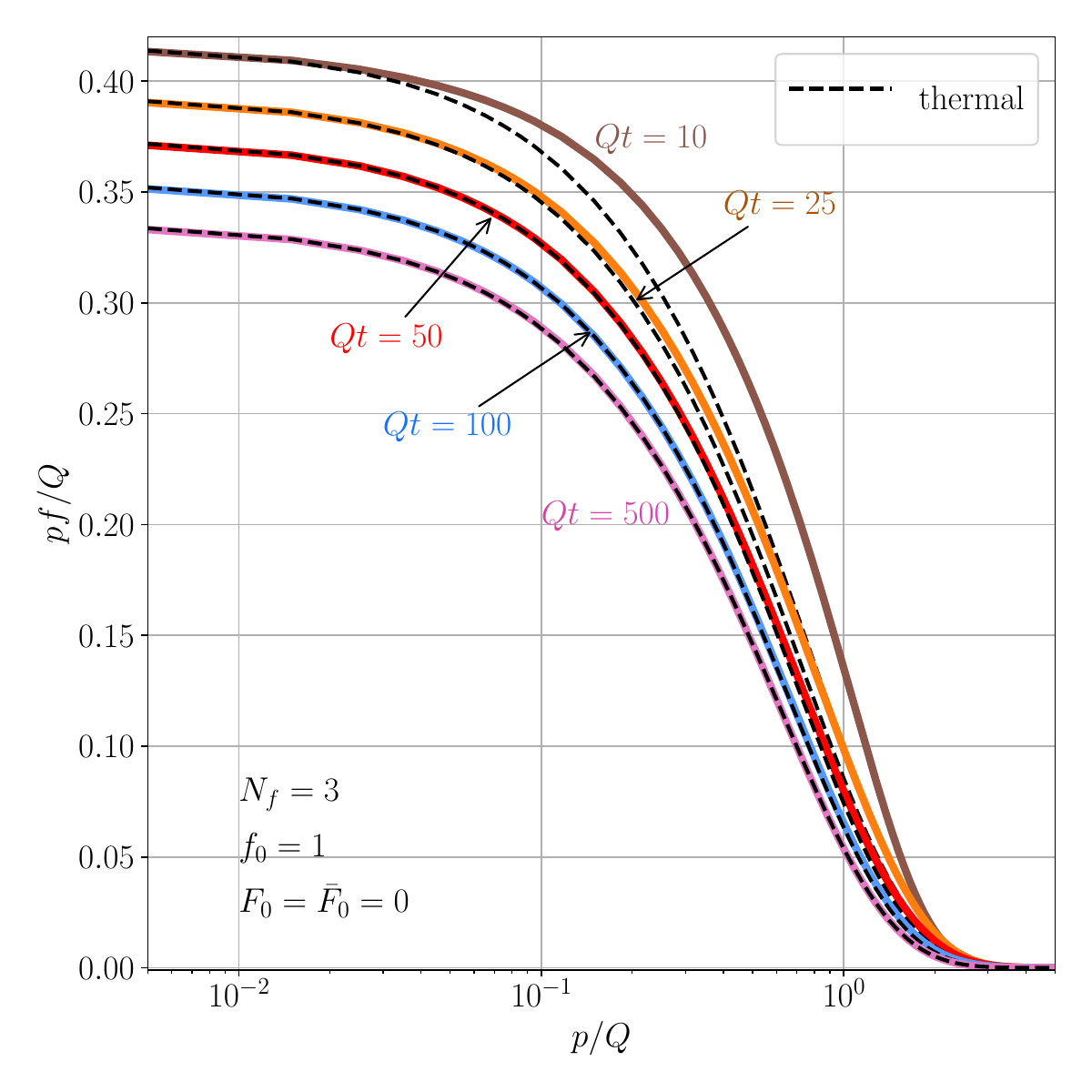}
    \caption{The top-down thermalization for $f_0=1$. The left panel shows how the same set of quantities as those in the lower left panels of fig.~\ref{fig:f0_0p01_thermalization} approach unity. The right panel shows the gluon distribution at different times, comparing to the Bose-Einstein distribution (thermal) with a time-dependent temperature $T_*$.}
    \label{fig:f0_1_thermalization}
\end{figure}

The system for $N_f=3$ also undergoes the top-down thermalization before it achieves fully thermalization. In terms of $T_*$, $n$ and $s$, one has $t_{eq}=106 Q^{-1}$ in comparison with $t^{N_f=0}_{eq}=23.2 Q^{-1}$ for $N_f=0$. However, the left panel of fig.~\ref{fig:f0_1_thermalization} shows that the energy density, number density and entropy density of gluons and quarks are still quite different from their final thermal equilibrium values around this time. That is, the system has not achieved chemical equilibration yet.  In contrast, the relative deviations of the energy density, number density and entropy density of gluons from their thermal equilibrium values at temperature $T_*$ are much smaller. Moreover, the right panel of fig.~\ref{fig:f0_1_thermalization} shows that the gluon distribution can already fit, with a relative error of 5\% or less, the Bose-Einstein distribution with temperature $T_*$ up to $p=3.4 T_*$ as early as $Qt=100$. Then, it takes only a little bit longer for their relative differences to all drop below 5\% around $t_{eq, g}=138 Q^{-1}$. At the end, in terms of the energy density, number density and entropy density of gluons and quarks, the final equilibration time is given by $t_{eq} = 229Q^{-1}$, which is about 9.9 times the equilibration time for $N_f=0$.

\subsubsection{Initially very dense systems}

\begin{figure}
    \centering
    \includegraphics[height = 0.45\textwidth]{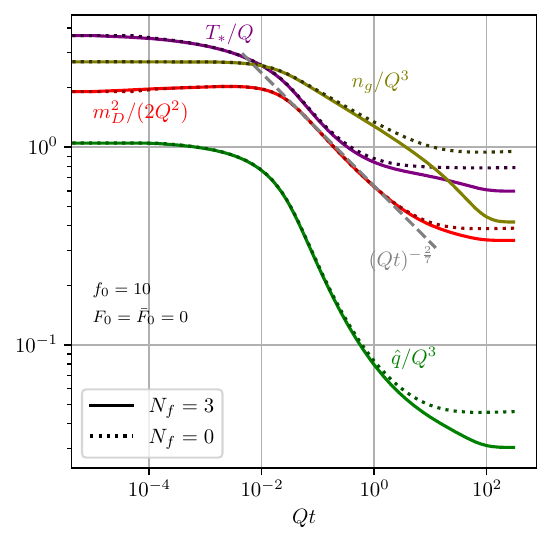}
    \includegraphics[height=0.45\textwidth]{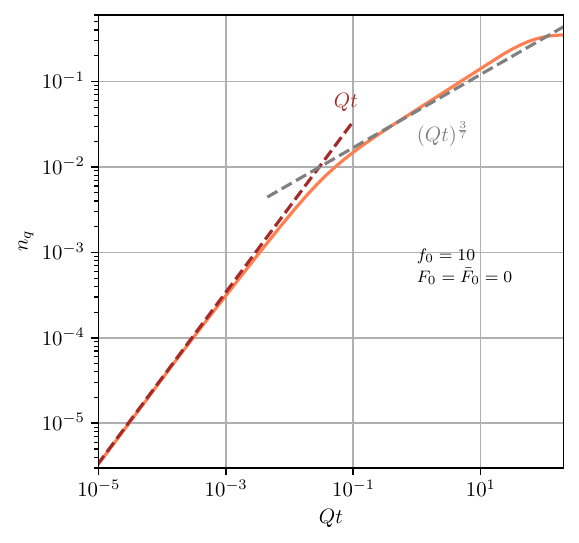}
    \caption{
   The time evolution of different macroscopic quantities for $f_0=10$. The left panel shows the results of $T_*$, $m_D^2$, $\hat{q}$ and $n_g$ for $N_f=0$ (dotted) and $N_f=3$ (solid). The right panel shows $n_q$ over time in comparison with the scaling behavior in the limit $f_0\gg 1$.}
    \label{fig:f0_10_macs}
\end{figure}

The time evolution of $\hat{q}$, $m_D^2$, $T_*$, $n_g$ and $n_q$ in the system with $f_0=10$ is shown in the two panels of fig. \ref{fig:f0_10_macs}. Similar to the cases studied previously, quarks and antiquarks play a negligible role at early times. In terms of $\hat{q}$, $T_*$ and $n_g$ shown in the left panel of fig. \ref{fig:f0_10_macs}, the system, regardless of the value of $N_f$, universally experiences a rapid cooling stage following a slowly varying overheating one. At $Qt\approx 1.4$, $\hat{q}$, $T_*$ and $n_g$ for $N_f=3$ all start to drop more than 5\% below their corresponding values in the pure gluon case while it takes about 10 times as much time for $m_D^2$ to reach a similar level of deviation. In the meanwhile, the production rate of $n_q$, as shown in the right panel of  Fig. \ref{fig:f0_10_macs}, slows down after a linear growth over time. However, these quantities except $m_D^2$ do not demonstrate convincingly the scaling behaviors in the cooling stage for $f_0\gg1$. 

\begin{figure}
    \centering
    \includegraphics[width=0.48\textwidth]{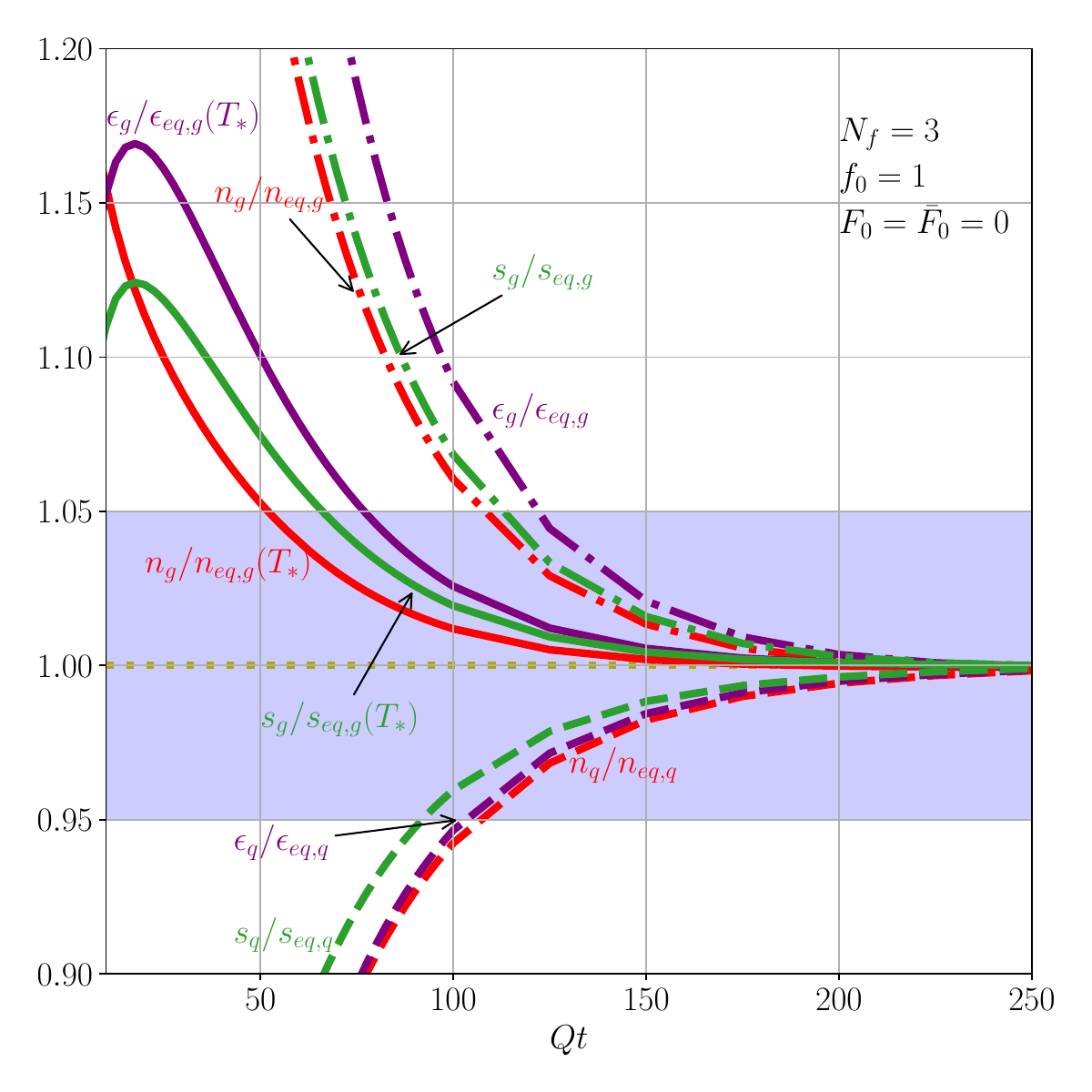}    \includegraphics[width=0.48\textwidth]{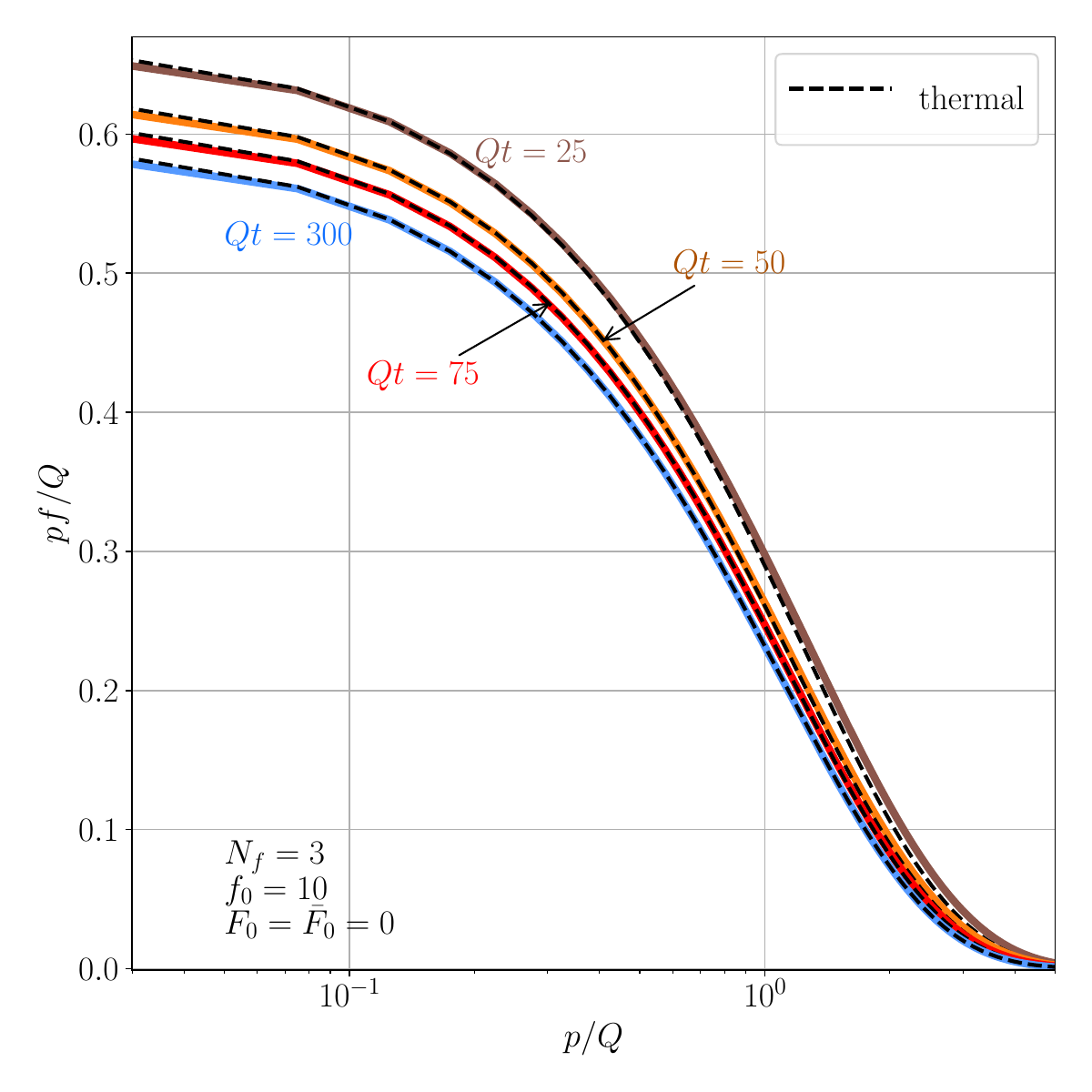}
    \caption{The top-down thermalization of gluons for $f_0=10$. The left panel shows how the same set of quantities as those in the lower left panel of fig.~\ref{fig:f0_0p01_thermalization} approach unity. The right panel shows the gluon distribution at different times, comparing to the Bose-Einstein distribution (thermal) with a time-dependent temperature $T_*$.}
    \label{fig:f0_10_thermalization}
\end{figure}

In the final stage of thermalization, gluons equilibrate first with a time-dependent temperature $T_*$ for $N_f=3$, as qualified by the quantities in the left panel of fig.~\ref{fig:f0_10_thermalization}. This is qualitatively the same as the cases for $f_0\geq 10
^{-4}$. Quantitatively, the equilibration time for gluons is determined to be $t_{eq,g}=76.9Q^{-1}$ by comparing the energy density, number density and entropy density of gluons to their thermal equilibrium values at temperature $T_*$. Even before this time, the gluon distribution can be very well fitted by the Bose-Einstein distribution with $T_*$ up to some maximum momentum higher than $T_*$, as exemplified in the right panel of fig.~\ref{fig:f0_10_thermalization}. For example, at $Qt=25$, the relative difference between $f$ and the corresponding thermal distribution with temperature $T_*$ is smaller than 5\% for all the values of $p\lesssim 2.0 T_*$. Eventually, gluons, quarks and antiquarks all equilibrate around $t_{eq} = 130Q^{-1}$, which is about 11.6 times that in the pure gluon case as determined by $T_*$, $n$ and $s$.

\begin{figure}
    \centering
    \includegraphics[height = 0.47\textwidth]{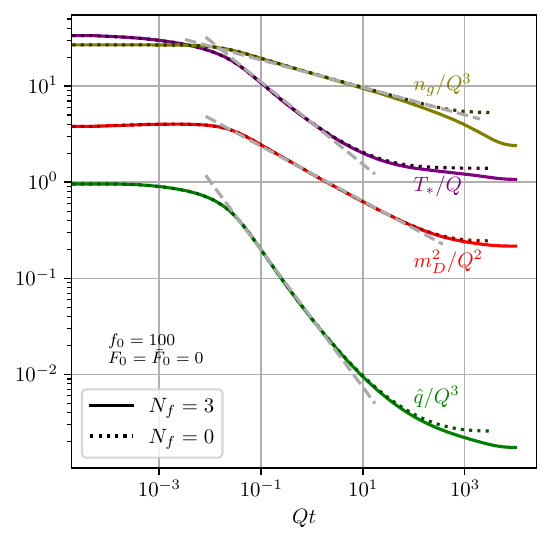}
    \includegraphics[height=0.47\textwidth]{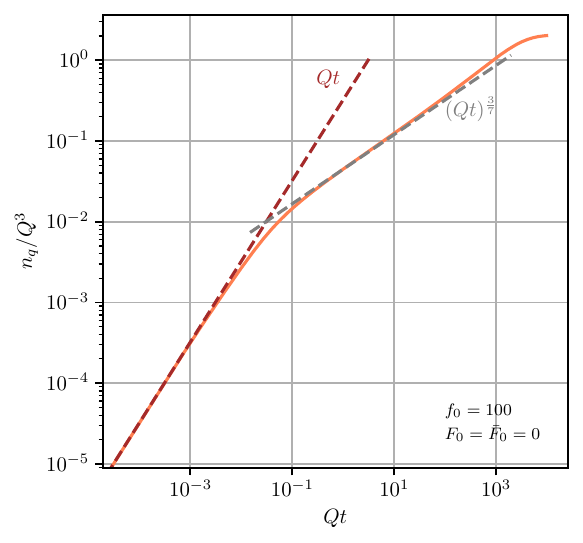}
    \caption{The time evolution of different macroscopic quantities for $f_0=100$ and $\alpha_s=0.01$. The left panel shows the results of $T_*$, $m_D^2$, $\hat{q}$ and $n_g$ for $N_f=0$ (dotted) and $N_f=3$ (solid). The right panel shows $n_q$ over time. In both panels, the expected scaling behaviors of these quantities (dashed) are also shown.}
    \label{fig:f0_100_macs}
\end{figure}

Let us now study the even denser system with $f_0 = 100$ and $\alpha_s=0.01$, where $\hat{q}$ and $m_D^2$ are known to exhibit the scaling behaviors in eq.~(\ref{eq:macsOverII}) for $N_f=0$~\cite{BarreraCabodevila:2022jhi}. As shown in the left panel of fig.~\ref{fig:f0_100_macs}, $\hat{q}$, $m_D^2$, $n_g$ and $T_*$ for $N_f=3$ first experience the slowly varying, overheating stage. Then, they approach the expected scaling behaviors. Before exiting the scaling regions, they are all barely distinguishable from those in the pure gluon case. Regarding $n_q$ as plotted in the right panel of Fig. \ref{fig:f0_100_macs}, it initially grows linear with time, and then approaches the anticipated scaling behavior of $(Qt)^{\frac{3}{7}}$.

\begin{figure}
    \centering
    \includegraphics[width=0.8\textwidth]{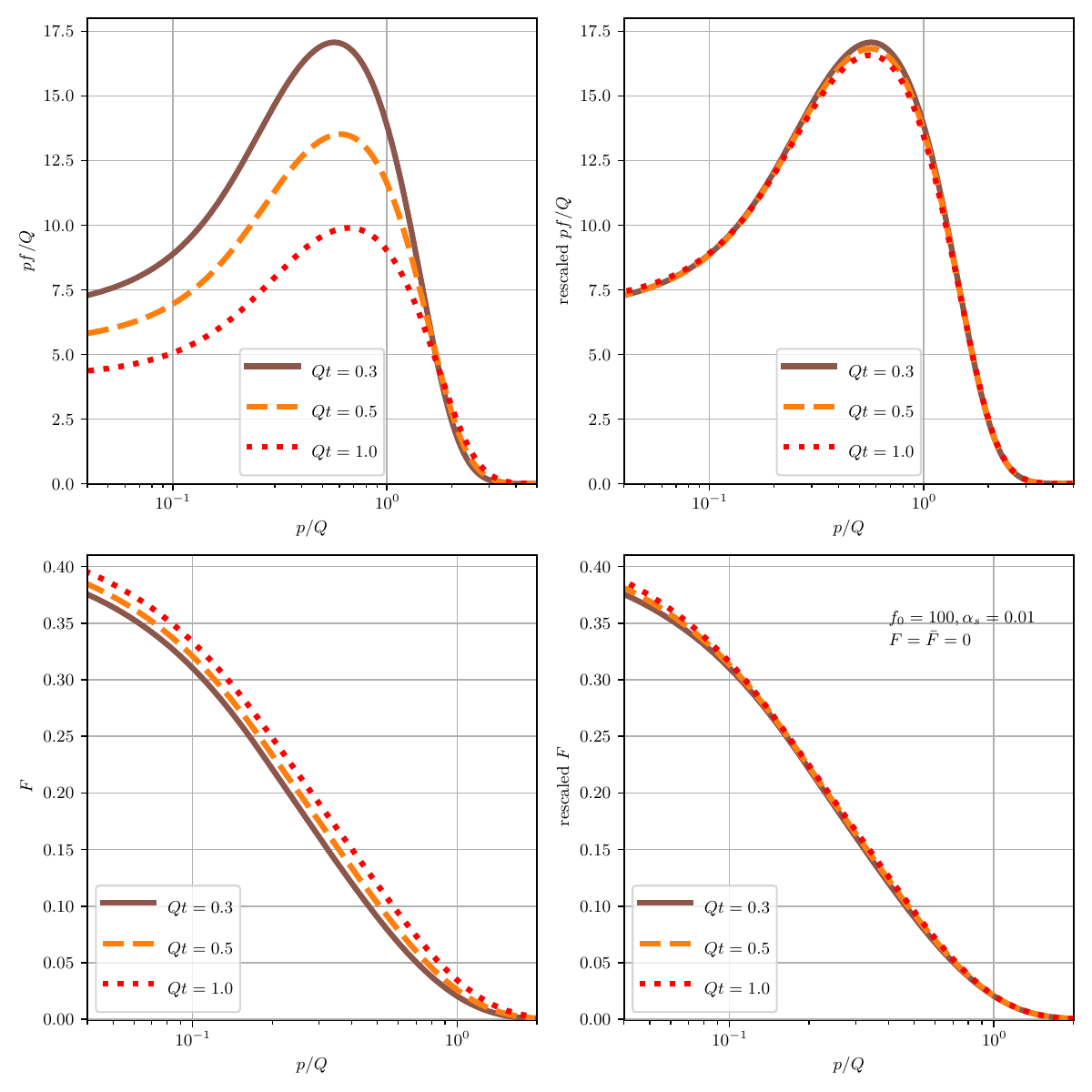}
    \caption{Self-similar scaling behaviors of the parton distributions for $f_0=100$ and $\alpha_s=0.01$. The left panels show the gluon (upper) and quark (lower) distributions at three different times in the range of $t$ where all the macroscopic quantities in fig.~\ref{fig:f0_100_macs} follow the scaling laws.  The right panels compare the distributions at $Qt=0.3$ and the others rescaled according to eq.~(\ref{eq:selfSimilar}).}
    \label{fig:f0_100_pf_selfSimilar}
\end{figure}

In our parametric estimates, it suffices to make the approximation: $pf\sim T_*$ and $F\sim 1/2$ up to the typical momentum $\bar{p} \equiv  (\alpha_s f_0)^{\frac{2}{7}}(Qt)^{\frac{1}{7}} Q$ in Stage 2. Keeping only the range of $p\lesssim \bar{p}$ yields
\begin{align}\label{eq:selfSimilarPara}
    {(Qt)^{-\frac{3}{7}}} D_s(p/\bar{p})\equiv pf/Q \sim T_*\sim \frac{(\alpha_s f_0)^{\frac{1}{7}}}{\alpha_s}{(Qt)^{-\frac{3}{7}}},\qquad F_s(p/\bar{p}) \equiv F\sim 1/2.
\end{align}
 And solving exactly the BEDA only fills in quantitative details that are not parametrically more important than the above parametric form. Consistent with eq.~(\ref{eq:selfSimilarPara}), keeping only the dominant terms in the collision kernels under the assumption $f\gg1$  yields the following self-similar solutions
 \begin{align}\label{eq:selfSimilar}
     f= {(Qt)^{-\frac{3}{7}}} \frac{Q}{p}D_s(p/\bar{p})\equiv {(Qt)^{-\frac{4}{7}}} f_s(\tilde{p}),\qquad F=F_s(\tilde{p})
 \end{align}
 with $\tilde{p}\equiv(Qt)^{-\frac{1}{7}}p/Q$. Here, $f_s$ is the same as the pure gluon case, which can be obtained based on the same argument as that in refs.~\cite{Blaizot:2011xf, Kurkela:2012hp, Schlichting:2019abc}. To justify the self-similar form of $F$, let us assume the ansatz $F=(Qt)^{\alpha}F_s((Qt)^{\beta} p/Q)$ and take the $1\leftrightarrow2$ kernel as a start. Keeping only the dominant term from the $g\to q\bar{q}$ splitting, one has $ C^{q}_{1\leftrightarrow2}=(Qt)^{-1}\tilde{C}^{q}(\tilde{p})$ with $\tilde{C}^{q}$ only dependent of $\tilde{p}$. Then, matching it to the time derivative of $F$ yields $\alpha=0$ and $\beta=-1/7$. One can easily see that including the $2\leftrightarrow2$ kernel does not change these two exponents. As illustrated in fig.~\ref{fig:f0_100_pf_selfSimilar}, our numerical solutions are indeed consistent with the self-similar solutions for some range of $t$ where all the quantities shown in fig.~(\ref{fig:f0_100_macs}) approach the expected scaling behaviors.

\begin{figure}
    \centering
    \includegraphics[width=0.48\textwidth]{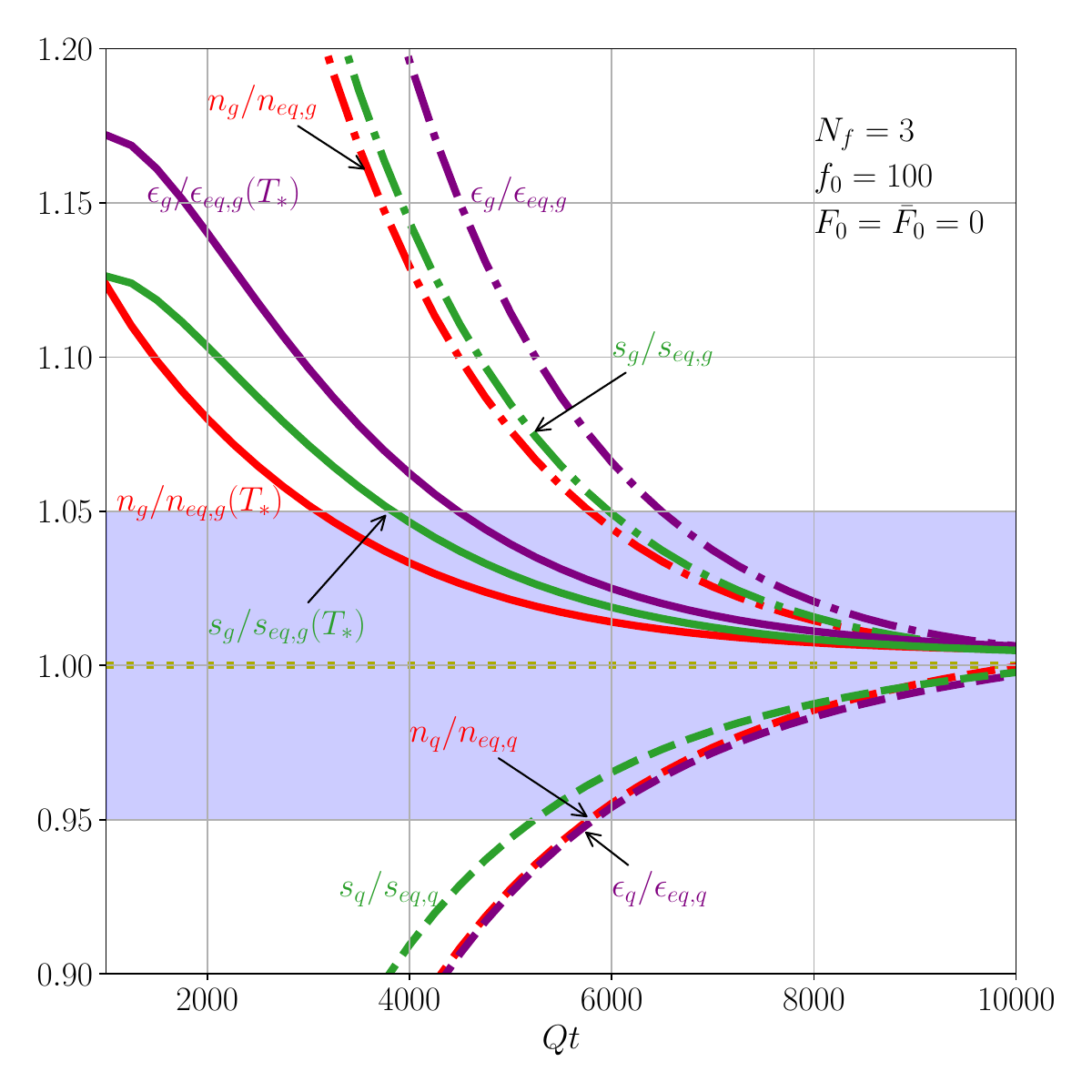}    \includegraphics[width=0.48\textwidth]{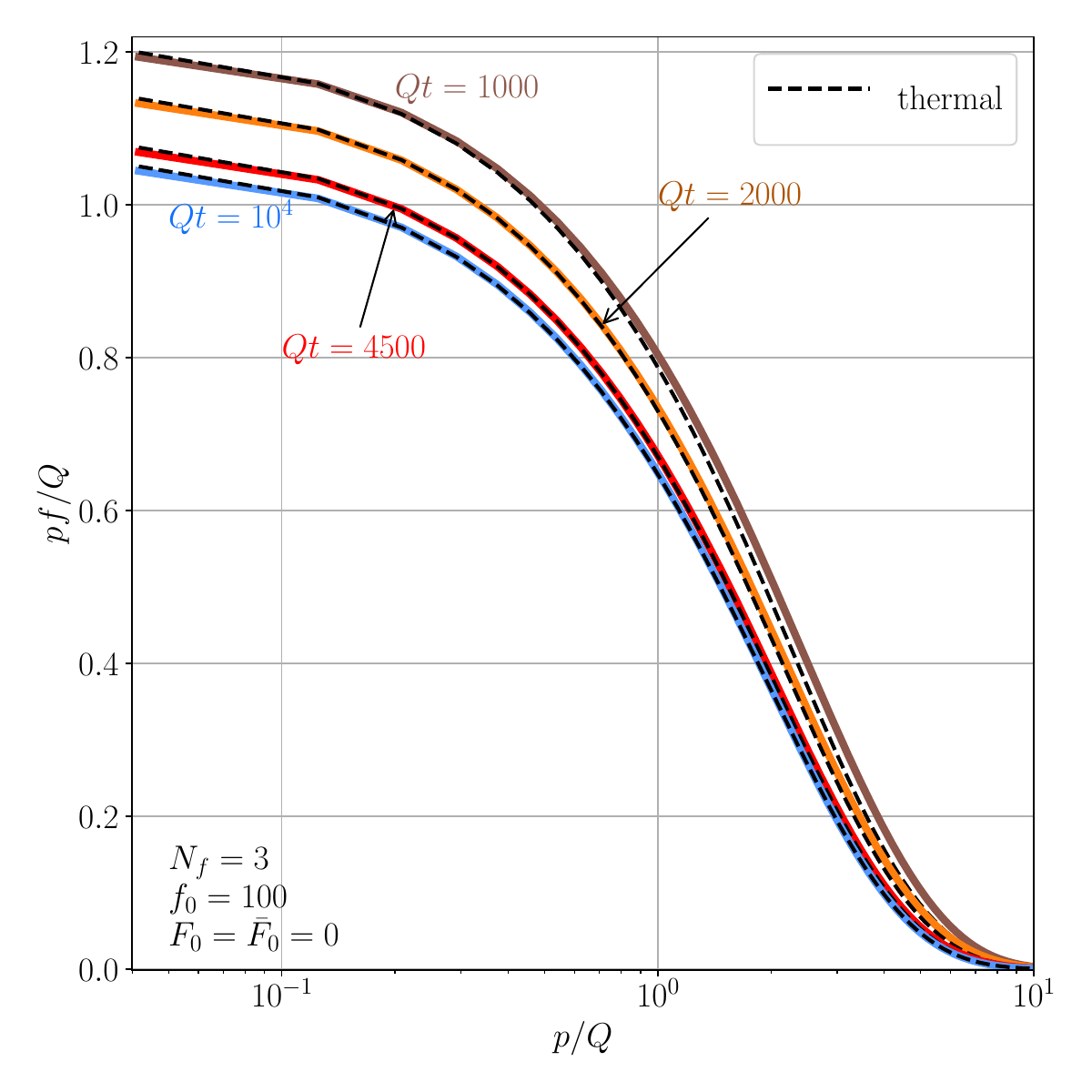}
    \caption{The top-down thermalization of gluons for $f_0=100$ and $\alpha_s=0.01$. The left panel shows how the same set of quantities as those in the lower left panel of fig.~\ref{fig:f0_0p01_thermalization} approach unity. The right panel shows the gluon distribution at different times, comparing to the Bose-Einstein distribution (thermal) with a time-dependent temperature $T_*$.}
    \label{fig:f0_100_thermalization}
\end{figure}

 After exiting the scaling phase,  the pure gluon system equilibrates around $t^{N_f=0}_{eq}\approx 634 Q^{-1}$ for $N_f=0$. For $N_f=3$, gluons undergo the top-down thermalization before achieving full thermalization. This is qualitatively similar to all the cases studied previously for $f_0\geq 10^{-4}$. As shown in the left panel of fig.~\ref{fig:f0_100_thermalization}, gluons first equilibrate around $t_{eq, g}=4479 Q^{-1}$. Even before this time, the gluon distribution can be very well fitted by the Bose-Einstein distribution with $T_*$ up to some maximum momentum higher than $T_*$, as exemplified in the right panel of fig.~\ref{fig:f0_100_thermalization}. Eventually, all the partons equilibrate around $t_{eq} = 6491Q^{-1}$, which is about 10.2 times that in the pure gluon case.

\section{Conclusions}

We have introduced the Boltzmann Equation in Diffusion Approximation (BEDA) to study the thermalization of a system of massless quarks and gluons with both $2\leftrightarrow 2$ and $1\leftrightarrow 2$ kernels. First, we have incorporated the $1\leftrightarrow2$ kernel using the LPM splitting rates~\cite{Baier:1996kr, Zakharov:1996fv, Baier:1998kq, Arnold:2008zu} into the BEDA while only the $2\leftrightarrow2$ kernel for gluons and $N_f$ flavors of massless quarks and antiquarks were included in ref.~\cite{Blaizot:2014jna}. In this way, we have assembled a full set of BEDA as an extension of that for pure gluon systems in ref.~\cite{Baier:2000sb}. Its collision kernels contain the following space-time dependent quantities, calculated from the phase-space distribution functions of partons: the jet quenching parameter $\hat{q}$, the effective temperature $T_*\propto \hat{q}/(\alpha_s m_D^2)$ with $m_D^2$ the screening mass squared, and two more for each quark flavor $i$, denoted by $\mathcal{I}^i_c$ and $\mathcal{\bar{I}}^i_c$ in eq.~(\ref{eq:IcIcb}). In terms of the latter two quantities, a definition of the effective net quark chemical potential $\mu_*^i$ has been given for flavor $i$ in eq.~(\ref{eq:mus}), which equals the net quark chemical potential after thermal equilibration.

Then, we have studied spatially homogeneous systems of gluons by solving the BEDA. To do so, we have first studied the low momentum limit of the BEDA exploring the possibility of the formation of Bose-Einstein Condensate. Similar to the observations in refs.~\cite{Blaizot:2016iir, BarreraCabodevila:2022jhi}, we find that both the gluon and quark distributions in the limit $p\to0$ are identical to those in thermal equilibrium with the temperature and the net quark chemical potential given by $T_*$ and $\mu_*^i$, respectively. As a result, the gluon flux at $p=0$, which is proportional to $(\lim\limits_{p\to0} p f^g - T_*)$ with $f^g$ the gluon distribution~\cite{Blaizot:2014jna}, always vanishes and, hence, no gluons are deposited into a BEC. 
 
After establishing the framework, we have applied it to investigate the thermalization process and quark production. First, we have focused on the initially underpopulated scenario with the initial occupation number $f_0 \ll 1$. Exactly like the pure gluon case~\cite{Schlichting:2019abc, BarreraCabodevila:2022jhi}, the system thermalizes through three stages for $N_f=3$. And we find that quark production does not modify the parametric forms of $\hat{q}$ and $m_D^2$ during each of these stages. In Stage 1, both $\hat{q}$ and $m_D^2$ are dominated by hard gluons. And the soft sector is filled with gluons, quarks and antiquarks according to the thermal distributions with an overheated temperature $T_*\sim \,Q$ and $\mu_*^i=0$ up to $p\sim p_s$. Here, $p_s$ is the typical soft momentum much smaller than $Q$. During this stage, most of quarks and antiquarks are hard and their number density grows linearly over time. And the contributions from soft gluons, quarks and antiquarks to $\hat{q}$ and $m_D^2$, albeit increasing over time, are all parametrically smaller than those from hard gluons. In Stage 2, $m_D^2$ receives dominant contributions from soft gluons and keeps increasing over time while $\hat{q}$ is still dominated by hard gluons, leading to a declining $T_*$. Around the moment when $T_*$ cools down to the thermal equilibrium temperature $T\sim f_0^{1/4}Q$, soft quarks and antiquarks become more abundant than hard ones. Later on, $T_*$ keeps cooling while the number density of quarks and antiquarks $\sim (\hat{q}t)^{3/2}$ grows more rapidly than that of soft gluons $\sim T_* \hat{q}t$. This stage ends when there are parametrically equal numbers of soft gluons, hard gluons, quarks and antiquarks. Moreover, the soft partons have enough time to thermalize into a quark-gluon plasma (QGP) with an overcooled temperature $T_*\sim f_0^{1/3}Q$. In Stage 3, both $\hat{q}$ and $m_D^2$ are predominantly determined by soft partons in the QGP with its temperature $T_*$ increasing over time. The system achieves final equilibration after hard partons are fully quenched, causing $T_*$ to be reheated up to $T$. That is, the system in this stage undergoes the bottom-up thermalization, qualitatively similar to the cases studied in refs.~\cite{Baier:2000sb, Kurkela:2014tea, Kurkela:2015qoa, Kurkela:2018oqw, Kurkela:2018xxd, Du:2020zqg, Du:2020dvp, BarreraCabodevila:2022jhi}.

We have also carried out parametric estimates for initially over-populated systems with $f_0\gg 1$. We have found that the two stages preceding thermalization in the pure gluon case~\cite{Schlichting:2019abc, BarreraCabodevila:2022jhi} are not affected by the production of quarks and antiquarks. And their contributions to $\hat{q}$ and $m_D^2$ are always parametrically smaller than those from gluons before the system achieves thermal equilibration. To sum up, the system during Stage 1 shares some similarities with that in the limit $f_0\ll 1$: $\hat{q}$ and $m_D^2$ are predominantly determined by hard gluons. Accordingly, the soft sector of the system looks the same as that in thermal equilibrium with an overheated temperature $T_*\sim f_0 Q$. And the number density of quarks and antiquarks  grows linearly over time. This stage ends when the typical momentum of soft partons is pushed up to $\sim Q$ by multiple elastic scatterings. In Stage 2, the typical momentum of partons continuously grows over time driven by multiple elastic scatterings. Since there are still parametrically much less quarks and antiquarks compared to gluons, the number density of gluons decreases as mandated by energy conservation. And all the quantities in the system including the number density of quarks and antiquarks evolve according to a set of scaling laws, manifest as a universal {scaling} %non-thermal fix-point 
solution previously discovered in both classical statistical field simulations~\cite{Berges:2012ev, Schlichting:2012es, Kurkela:2012hp} and kinetic theory~\cite{Blaizot:2011xf, AbraaoYork:2014hbk, Kurkela:2014tea, Du:2020dvp, BarreraCabodevila:2022jhi}. Full thermalization is established when all the quantities exit their scaling regions and approach their corresponding thermal equilibrium values.

In order to complement our parametric descriptions, we have also presented quantitative results for several scenarios with different initial occupancies: $f_0=10^{-6},\, 10^{-4},\, 10^{-2},\, 1,\, 10$ and $100$. Here, we have shown several examples in which the thermalization time becomes longer when quarks and antiquarks are allowed to participate in the system's evolution. Nevertheless, this is true when one compares directly thermalization times between systems with identical initial conditions. In phenomenological studies, one might be more interested in looking at systems with the same final configuration, that is, the same temperature, viscosity, etc. In this case, one can naively re-scale the time variable with the equilibrium temperature of the system $T$, such that $t_{th} \rightarrow t_{th} T$. The value of $T$ can be computed using eq.~\eqref{eq:temperature_eq}, yielding $T(N_f=0) \approx 1.31 T(N_f=3)$. Therefore, this is irrelevant for over-occupied systems, where the thermalization time varies around one order of magnitude. For the under-occupied scenario, this is not relevant either, since this is the case where the thermalization times for $N_f=0$ and $N_f=3$ are really close and the scaling factor of $T$ barely changes the result.

Another more complex option has been used in previous works~\cite{Kurkela:2015qoa, Kurkela:2018oqw, Kurkela:2018xxd, Du:2020zqg, Du:2020dvp} with the main purpose of studying the isotropization of the system. In this case, the re-scaling variable is $t \rightarrow \frac{t T}{4 \pi \eta / s}$. Here, the entropy density $s$ can be extracted from eq.~\eqref{eq:nseq} and the values of the viscosity $\eta$ can be calculated for the scenarios that we have studied according to ref.~\cite{Arnold:2000dr}. In this way, we observe no significant reduction of the thermalization time for over-populated systems. However, this contribution became more important when the initial system has a small population, where the thermalization time of the scenarios approach faster with the re-scaled time than with our previous set-up.

One universal feature for the cases with $f_0\geq 10^{-4}$ that could not be revealed by the above parametric estimates is as follows. Before achieving full thermalization, gluons undergo {\it top-down thermalization}, meaning that they first equilibrate at temperature $T_*$ higher than $T$, and then the thermal bath of gluons cools down to $T$ via generating quarks and antiquarks. The whole thermalization process is hence delayed compared to the pure gluon cases. Among the aforementioned values of $f_0$, the reheating stage is only observed for $f_0=10^{-6}$ while the self-similarity in both gluon and quark distributions and the scaling laws are only discernible for $f_0=100$. These observations qualitatively agree with those using EKT in refs.~\cite{Kurkela:2018oqw, Kurkela:2018xxd, Du:2020zqg, Du:2020dvp}, Thus, the BEDA provides us a framework which retains the essential physics of EKT while offering two advantages owing to its simplicity. Firstly, it is more cost-effective to implement numerically, enabling exploration of more realistic geometries. Secondly, the BEDA is easier to handle for analytical estimations. A quantitative comparison between the BEDA and EKT will be presented in an forthcoming publication.

\acknowledgments

We thank Jean-Paul Blaizot and Xiaojian Du for useful  discussions. This work is supported by European Research Council project ERC-2018-ADG-835105 YoctoLHC; by Spanish Research State Agency under project PID2020-119632GB- I00; and by Xunta de Galicia (Centro singular de investigaci\'{o}n de Galicia accreditation 2019-2022), by European Union ERDF.  B.W. acknowledges the support of the Ram\'{o}n y Cajal program with the Grant No. RYC2021-032271-I and the support of Xunta de Galicia under the ED431F 2023/10 project. S.B.C. acknowledges the support of the Axudas de apoio \' {a} etapa predoutoral program (Ref. ED481A 2022/279).

\appendix

\section{The $1\to2$ splitting rates in the deep LPM regime}
\label{sec:1to2}
The splitting rates for $a\to bc$ in the deep LPM regime can be read out from ref.~\cite{Arnold:2008zu}:
\begin{align}
    \frac{dI_{a\to bc}(p)}{dx dt}=\frac{(2\pi)^3}{\nu_a p} \gamma_{a\leftrightarrow bc}(p;xp, (1-x)p)
\end{align}
where
\begin {subequations}
\begin {align}
   \gamma_{g\leftrightarrow gg}(p; xp, (1-x)p)
   &= \frac{d_A C_A \alpha_s}{(2\pi)^4 }
     \, m_D^2 \, \hat\mu_\perp^2(1,x,1{-}x;{\rm A},{\rm A},{\rm A}) \,
     \frac{1+x^4+(1-x)^4}{x^2(1-x)^2} \,,
\\
   \gamma_{q\leftrightarrow gq}(p; xp, (1-x)p)
   &= \frac{d_F C_F \alpha_s}{(2\pi)^4 }
     \, m_D^2 \, \hat\mu_\perp^2(1,x,1{-}x;{\rm F},{\rm A},{\rm F}) \,
     \frac{1+(1-x)^2}{x^2(1-x)} \,,
\\
   \gamma_{g\leftrightarrow q\bar q}(p; xp, (1-x)p)
   &= \frac{d_F C_F \alpha_s}{(2\pi)^4}
     \, m_D^2 \, \hat\mu_\perp^2(1,x,1{-}x;{\rm A},{\rm F},{\rm F}) \,
     \frac{x^2+(1-x)^2}{x(1-x)}
\end {align}
\end {subequations}
with $d_F=N_c$ and $d_A=N_c^2-1$, and
\begin {align}
  \hat\mu_\perp^2(x_1,x_2,x_3;s_1,s_2,s_3)
  =&
  \frac{gT}{m_D}
  \left[
    \frac{1}{\pi} \,
    x_1 x_2 x_3 \,
    \frac{p}{T}\mathcal{L}
  \right]^{1/2}
    \Biggl[
       \tfrac12 (C_{s_2}+C_{s_3}-C_{s_1}) x_1^2\notag\\
       &+ \tfrac12 (C_{s_3}+C_{s_1}-C_{s_2}) x_2^2
       + \tfrac12 (C_{s_1}+C_{s_2}-C_{s_3}) x_3^2
    \Biggr]^{1/2} .
\label {eq:muLL}
\end {align}
Eliminating $m_D^2$ and $T$ in terms of $\hat{q} = \alpha_s \mathcal{L} m_D^2 C_A T$ in the above equations yields the splitting rates in eq.~(\ref{eq:splitting_rates}). And one can easily check that the rates for $g\to gg$ and $q\to gq$ agree with those in ref.~\cite{Baier:1998kq} in the deep LPM regime. Note that these splitting rates are derived for a static medium. In a generic medium, they remain applicable when the formation time of the splitting is shorter than the typical time scale during which $\hat{q}$ undergoes substantial changes (see \cite{Iancu:2018trm} for a detailed discussion in longitudinally boost-invariant systems).

\section{Numerical method}
\label{sec:numerics}
In the spatially homogeneous case, eq. \eqref{eq:Boltzmann} reduces to 
\begin{align}
    \partial_t f^a =&C^a_{2\leftrightarrow2} + C^a_{1\leftrightarrow2}.
\end{align}
In order to solve it numerically, one can use the usual Euler integration method. That is, given $f^a$ at $t$, the distribution at the next time step $t + \Delta t$ is given by
\begin{align}
    f^a(t+\Delta t) = f^a(t) + \Delta t \lbrace C^a_{2\leftrightarrow2}[f(t)] + C^a_{1\leftrightarrow2}[f(t)] \rbrace.
\end{align}
And the discretization of the $2 \leftrightarrow 2$ and $1 \leftrightarrow 2$ kernels in our code is as follows.

\subsection{Discretization of the $2 \leftrightarrow 2$ kernel}

As shown in eq.~(\ref{eq:C2to2}),  in diffusion approximation the $2 \leftrightarrow 2$ kernel can be expressed as that in the Fokker-Planck equation plus some source term written in eq. \eqref{eq:drag_kern}. Here, one only needs to calculate derivatives and a few integrals including $\hat{q}$, $m_D^2$, $\mathcal{I}_c$ and $\bar{\mathcal{I}}_c$, which can be easily computed with, e.g., the trapezoidal rule. In order to compute the derivatives, we define two different momentum grids: $p_s$ and $p$ with the $i$-th point $p_i = (p_{s,i} + p_{s,i-1}) / 2$ and $i=0, 1,\cdots N$. Note $p_{s,-1}=0$ is not include in the grid $p_s$. The distribution function $f$ is defined over the grid $p$ while $J$, over the grid $p_s$. The value of the function distribution on the grid $p_s$, denoted by $f_s$, is computed via a linear interpolation of $f$ on $p$. For example, this definition allows one to compute the first order derivative at the $i$-th point of the $p_s$ grid as
\begin{align}
    \frac{\partial f}{\partial p} \bigg |_{p_{s,i}} = \frac{f_{i+1} - f_i}{p_{i+1}-p_i}
\end{align}
for a uniform grid. Then, the current of the Fokker-Planck term, $p^2J$, is easily computed on the $p_s$ grid. In terms of $J$ on $p_s$ and $\mathcal{S}$ on $p$, one has
\begin{align}
    C^a_{2\leftrightarrow2, i}= \frac{p_{s,i}^2 J_{s, i}-p_{s,i-1}^2 J_{s, i-1}}{\Delta V_i} + \mathcal{S}_i
\end{align}
with $\Delta V_i = \frac{1}{3}(p_{s,i}^3-p_{s,i-1}^3)$.

One also needs to specify two boundary conditions:
\begin{align}
    p_{s,-1}^2 J_{s, -1} = 0 , \qquad p_{s, N}^2 J_{s,N} = 0
\end{align}
with $p_{s,-1}=0$. In initially over-populated cases, the first term can be set without conflict only if the inelastic kernel is included. In this case, the distributions of both quarks and gluons approach an equilibrium profile with temperature $T_*$ at small $p$, as shown in eq.~(\ref{eq:fslowp}). The second boundary condition just approximates the restriction that there can not be particle flux at $p=\infty$.

\subsection{Discretization of the $1 \leftrightarrow 2$ kernel}

Evaluation of this kernel consumes most of the computation time since it involves an integral of the form of eq. \eqref{eq:inel_kern} for each momentum grid element $p_i$. That is, we need to compute
\begin{align}
    C^{1 \leftrightarrow 2} (p_i) = \int_0^1 dx \mathcal{K} (p_i, x), 
\end{align}
where $\mathcal{K} (p_i, x)$ includes the collinear splitting functions and the statistical factors given in eq.~\eqref{eq:inel_kern3}. In our code, the $x$-integration is carried out by using the trapezoidal rule.

Computing the above integral involves evaluating the distribution functions at momentum $p$ different from the momentum grid points. To handle this, we employ the third order Lagrange interpolation (for $pf$ and $F/\bar{F}$) by using four adjacent grid points: two before and two after $p$ if it does not lie between the first or the last two grid points. Otherwise, the linear interpolation is used.

The evaluation of $C^{1 \leftrightarrow 2} (p)$ at each time step is amenable to parallel computation. We implement this calculation using both the GPU and CPU programming. Given our current computational resources, the GPU code is found to be about one order of magnitude faster than the CPU code for the same set of initial conditions and parameters.

\bibliographystyle{JHEP}
\bibliography{bulk.bib}

\end{document}